\documentclass[nopreprintline,11pt]{elsarticle}
\pdfoutput=1
\usepackage{geometry}
\usepackage{amsfonts}
\usepackage{amsmath}
\usepackage{amssymb}
\usepackage{amsthm}
\usepackage[english]{babel}
\usepackage{enumerate}
\usepackage[svgnames]{xcolor}
\usepackage{graphicx}
\usepackage[hidelinks]{hyperref}

\usepackage{tikz}
\IfFileExists{SkipFigure}
{\usetikzlibrary{calc,angles,arrows}}
{\usetikzlibrary{calc,angles,quotes,arrows}}

\usepackage{ifthen} 
\newcommand{\optionalargumentinbrackets}[1]{\ifthenelse{\equal {#1} {} }{}{~(#1)}}

\newcounter{theoremcnt}[section]
\renewcommand{\thetheoremcnt}{\thesection.\arabic{theoremcnt}}

\newenvironment{corollary}[1][]%
{\begin{trivlist}\refstepcounter{theoremcnt}
\item[]\textbf{Corollary~\thetheoremcnt \optionalargumentinbrackets{#1}.} }{\end{trivlist}}

\newenvironment{definition}[1][]
{\begin{trivlist}\refstepcounter{theoremcnt}
\item[]\textbf{Definition~\thetheoremcnt \optionalargumentinbrackets{#1}.} }{\end{trivlist}}

\newenvironment{example}[1][]%
{\begin{trivlist}\refstepcounter{theoremcnt}
\item[]\textbf{Example~\thetheoremcnt \optionalargumentinbrackets{#1}.} }{\end{trivlist}}

\newenvironment{lemma}[1][]%
{\begin{trivlist}\refstepcounter{theoremcnt}
\item[]\textbf{Lemma~\thetheoremcnt \optionalargumentinbrackets{#1}.} }{\end{trivlist}}

\newenvironment{theorem}[1][]%
{\begin{trivlist}\refstepcounter{theoremcnt}
\item[]\textbf{Theorem~\thetheoremcnt \optionalargumentinbrackets{#1}.} }{\end{trivlist}}

\makeatletter
\newcommand{\xRightarrow}[2][]{\ext@arrow
  0359\Rightarrowfill@{#1}{#2}}
\makeatother

\newcommand{\Arrow}[1]{\xRightarrow{\, {#1} \; \,}}
\newcommand{\AXd}{\mathit{A \mkern-2mu X}}
\newcommand{\AXdb}{\mathit{A \mkern-2mu X}^b}
\newcommand{\AXp}{\mathit{A \mkern-2mu X}_{\mkern-3mu p}}
\newcommand{\AXpI}{\mathit{A \mkern-2mu X}^{I}_{\mkern-3mu p}}
\newcommand{\AXpb}{\mathit{A \mkern-2mu X}^b_{\mkern-3mu p}}
\newcommand{\DistrE}{\Distr(\calE)}
\newcommand{\DistrEconc}{\Distr(\calE_{\mathit{cc}})}
\newcommand{\DistrX}{\Distr(X)}

\newcommand{\alphaarrow}{\arrow{\alpha}}

\newcommand{\alphahidearrow}{\arrow{(\alpha)}}
\newcommand{\arrow}[1]{\mkern1mu \xrightarrow{\raisebox{0ex}[0.1ex][-0.1ex]{\scriptsize $\,#1\,$}}}
\newcommand{\astop}{a \pref \partial(\bfzero)}

\newcommand{\bfzero}{\normalfont{\textbf{0}}}

\newcommand{\bigoplusiinI}{\textstyle{\bigoplus_{i \mathord{\in} I}}}
\newcommand{\bigoplusjinJ}{\textstyle{\bigoplus_{j \mathord{\in} J}}}
\newcommand{\bigoplusjinJi}{\textstyle{\bigoplus_{j \mathord{\in} J_i}}}

\newcommand{\bigopluskinK}{\textstyle{\bigoplus_{k \mathord{\in} K}}}

\newcommand{\bisim}{\mathrel{\,%
  \raisebox{.3ex}{$\underline{\makebox[.7em]{$\leftrightarrow$}}$}\,}}
\newcommand{\bnfeq}{\mathrel{{:}{:}{=}}}
\newcommand{\brbisim}{\mathrel{\,%
  \raisebox{.3ex}{$\underline{\makebox[.8em]{$\leftrightarrow$}}_{\,
      b}$}\mkern-1mu}}
\newcommand{\bstop}{b \pref \partial(\bfzero)}
\newcommand{\ccR}{\ensuremath{\mathrel{\mathit{cc}(\calR)}}}
\newcommand{\cstop}{c \pref \partial(\bfzero)}
  
\newcommand{\der}{\textsl{der} \mkern1mu}  

\newcommand{\half}{\textstyle{\frac12}}
\newcommand{\mydot}{{\cdot}}
\newcommand{\myeq}{{} \, = \:}

\newcommand{\lc}{\lbrace \:}
\newcommand{\mopcalR}{\,{\calR}\,}   
\newcommand{\mycdot}{\mathop{\cdot}}
\newcommand{\myfrac}[2]{\textstyle{\frac{#1}{#2}}}
\newcommand{\myoh}{\textrm{o}}
\newcommand{\mytinyfrac}[2]{\mbox{\tiny $\textstyle{\frac{#1}{#2}}$}}

\newcommand{\nbrbisim}{\mkern6mu {\not\mkern-2.3mu\brbisim} \mkern4mu}

\newcommand{\oplusr}{\oplus_r}
\newcommand{\pair}[2]{\langle {#1}, {#2} \rangle}

\newcommand{\pref}{\mathop{.}}

\newcommand{\rc}{\: \rbrace}

\newcommand{\rijE}{r_{i,j}^E}
\newcommand{\rootedbrbisim}{\mathrel{\,%
  \raisebox{.3ex}{$\underline{\makebox[.7em]{$\leftrightarrow$}}_{\,
      rb}$}\mkern-1mu}}
\newcommand{\singleton}[1]{\lbrace {#1} \rbrace}
  
\newcommand{\spt}{\mathrm{spt}}
\newcommand{\sosrule}[2]{%
  \def\arraystretch{0.50}
  \begin{array}{c} {#1}\rule[-4pt]{0pt}{11pt} \\ \hline
    \rule{0pt}{13pt}{#2} \end{array}
  \def\arraystretch{1.0}}
\newcommand{\sumiinI}{{\textstyle\sum_{i {\in} I}}}
\newcommand{\sumjinJ}{{\textstyle\sum_{j {\in} J}}}
\newcommand{\sumkinK}{\textstyle\sum_{k {\in} K}}

\newcommand{\tauarrow}{\arrow{\tau}}
\newcommand{\tauhidearrow}{\arrow{(\tau)}}

\newcommand{\Ebar}{\bar{E}}
\newcommand{\Ehat}{\hat{E}}
\newcommand{\Fbar}{\bar{F}}
\newcommand{\Fhat}{\hat{F}}
\newcommand{\Ftilde}{\tilde{F}}
\newcommand{\Gbar}{\bar{G}}
\newcommand{\Pbar}{\bar{P}}
\newcommand{\Qbar}{\bar{Q}}
\newcommand{\etabar}{\bar{\eta}}

\newcommand{\mubar}{\bar{\mu}}
\newcommand{\muhat}{\hat{\mu}}
\newcommand{\nubar}{\bar{\nu}}

\newcommand{\varrhobar}{\bar{\varrho}}
\newcommand{\varrhohat}{\hat{\varrho}}

\newcommand{\Distr}{\textsl{Distr}\mkern1mu}

\newcommand{\calA}{\mathcal{A}}

\newcommand{\calE}{\mathcal{E}}

\newcommand{\calEconc}{\mathcal{E}_{\mathit{cc}}}
\newcommand{\calErigid}{\mathcal{E}_{\hspace{.7pt}r}}

\newcommand{\calP}{\mathcal{P}}
\newcommand{\calPconc}{\mathcal{P}_{\!\mathit{cc}}}
\newcommand{\calR}{\mathcal{R}}
\newcommand{\calS}{\mathcal{S}}

\newcommand{\blankline}{\vspace*{1.0\baselineskip}}
\newcommand{\halflineup}{\vspace*{-0.5\baselineskip}}

\colorlet{colorA}{LightYellow}
\colorlet{colorB}{Pink}
\colorlet{colorC}{LightBlue}
\colorlet{colorD}{Plum}
\colorlet{colorE}{Gold!50}
\colorlet{colorF}{Orange!50}
\colorlet{colorG}{Red!50}
\colorlet{colorH}{Green!50}
\colorlet{colorI}{NavyBlue!50}
\colorlet{colorJ}{Brown!50}
\colorlet{colorK}{Coral}
\colorlet{colorL}{RoyalBlue!50}
\colorlet{colorM}{OliveDrab!50}
\colorlet{colorX}{Gray!50}

\newcommand{\bisimilarity}{bisimilarity}
\newcommand{\probc}[1]{\mathbin{\mbox{${}_{\scriptscriptstyle #1 \mkern1mu} \hspace{-.8pt}\oplus\,$}}}
\renewcommand{\oplusr}{\probc{r}}
\newcommand{\den}[1]{\mbox{$[\hspace{-1.6pt}[$}#1\mbox{$]\hspace{-1.6pt}]$}}  
\newcommand{\plat}[1]{\raisebox{0pt}[0pt][0pt]{#1}}
\newfont{\bbb}{bbm10 scaled 1100}                     
\newcommand{\IN}{\mbox{\bbb N}}                       
\newcommand{\IR}{\mbox{\bbb R}}                       
\newcommand{\stardot}{\mycdot}

\definecolor[named]{darkorange}{cmyk}{0,0.82,1,0.01}
\definecolor[named]{amethyst}{rgb}{0.6, 0.4, 0.8}
\definecolor[named]{kellygreen}{rgb}{0.3, 0.73, 0.09}

\newcommand{\dg}[1]{\textcolor{kellygreen}{#1}}
\newcommand{\dor}[1]{\textcolor{darkorange}{#1}}
\newcommand{\mg}[1]{\textcolor{amethyst}{#1}}

\title{A Complete Axiomatisation of \texorpdfstring{\\}{}
  Distribution-Based Branching Bisimilarity\texorpdfstring{\\}{}
  for Non-Deterministic and Probabilistic Choice}
 
\author{R.J.~van Glabbeek\texorpdfstring{\corref{support}}{}}
\ead{rvg@cs.stanford.edu}
\address{School of Informatics, University of Edinburgh, UK \\
  School of Computer Science and
  Engineering, University of New South Wales, Australia}

\author{J.F.~Groote}
\ead{J.F.Groote@tue.nl}
\address{Department of Mathematics and Computer Science,
  Eindhoven University of Technology, The Netherlands}

\author{E.P.~de Vink
  \texorpdfstring{\corref{cor1}}{}%
}
\ead{evink@win.tue.nl}
\address{Department of Mathematics and Computer Science,
  Eindhoven University of Technology, The Netherlands}

\cortext[support]{Supported by Royal Society Wolfson Fellowship RSWF\textbackslash R1\textbackslash 221008.}
\cortext[cor1]{Corresponding author.}

\date{}

\begin{document}

\begin{abstract}
  \noindent
  This paper proposes a notion of branching {\bisimilarity} for
  non-deterministic probabilistic processes. In order to
  characterise the corresponding notion of rooted branching
  probabilistic {\bisimilarity}, an equational theory is proposed for
  a basic, recursion-free process language with non-deterministic as
  well as probabilistic choice. The proof of completeness of the
  axiomatisation builds on the completeness of strong probabilistic
  {\bisimilarity} on the one hand and on the notion of a concrete
  process, i.e.\ a process that does not display (partially)
  inert $\tau$-moves, on the other hand. The approach is first
  presented for the non-deterministic fragment of the calculus and
  next generalised to incorporate probabilistic choice, too.
\end{abstract}

\maketitle



\section{Introduction}

\noindent
In~\cite{DP07:tcs}, in a setting of a process language featuring both
non-deterministic and probabilistic choice, Yuxin Deng and Catuscia
Palamidessi propose an equational theory for a notion of weak
{\bisimilarity} and prove its soundness and completeness. Not
surprisingly, the axioms dealing with a silent step are reminiscent to
the well-known $\tau$-laws of Milner~\cite{Mil89:phi,Mil89:ic}. The
process language treated in~\cite{DP07:tcs} includes recursion,
thereby extending the calculus and axiomatisation
of~\cite{BS01:icalp}. While the weak transitions of~\cite{DP07:tcs}
can be characterised as finitary, infinitary semantics is treated
in~\cite{FG19:jlamp}, providing a sound and complete axiomatisation
also building on the seminal work of Milner~\cite{Mil89:ic}.

In this paper we focus on branching {\bisimilarity} in the sense
of~\cite{GW96:jacm}, rather than on weak {\bisimilarity} as
in~\cite{BS01:icalp,DP07:tcs,FG19:jlamp}. 
In the non-probabilistic setting branching {\bisimilarity} has the
advantage over weak {\bisimilarity} that it has substantially more
efficient algorithms~\cite{GJKW17:tcl,GV90:icalp}. Furthermore,
branching bisimilarity has a strong logical
underpinning~\cite{DV95:jacm}. It would be attractive to have these
advantages available also in the probabilistic case, in particular for
the support of model checking which is more demanding in the
probabilistic setting than in the non-deterministic one. See also the
initial work reported in~\cite{GV17:festschrift}.

For a similarly basic process language as in~\cite{DP07:tcs}, without
recursion though, we propose a notion of branching probabilistic
{\bisimilarity} as well as a sound and complete equational
axiomatisation of its rooted version. Hence, instead of lifting all
$\tau$-laws to the probabilistic setting, we need to do this here for
the well-known B-axiom of~\cite{GW96:jacm}, the single axiom capturing
inert silent steps. For what is referred to as the alternating
model~\cite{HJ90:rtss}, branching probabilistic {\bisimilarity} has
been studied
in~\cite{AGT12:tcs,AW06:tcs}. Also~\cite{Seg95:thesis,SL94:concur}
discuss branching probabilistic {\bisimilarity}. However, the
proposed notions of branching {\bisimilarity} are either no congruence
for the parallel operator, or they invalidate the identities below
which we desire. The paper~\cite{AG09:sofsem} proposes a complete
theory for a variant of branching {\bisimilarity} that is not
consistent with the first $\tau$-law unfortunately.

In view of the above, our investigation is led by the wish to identify
the three processes below, involving as a subprocess a probabilistic
choice between probabilistic processes $P$ and~$Q$ with equal
probability for each. Essentially, ignoring the occurrence of the
initial action~$a$, the three processes represent (i)~a probabilistic
choice between two instances of the subprocess mentioned, the one with
probability~$\frac34$, the other with probability~$\frac14$, (ii)~the
subprocess on its own, and (iii)~a probabilistic choice between the
subprocess guarded by a $\tau$-prefix with probability~$\frac13$ and a
rescaling of the subprocess, selected with probability~$\frac23$ for
$P$ and~$Q$ together, thus with probability~$\frac13$ for $P$ and~$Q$
individually.

\bigskip

\centerline{%
\begin{tikzpicture}[> = stealth', semithick, inner sep = 1pt, scale=0.45]
  \tikzstyle{state} = [circle, draw=Black, fill=LightYellow, minimum size=0.0mm]
  \tikzstyle{split} = [circle, draw=black, fill=black]
  \tikzstyle{dtrans} = [>=]
  \tikzstyle{ptrans} = [->, shorten >=0.25mm]
  \node (top) at (2,10) {} ;
  \node (name) at (1.5,9.2) {$\mathbf{G_0}$} ;
  \node [state] (t0) at (2.5,9) {} ; 
  \node [split] (y0) at (2.5,7.25) {} ;
  \node [state] (t1) at (1.0,6) {}; 
  \node [split] (y1) at (1.0,4.25) {}; 
  \coordinate (t3) at (-0.25,3); 
  \coordinate (t4) at (2.0,3); 
  \node [state] (t2) at (4.0,6) {}; 
  \node [split] (y2) at (4.0,4.25) {}; 
  \coordinate (t5) at (3.0,3); 
  \coordinate (t6) at (5.25,3); 
  \path [>=] (t0) edge node [midway, left=1pt] {$a$} (y0);
  \path [->] (y0) edge node [midway, above left] {$\mytinyfrac{3}{4}$} (t1);
  \path [->] (y0) edge node [midway, above right] {$\mytinyfrac{1}{4}$} (t2);
  \draw [>=] (y0) ++(220:.3) arc (217.5:322.5:.3);
  \path [>=] (t1) edge node [midway, left=0.5pt] {$\tau$} (y1);
  \path [->, shorten >=0.25mm] (y1) edge node [midway, above left] {$\mytinyfrac{1}{2}$} (t3);
  \path [->, shorten >=0.25mm] (y1) edge node [midway, above right] {$\mytinyfrac{1}{2}$} (t4);
  \draw [>=] (y1) ++(220:.3) arc (217.5:315:.3);
  \fill [fill=Pink, draw=Black, >=] (t3) -- ++(-1.75,-2.0) -- ++(2.0,0.0)
  -- (t3);  
  \node at ([shift={(-0.35,-1.4)}] t3) {P};
  \fill [fill=LightBlue, draw=Black, >=] (t4) -- ++(-1.75,-2.0) --
  ++(2.0,0.0) -- (t4); 
  \node at ([shift={(-0.45,-1.4)}] t4) {Q};
  \path [>=] (t2) edge node [midway, right=0.5pt] {$\tau$} (y2);
  \path [->, shorten >=0.25mm] (y2) edge node [midway, above left] {$\mytinyfrac{1}{2}$} (t5);
  \path [->, shorten >=0.25mm] (y2) edge node [midway, above right] {$\mytinyfrac{1}{2}$} (t6);
  \draw [>=] (y2) ++(227.5:.3) arc (217.5:320:.3);
  \fill [fill=Pink, draw=Black, >=] (t5) -- ++(-0.25,-2.0) -- ++(2.0,0.0)
  -- (t5);  
  \node at ([shift={(+0.45,-1.4)}] t5) {P};
  \fill [fill=LightBlue, draw=Black, >=] (t6) -- ++(-0.25,-2.0) --
  ++(2.0,0.0) -- (t6); 
  \node at ([shift={(+0.45,-1.4)}] t6) {Q};
  \node (S) at (2.5,-0.25) {\small
    $a \pref 
    \bigl( 
    \partial \bigl( \tau \pref (P \probc{\frac12} Q) \bigr) 
    \probc{\frac34} 
    \partial \bigl( \tau \pref (P \probc{\frac12} Q) \bigr)
    \mkern-2mu \bigr)$} ;
\end{tikzpicture}
\hspace*{-0.75cm}
\begin{tikzpicture}[> = stealth' , semithick, inner sep = 1pt, scale=0.45]
  \tikzstyle{state} = [circle, draw=Black, fill=LightYellow, minimum size=0.0mm]
  \tikzstyle{split} = [circle, draw=black, fill=black]
  \node (top) at (2,10) {} ;
  \node (name) at (1.5,9.2) {$\mathbf{G_1}$} ;
  \node [state] (u0) at (2.5,9) {}; 
  \node [split] (x0) at (2.5,7.25) {};
  \coordinate (u1) at (1.0,6); 
  \coordinate (u2) at (4.0,6); 
  \coordinate (u3) at (4.0,1); 
  \path (u2) -- (u3);
  \path [>=] (u0) edge node [midway, left=1pt] {$a$} (x0);
  \path [->, shorten >=0.25mm] (x0) edge node [midway, above left] {$\mytinyfrac{1}{2}$} (u1);
  \path [->, shorten >=0.25mm] (x0) edge node [midway, above right] {$\mytinyfrac{1}{2}$} (u2);
  \draw [>=] (x0) ++(217.5:.3) arc (217.5:322.5:.3);
  \fill [fill=Pink, draw=Black, >=] (u1) -- ++(-1.0,-2.0) --
  ++(2.0,0.0) -- (u1);  
  \node at ([shift={(0,-1.4)}] u1) {P};
  \fill[fill=LightBlue, draw=Black, >=] (u2) -- ++(-1.0,-2.0) -- ++(2.0,0.0) -- (u2);
  \node at ([shift={(0,-1.4)}] u2) {Q};
  \node (T) at (2.5,2.75) {\small
    $a \pref (P \probc{\mytinyfrac12} Q)$} ; 
  \node (T') at (2.5,-0.5) {} ; 
\end{tikzpicture}
  \hspace*{-0.75cm}
\begin{tikzpicture}[> = stealth', semithick, inner sep = 1pt, scale=0.45]
  \tikzstyle{state} = [circle, draw=Black, fill=LightYellow, minimum size=0.0mm]
  \tikzstyle{split} = [circle, draw=Black, fill=black]
  \node (top) at (2.5,10) {} ;
  \node (name) at (1.5,9.2) {$\mathbf{G_{2}}$} ;
  \node [state] (s0) at (2.5,9) {}; 
  \node [split] (x0) at (2.5,7.25) {};
  \node [state] (s1) at (1.0,6) {}; 
  \node [split] (x1) at (1.0,4.25) {}; 
  \coordinate (s2) at (3,6); 
  \coordinate (s3) at (5.5,6); 
  \coordinate (s4) at (-0.5,3); 
  \coordinate (s5) at (2.5,3); 
  \path [>=] (s0) edge node [midway, left=1pt] {$a$} (x0);
  \path [->] (x0) edge node [midway, above left] {$\mytinyfrac{1}{3}$} (s1);
  \path [->] (x0) edge node [midway, right=2pt] {$\mytinyfrac{1}{3}$} (s2);
  \path [->] (x0) edge node [midway, above right] {$\mytinyfrac{1}{3}$} (s3);
  \draw [>=] (x0) ++(217.5:.3) arc (217.5:340:.3);
  \path [>=] (s1) edge node [midway, left =0.5pt] {$\tau$} (x1);
  \fill [fill=Pink, draw=Black, >=] (s2) -- ++(-0.5,-2.0) -- ++(2.0,0.0) -- (s2); 
  \node at ([shift={(+0.35,-1.4)}] s2) {P};
  \fill [fill=LightBlue, draw=Black, >=] (s3) -- ++(-0.5,-2.0) --
  ++(2.0,0.0) -- (s3); 
  \node at ([shift={(+0.35,-1.4)}] s3) {Q};
  \draw [>=] (x1) ++(217.5:.3) arc (217.5:322.5:.3);
  \path [->] (x1) edge node [midway, above left] {$\mytinyfrac{1}{2}$} (s4);
  \path [->] (x1) edge node [midway, above right] {$\mytinyfrac{1}{2}$} (s5);
  \fill [fill=Pink, draw=Black, >=] (s4) -- ++(-1.0,-2.0) -- ++(2.0,0.0) -- (s4); 
  \node at ([shift={(0,-1.4)}] s4) {P};
  \fill [fill=LightBlue, draw=Black, >=] (s5) -- ++(-1.0,-2.0) --
  ++(2.0,0.0) -- (s5); 
  \node at ([shift={(0,-1.4)}] s5) {Q};
  \node (U) at (2.5,-0.25) {\small
    $a \pref 
    \bigl( 
    \partial \bigl( \tau \pref (P \probc{\frac12} Q) \bigr)
    \probc{\frac13} 
    (P \probc{\frac12} Q)
    \bigr)$} ;
\end{tikzpicture}
}

\bigskip

\noindent
In our view, the three processes starting from $\mathbf{G_0}$,
$\mathbf{G_1}$ and~$\mathbf{G_2}$ are to be identified. The behaviour
that can be observed from them when ignoring $\tau$-steps and
exploiting coin tosses to resolve probabilistic choices is the
same. This leads to a definition of probabilistic branching
{\bisimilarity} that hitherto was not proposed in the literature and
appears to be the pendant of weak distribution {\bisimilarity} defined
by~\cite{EHKTZ13:qest}.

As for~\cite{DP07:tcs} we seek to stay close to the treatment of the
non-deterministic fragment of the process calculus at hand. However,
as an alternate route in proving completeness, we rely on a novel
definition, viz.\ that of a concrete process.  We first apply the
approach for strictly non-deterministic processes and \emph{mutatis
  mutandis} for the whole language allowing processes that involve
both non-deterministic and probabilistic choice. For now, let's call a
process concrete if it doesn't exhibit inert transitions, i.e.\
$\tau$-transitions that don't change the potential behaviour of the
process essentially. The approach we follow first establishes
soundness for branching (probabilistic) {\bisimilarity} and soundness
and completeness for strong (probabilistic) {\bisimilarity}. Because
of the non-inertness of the silent steps involved, strong and
branching {\bisimilarity} coincide for concrete processes. The trick
then is to relate a pair of branching (probabilistically) bisimilar
processes to a corresponding pair of concrete processes. Since these
are also branching (probabilistically) bisimilar, as argued, they are
consequently strongly (probabilistically) bisimilar, and, voil\`a,
provably equal by the completeness result for strong (probabilistic)
{\bisimilarity}.

\paragraph{Related work}
Our branching probabilistic bisimulation semantics is essentially
\emph{distribution-based}, in the sense that it employs bisimulation
relations between probability distributions of states that cannot be
reduced to relations between states.  This feature makes it coarser,
or less discriminating, than the state-based branching probabilistic
bisimulation semantics of Segala \& Lynch \cite{SL94:concur}, which
distinguishes the processes $\mathbf{G_0}$, $\mathbf{G_1}$
and~$\mathbf{G_2}$.  Distribution-based notions of weak probabilistic
simulation and bisimulation semantics were explored in \cite{DGHM09}
and \cite{EHKTZ13:qest}.
  The standard pair of processes $a \pref (\tau \pref b + c)$ vs.\
  $a \pref (\tau \pref b + c) + a \pref b$ that are equivalent for
  weak bisimilarity but inequivalent for branching bisimulation (see
  e.g.~\cite{GM14:mit}) yields the same observation when translated to
  the probabilistic setting: identified by weak distribution
  equivalence as introduced in~\cite{EHKTZ13:qest} but different with
  respect to branching probabilistic bisimulation proposed
  here. However, the paper~\cite{EHKTZ13:qest}
  additionally presents a
  decision procedure for their probabilistic equivalence based on the
  notion of maximal end components and on earlier results reported
  in~\cite{HT12:fsttcs}.

A state-based approach to branching probabilistic bisimilarity by
Andova and Wil\-lem\-se is presented in~\cite{AW06:tcs}. The focus in
the paper is on the handling of $\tau$-loops in probabilistic
processes represented by finite bipartite graphs.
  The ensuing behavioural notion can be characterised in terms of concrete coloured traces involving so-called blends
  of colours. The technique of~\cite{AW06:tcs} generalises the
  paradigm of coloured traces for the non-deterministic setting
  originating from~\cite{GW96:jacm} to the setting of mixed
  non-deterministic and probabilistic processes. However, equivalence
  based on concrete coloured traces is finer grained than the
  branching bisimilarity as proposed here. For
  example, referring forward to Figure~\ref{fig:examples}, the
  processes $\mathbf{T}_1$ to~$\mathbf{T}_2$ are not identified in the
  setting of~\cite{AW06:tcs}, contrary to the present paper, because
  in~\cite{AW06:tcs} there is no means to identify the equal blend of
  the two colours $b \pref ( \frac13 P \oplus \frac23 Q)$ and
  $b \pref ( \frac23 P \oplus \frac13 Q)$ with the single colour blend
  of $b \pref (\frac12 P \oplus \frac12 Q)$.

  The comprehensive account~\cite{BDH2020:acta} of the `probabilistic
  bisimulation spectrum with silent moves' focuses on the analysis of
  branching, delay,~$\eta$, vs.\ weak bisimilarity (in various
  stuttering and divergence flavours) for which corresponding
  Hennessy/Milner style modal logics are shown to be sound and
  complete. The paper predominantly considers state-based behavioural
  equivalences. As argued in~\cite{EHKTZ13:qest}, state-based
  equivalences with equivalences on distributions induced by so-called
  lifting do not deal with silent moves in the way aimed for
  in~\cite{EHKTZ13:qest}---and here. The processes $\mathbf{G_0}$,
  $\mathbf{G_1}$, and~$\mathbf{G_2}$ above are all distinguished
  in these state-based equivalences
  because the distributions involved assign different probabilities to
  (the non-deterministic processes in the support of) $P$~and $Q$, and
  $\tau \pref (\frac12 P \oplus \frac12 Q)$.

Our syntax for processes is two-sorted, and comprises
  nondeterministic processes, ranged over by~$E,F$, and probabilistic
  processes, ranged over by~$P,Q$. Following~\cite{BS01:icalp,DGHM09},
  nondeterministic processes denote states in a probabilistic labelled
  transition system, whereas probabilistic processes denote
  probability distributions of states.  They are connected through the
  operator~$\partial$, which turns a nondeterministic process~$E$ into
  a probabilistic processes~$\partial(E)$, denoting the Dirac
  distribution~$\delta(E)$, which assigns all its probabilistic weight
  to the state~$E$. In~\cite{DGHM09} the operator~$\partial$ was
  elided, thereby identifying the processes $E$ and~$\partial(E)$.
  This practice naturally leads to semantic equivalences~$\sim$
  satisfying
\begin{equation}
  \label{identification}
  E \sim F \Leftrightarrow \partial(E) \sim \partial (F) \; .
\end{equation}
In fact, this property holds for any semantic equivalence~$\sim$ from
the literature that we know of, and for which both sides are defined.

In~\cite{GGV19:fc}, which is an extended abstract of an earlier
version of the present paper, we present a complete axiomatisation of
a form of branching probabilistic bisimulation semantics that on
nondeterministic processes coincides with the version in the present
paper, but on probabilistic processes it is strictly finer.  Whereas that
semantics satisfies (\ref{identification}), our present semantics
merely has the congruence property
$E \sim F \Rightarrow \partial(E) \sim \partial (F)$.  This is because
for nondeterministic processes we impose a root condition, making
branching probabilistic bisimilarity into a congruence for the CCS
choice operator $+$, whereas we refrain from doing so for
probabilistic processes.

\bigskip

\noindent
The remainder of the paper is organised as follows. In
Section~\ref{sec-prelim} we gather some notation regarding probability
distributions. For illustration purposes Section~\ref{sec-nondet}
treats the simpler setting of non-deterministic processes, reiterating
the completeness proof for the equational theory of~\cite{GW96:jacm}
for rooted branching {\bisimilarity}. Next, after introducing
probabilistic processes in Section~\ref{sec-probabilistic} and
branching probabilistic {\bisimilarity} and some of its fundamental
properties in Section~\ref{sec-bpb}, we prove in
Section~\ref{sec-completeness} the main result, viz.\ the completeness
of an equational theory for rooted branching probabilistic
{\bisimilarity}, exploiting the notion of a concrete process and
following the same lines set out in Section~\ref{sec-nondet}. In
Section~\ref{sec-concl} we wrap up and make concluding remarks.



\section{Preliminaries}

\label{sec-prelim}

\noindent
Let $\DistrX$ be the set of probability distributions of finite
support over the set~$X$. The support of a distribution $\mu$ is
denoted as $\mathit{spt}(\mu)$, thus
$\mathit{spt}(\mu) = \lc x \in X \mid \mu(x) > 0 \rc$.  Each
distribution~$\mu \in \DistrX$ can be represented in the form
$\mu = \bigoplusiinI \: p_i \stardot x_i$ where $\mu(x_i) = p_i$ for
$i \in I$ and $\sumiinI \: p_i = 1$ for some finite index set~$I$.
For convenience later, we do not require $x_i \neq x_{i'}$ for
$i \neq i'$ nor $p_i > 0$ for $i, i' \in I$. So, the representation
of~$\mu$ is not unique.

We use $\delta(x)$ to denote the Dirac distribution for~$x \in X$ with
$\delta(x)(x) = 1$ and $\delta(x)(y) = 0$ for~$y \in X$,~$y \neq x$.
For $\mu, \nu \in \DistrX$ and $r \in [0,1]$ we define $\mu \oplusr
\nu \in \DistrX$ by $(\mu \oplusr \nu)(x) = r \mycdot \mu(x) + (1{-}r)
\mycdot \nu(x)$. By definition $\mu \probc0 \nu = \nu$ and $\mu
\probc1 \nu = \mu$. 
For an index set~$I$, $p_i \in [0,1]$ and $\mu_i \in \DistrX$, we
define $\bigoplusiinI \: p_i \mycdot \mu_i \in \DistrX$ by
$(\bigoplusiinI \: p_i \mycdot \mu_i)(x) = \sumiinI \: p_i \mycdot
\mu_i(x)$ for~$x \in X$. For
$\mu = \bigoplusiinI \: p_i \mycdot \mu_i$,
$\nu = \bigoplusiinI \: p_i \mycdot \nu_i$, and $r \in [0,1]$ it holds
that $\mu \oplusr \nu = \bigoplusiinI \: ( \mu_i \oplusr \nu_i )$.
For a subset~$X' \subseteq X$ and distribution~ $\mu \in \DistrX$, we
occasionally write~$\mu[X' \mkern1mu
]$ for $\sum_{x {\in} X'} \: \mu(x)$.

Given a binary relation $\calR \subseteq \DistrX \times \DistrX$
between distributions, its \emph{convex closure}~$\ccR$ is defined by
\(\ccR := \{(\bigoplusiinI \: p_i \mycdot \mu_i,\bigoplusiinI \: p_i
\mycdot \nu_i) \mid \forall {i \mathbin\in I}.~(\mu_i,\nu_i)\in\calR
\wedge \sum_{i \in I}p_i=1\}\).

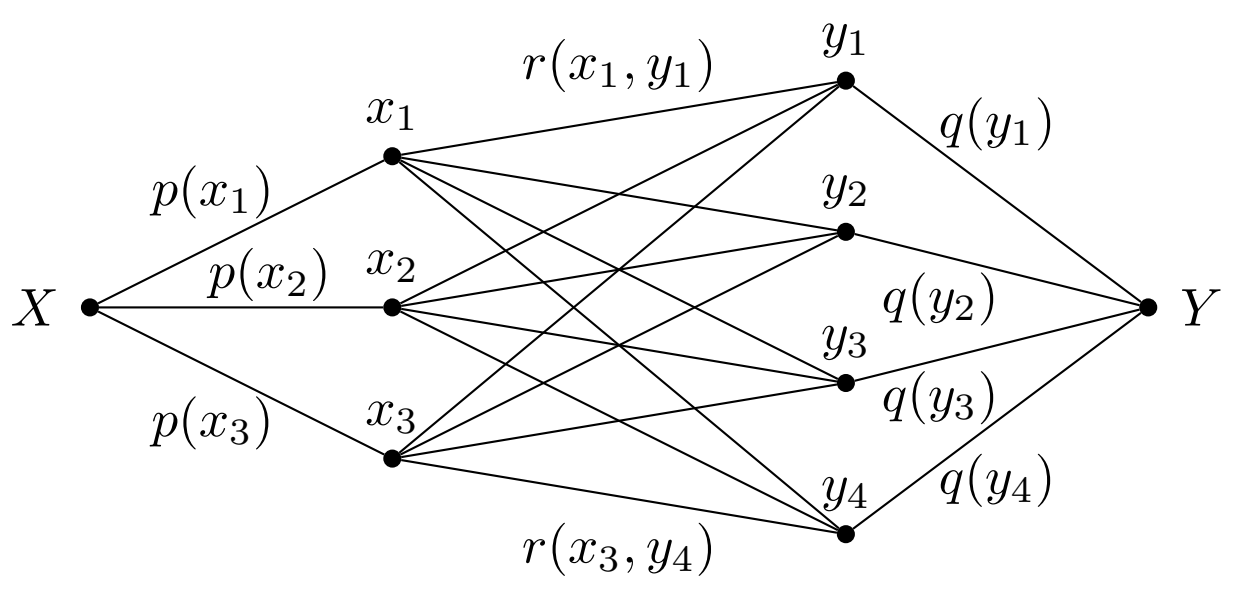
\begin{figure}
  \label{fig-flow-graph}
  \centering
  \begin{tikzpicture}
    \tikzstyle{every node}=[circle, draw, fill=black, inner sep=0pt, minimum
      width=3pt]
    \node [label={[label distance=1mm]left:{\small $X$}}] (source) at (0,2) {} ;
    \node [label={\small $x_1$}] (x1) at (2,3) {} ;
    \node [label={\small $x_2$}] (x2) at (2,2) {} ;
    \node [label={\small $x_3$}] (x3) at (2,1) {} ;
    \node [label={\small $y_1$}] (y1) at (5,3.5) {} ;
    \node [label={\small $y_2$}] (y2) at (5,2.5) {} ;
    \node [label={\small $y_3$}] (y3) at (5,1.5) {} ;
    \node [label={\small $y_4$}] (y4) at (5,0.5) {} ;
    \node [label={[label distance=1mm]right:{\small $Y$}}] (sink) at (7,2) {} ;

    \tikzstyle{every node}=[draw=none, fill=none, inner sep=0pt,
    minimum width=0pt] 

    \draw (source) -- node [above=5pt, pos=0.4] {\small $p(x_1)$} (x1) ;
    \draw (source) -- node [above=1pt, pos=0.6] {\small $p(x_2)$} (x2) ;
    \draw (source) -- node [below=5pt, pos=0.4] {\small $p(x_3)$} (x3) ;
    
    \foreach \xnode in {x1,x2,x3} {
      \foreach \ynode in {y1,y2,y3,y4} {
        \draw (\xnode) -- (\ynode) ;
      }
    }
    \node (z11) at (3.5,3.6) {\small $r(x_1,y_1)$} ;
    \node (z34) at (3.5,0.4) {\small $r(x_3,y_4)$} ;

    \draw (y1) -- node [above=8pt] {\small $q(y_1)$} (sink) ;
    \draw (y2) -- node [below=3pt, pos=0.3] {\small $q(y_2)$} (sink) ;
    \draw (y3) -- node [below=2pt, pos=0.3] {\small $q(y_3)$} (sink) ;
    \draw (y4) -- node [below=6pt] {\small $q(y_4)$} (sink) ;
  \end{tikzpicture}
  \caption{Distributions $p$ on~$X$, $q$ on~$Y$, and~$r$ on~$X \times Y$.}
\end{figure}

In the sequel we make frequent use of the following property. If
$p : X \to \IR$ and $q : Y \to \IR$ are (weight)
functions of finite support such that $p[X] = q[Y]$, then there is an
$r : X \times Y \to \IR$ satisfying
$p(x) = \sum_{y {\in} Y} \: r(x,y)$ for all~$x \in X$ and
$q(y) = \sum_{x {\in} X} \: r(x,y)$ for all~$y \in Y$. See
Figure~\ref{fig-flow-graph}. For $p$ and~$q$ distributions one can
define $r(x,y) = p(x) q(y)$.
A direct proof is by application of the max-flow min-cut theorem
assigning capacity~$1$ to the edges~$(x_i,y_j)$, cf.~\cite{FF56:cjm}. 

\section{Concrete processes}
\label{sec-nondet}

\noindent
In this section we present an approach to prove completeness of an
axiomatic theory for branching {\bisimilarity} exploiting the notion
of a concrete process. To explicate the concept, we consider a basic
process language with inaction, prefix, and non-deterministic choice
only. The result is not new, but the approach is non-standard. In
the remainder of the paper we extend the approach to establish the
completeness of an axiomatisation of branching probabilistic
bisimulation.

\blankline

\noindent
We assume to be given a set of actions~$\calA$ including the so-called
silent action~$\tau$. The process language we consider is called a
Minimal Process Language in~\cite{BBR10:cup}. It provides a constant
for inaction~$\bfzero$, a prefix construct for each action
$\alpha \mathbin\in \calA$, and non-deterministic choice between to
processes.

\begin{definition}[Syntax]
  \label{df-non-det-processes}
  The class~$\calE$ of non-deterministic processes over~$\calA$, with
  typical element~$E$, is given by\vspace{-.5ex}
  \[
    E \bnfeq \bfzero \mid \alpha \pref E \mid E + E\vspace{-1.5ex}
  \]
  with actions~$\alpha$ from~$\calA$.
\end{definition}

\noindent
As technical aid below, we define for $E \in \calE$ its
complexity~$c(E)$ by $c(\bfzero) = 0$, $c(\alpha \pref E) = c(E) + 1$,
and $c(E + F) = c(E) + c(F)$.

The behaviour of processes in~$\calE$ is as usual and given by a
structural operational semantics going back to~\cite{HM80:icalp}.

\begin{definition}[Operational semantics]
  \label{def-nd-transition-relation}
  The transition relation ${\rightarrow} \subseteq \calE \times \calA
  \times \calE$ is
  given by
  \halflineup
  \begin{displaymath}
    \begin{array}{c}
      \sosrule{}{\alpha \pref E \arrow{\alpha} E}
      \: \textsc{\small (pref)}
     \medskip \\
      \sosrule{E_1 \arrow{\alpha} E_1}{E_1 + E_2 \arrow{\alpha} E_1}
      \: \textsc{\small (nd-choice\,1)}
      \qquad
      \sosrule{E_2 \arrow{\alpha} E_2}{E_1 + E_2 \arrow{\alpha} E_2}
      \: \textsc{\small (nd-choice\,2)}
   \end{array}
  \end{displaymath}
\end{definition}

\medskip

\noindent
Thus, the process~$\bfzero$ cannot perform any action, $\alpha \pref E$ can
perform action $\alpha$ and subsequently behave as $E$, and $E_1+E_2$
represents the choice in behaviour between that of $E_1$ and of~$E_2$.

We have auxiliary definitions and relations derived from the
transition relation of Definition~\ref{def-nd-transition-relation}.
A process $E' \in \calE$ is called a derivative of a
process~$E \in \calE$ iff $n \geqslant 0$,
$E_0, \ldots, E_n \in \calE$, and $\alpha_1, \ldots, \alpha_n$ exist
such that $E = E_0$, $E_{i{-}1} \arrow{\alpha_i} E_i$ for
$1 \leqslant i \leqslant n$, and $E_n = E'$.
We define $\der(E) = \lc E' \in \calE \mid \mbox{$E'$ is a derivative of~$E$}
\rc$.
\vspace{2pt}
Furthermore, for $E, E' \in \calE$ and $\alpha \in \calA$ we write
$E \alphahidearrow E'$ iff $E \alphaarrow E'$, or $\alpha = \tau$ and
$E = E'$. We use $\Arrow{\:}$ to denote the reflexive transitive
closure of $\arrow{(\tau)}$.

\blankline

\noindent
The definitions of strong and branching {\bisimilarity} for~$\calE$ are
standard and adapted from~\cite{GW96:jacm,Mil89:phi}.

\begin{definition}[Strong and branching {\bisimilarity}]
  \label{def-branching-bisimilar}
  \begin{itemize}
  \item [(a)] A symmetric relation
    $\calR \subseteq \calE \times \calE$ is called a \textit{strong
      bisimulation relation} iff for all $E, E', F \in \calE$ such
    that $E \mopcalR F$ and $E \arrow{\alpha} E'$ there is an
    $F' \in \calE$ such that
    \begin{displaymath}
      F \alphaarrow F' \ \text{and} \ 
      E' \mopcalR F'.
    \end{displaymath}
  \item [(b)] A symmetric relation
    $\calR \subseteq \calE \times \calE$ is called a \textit{branching
      bisimulation relation} iff for all $E, E', F \in \calE$ such
    that $E \mopcalR F$ and $E \arrow{\alpha} E'$ there are
    $\Fbar, F' \in \calE$ such that
    \begin{displaymath}
      F \Arrow{\:} \Fbar, \ 
      \Fbar \alphahidearrow F', \ 
      E \mopcalR \Fbar, \ \text{and} \ 
      E' \mopcalR F'.
    \end{displaymath}
  \item [(c)] Strong {\bisimilarity}, denoted by~${\bisim} \subseteq \calE
    \times \calE$, and branching {\bisimilarity}, written as~${\brbisim}
    \subseteq \calE \times \calE$, are defined as the largest strong
    bisimulation relation on~$\calE$ and the largest branching
    bisimulation relation on~$\calE$, respectively.
  \end{itemize}
\end{definition}

\noindent
Clearly, in view of the definitions, strong {\bisimilarity} between two
processes implies branching {\bisimilarity} between the two
processes. Thus~${\bisim} \subseteq {\brbisim}$.

\blankline

\noindent
The notion of an \emph{inert} transition is well-known,
see~\cite{GW96:jacm}: if for a transition $E \arrow{\tau} E'$ we have
that $E \brbisim E'$, the transition is called \textit{inert}. Much
less well-known is the derived notion of a \emph{concrete} process,
going back to~\cite{GL98:cwi}: a
process~$\Ebar$ is called concrete iff it has no inert transitions,
i.e., if $\Ebar' \in \der(\Ebar)$ and $\Ebar' \arrow{\tau} \Ebar''$,
then~$\Ebar' \nbrbisim \Ebar''$.
Note, for $\Ebar' \in \der(\Ebar)$ we have $\der(\Ebar') \subseteq
\der(\Ebar)$, thus $\Ebar'$ is concrete if $\Ebar$~is.
A straightforward example of a concrete
process is $a \pref \bfzero + b \pref \bfzero$ in which no
$\tau$-actions occur at all, but also
$a \pref \bfzero + \tau \pref b \pref \bfzero$ is a concrete process
since
$a \pref \bfzero + \tau \pref b \pref \bfzero \nbrbisim b \pref
\bfzero$ (assuming $a \neq b$). We write
$\calEconc = \lc \Ebar \in \calE \mid \mbox{$\Ebar$ concrete} \rc$.

\blankline

\noindent
Next we introduce a restricted form of branching {\bisimilarity},
called rooted branching {\bisimilarity}, instigated by the fact that
branching {\bisimilarity} itself is not a congruence for the choice
operator, cf.~\cite{Fok00:springer}. This makes branching
{\bisimilarity} unsuitable for equational reasoning where it is
natural to replace subterms by equivalent terms. Note that weak
{\bisimilarity} has the same problem, cf.~\cite{Mil89:phi}.

For example, we have for any process~$E \in \calE$ that $E$
and~$\tau \pref E$ are branching bisimilar, but in the context of a
non-deterministic alternative they may not, i.e., it is not
necessarily the case that $E + F \brbisim \tau \pref E + F$. More
concretely, although $\bfzero \brbisim \tau \pref \bfzero$, it doesn't
hold that
$\bfzero + {b \pref \bfzero} \brbisim {\tau \pref \bfzero} + {b \pref
  \bfzero}$. The $\tau$-move of $\tau \pref \bfzero + b \pref \bfzero$
to~$\bfzero$ has no counterpart in $\bfzero + b \pref \bfzero$ because
$\bfzero + b \pref \bfzero \nbrbisim \bfzero$. Intuitively, the
summand $\tau \pref \bfzero$ can preempt non-deterministic
alternatives whereas the summand~$\bfzero$ cannot.

\begin{definition}
  \label{def-rooted-branching-bisimilarity}
  Two nondeterministic processes $E,F \in\calE$ are \emph{rooted
    branching bisimilar}, notation $E \rootedbrbisim F$, if for each
  transition $E \arrow\alpha E'$ with $\alpha \in \calA$ there is a
  transition $F \arrow\alpha F'$ with $E' \brbisim F'$,
  and, vice versa, for each transition $F \arrow\alpha F'$ there is a
  transition $E \arrow\alpha E$ with $E' \brbisim F'$.
\end{definition}

\noindent
When comparing processes~$E, F \in \calE$, rooted branching
bisimilarity requires {\lq strong\rq} {\bisimilarity} at the root,
i.e.\ for $E$ and~$F$ themselves, and allows branching {\bisimilarity}
for the derivatives of $E$ and~$F$.  
 
Direct from the definitions we obtain
${\bisim} \subseteq {\rootedbrbisim} \subseteq {\brbisim}$. As eluded
upon, we have a congruence result for rooted branching
{\bisimilarity}. A proof is given in~\cite{GW96:jacm}.

\begin{lemma}
  \label{lemma-congruence-nondet}
  $\rootedbrbisim$ is a congruence on~$\calE$ for the operators
  $\pref$ and~$+$.
  \qed
\end{lemma}

\noindent
In \cite{GW96:jacm} it is also shown that $\rootedbrbisim$ is the
coarsest, or largest, congruence included in $\brbisim \mkern1mu$.  It
is well-known that both strong {\bisimilarity} and branching
{\bisimilarity} for~$\calE$ can be equationally
characterised~\cite{BBR10:cup,GW96:jacm,Mil89:phi}. In line with the
so-called root condition of
Definition~\ref{def-rooted-branching-bisimilarity}, a prefix context
$\alpha \pref \mbox{\textvisiblespace}$ is added in the B-axiom.

\begin{definition}[Axiomatization of $\bisim$ and~$\rootedbrbisim$]
  The theory~$\AXd$ is given by the axioms A1 to~A4 listed in
  Table~\ref{table-axiomatization-of-branching-bisimilarity}.
  The theory~$\AXdb$ contains in addition the axiom~\hyperlink{B}{B}.
\end{definition}

\begin{table}
  \vspace{-4ex}
  \centering
  \def\arraystretch{1.1}
  \begin{tabular}{|@{\;}l@{\ \;}l|}
    \hline
    A1 & $E + F = F + E$ \rule{0pt}{12pt} \\
    A2 & $(E + F) + G = E + ( F + G)$ \\
    A3 & $E + E = E$ \\
    A4 & $E + \bfzero = E$ 
    \rule[-5pt]{0pt}{12pt}
    \\ \hline
    \hypertarget{B}{B} & $\alpha \pref ( \, \tau \pref ( E + F ) + F
    \, ) = \alpha 
    \pref ( E + F )$
    \rule{0pt}{12pt}\rule[-5pt]{0pt}{12pt} \\ 
    \hline
  \end{tabular}
  \def\arraystretch{1.0}

  \smallskip

  \caption{Axioms for strong and branching {\bisimilarity}}
  \label{table-axiomatization-of-branching-bisimilarity}
\end{table}

\noindent
If two processes are provably equal with respect to the theory~$\AXdb$,
then they are rooted branching bisimilar as well.

\begin{theorem} [$\AXdb$ sound for $\rootedbrbisim$]
  \label{theorem-soundness-base}
  For all $E,F \in \calE$, if $\AXdb \vdash E = F$ then $E
  \rootedbrbisim F$.
\end{theorem}

\begin{proof}(Sketch) It is easy to see that the left-hand side and
  the right-hand side of the axioms of~$\AXdb$ are rooted branching
  bisimilar. For example, for $\alpha \in \calA$ and~$E,F \in \calE$, the
  processes $\alpha \pref ( F + {\tau \pref (E+F)} )$ and $\alpha \pref (E+F)$
  in axiom~B both have an $\alpha$-transition as their only transition,
  which yield the branching bisimilar processes
  $F + {\tau \pref (E+F)}$ and $E+F$, respectively.  To finish the proof of this 
  lemma, one
  observes that rooted branching {\bisimilarity} is a congruence, see
  Lemma~\ref{lemma-congruence-nondet}.
\end{proof}

\noindent
In order to obtain a completeness result for~$\AXdb$
modulo~$\rootedbrbisim$ we deviate from the usual approach to prepare
the stage for the proof of completeness with respect to branching
probabilistic {\bisimilarity} as presented in
Section~\ref{sec-completeness}. Here, we will proceed along the
following route.
\begin{enumerate}
\item First we observe that the theory~$\AXd$, without the B-axiom, is
  (sound and) complete with respect to strong {\bisimilarity}; this is
  Theorem~\ref{theorem-AXd-sound-and-complete}.
\item Next, we obtain a result stating that for concrete processes,
  which have by definition no inert $\tau$-transitions, branching
  bisimilarity implies strong bisimilarity, and hence, the axiom
  set~$\AXd$ is complete for branching bisimilarity if processes are
  concrete, see
  Lemma~\ref{lemma-branching-implies-equal-for-concrete-processes}.
\item Then we have a technical result, relating to each
  process~$E \in \calE$ a concrete and branching bisimilar process
  $\Ebar \in \calEconc$ such that $E$ and~$\Ebar$ are equal
  for~$\AXdb$. This is
  Lemma~\ref{lemma-branching-bisim-vs-equal-under-prefix}.
\item Finally we combine the above results to obtain completeness
  of~$\AXdb$ for~$\rootedbrbisim$ as stated by
  Theorem~\ref{theorem-completeness}.
\end{enumerate}
So, we start off by noting that strong {\bisimilarity} is equationally
characterised by the axioms A1 to~A4 of
Table~\ref{table-axiomatization-of-branching-bisimilarity}.  See for
example~\cite[Section~7.4]{Mil89:phi} for a proof.

\begin{theorem}[$\AXd$ sound and complete for~$\bisim$]
  \label{theorem-AXd-sound-and-complete}
  For all processes $E, F \in \calE$ it holds that $\AXd \vdash E =
  F$ iff $E \bisim F$.
  \qed
\end{theorem}

\noindent
Next we consider concrete processes and branching {\bisimilarity}.
Since by their definition, concrete processes have no inert
transitions, it holds that branching {\bisimilarity} and strong
{\bisimilarity} coincide. Hence, by the above theorem,
Theorem~\ref{theorem-AXd-sound-and-complete}, branching
{\bisimilarity} implies equality for~$\AXd$.

\begin{lemma}
  \label{lemma-branching-implies-equal-for-concrete-processes}
  For all concrete $\Ebar,\Fbar \in \calEconc$, if
  $\Ebar \brbisim \Fbar$ then $\Ebar \bisim \Fbar$ and
  $\AXd \vdash \Ebar = \Fbar$.
\end{lemma}

\begin{proof}[Sketch]
  Consider $\Ebar, \Fbar \in \calEconc$ such that
  $\Ebar \brbisim \Fbar$. Define the
  relation~$\calR \subseteq {\calE \times \calE}$ to be the
  restriction of branching {\bisimilarity}~$\brbisim$ to the
  derivatives of $\Ebar$ and~$\Fbar$, i.e.,
  $\calR = {\brbisim} \mkern1mu \cap ( (\der(\Ebar) \times
  \der(\Fbar)) \cup (\der(\Fbar) \times \der(\Ebar)) )$. Then $\calR$
  is a strong bisimulation relation, since none of the processes
  involved admits an inert $\tau$-transition. In more detail, if
  $\Ehat \mopcalR \Fhat$ and $\Ehat \arrow{\alpha} \Ehat'$, then
  $\Fhat \Arrow{\:} \Ftilde$ and $\Ftilde \arrow{(\alpha)} \Fhat'$ for
  suitable processes $\Ftilde$ and~$\Fhat'$ with
  $\Ftilde \mopcalR \Fhat$ and~$\Fhat \brbisim \Ehat'$. Since
  $\Fhat \in \der(\Ebar) \cup \der(\Fbar)$~is a concrete process as
  well, we must have $\Fhat = \Ftilde$. Thus
  $\Fhat \arrow{(\alpha)} \Fhat'$. If $\alpha = \tau$ and $\Fhat$
  and~$\Fhat'$ would coincide, it follows that
  $\Ehat \arrow{\tau} \Ehat'$ while
  $\Ehat \brbisim \Fhat \brbisim \Fhat' \brbisim \Ehat'$, which cannot
  be since $\Ehat$ is concrete. Thus $\Fhat \arrow{\alpha} \Fhat'$ for
  $\Fhat' \brbisim \Ehat'$ and showing that $\calR$ is a strong
  bisimulation relation and~$\Ebar \bisim \Fbar$.

  To finish the proof of the lemma, by the completeness of~$\AXd$ with
  respect to strong bisimulation, see
  Theorem~\ref{theorem-AXd-sound-and-complete}, it follows that
  $\AXd \vdash \Ebar = \Fbar$.
\end{proof}

\noindent
We have now arrived at the main technical lemma of this section, viz.\
the result that branching {\bisimilarity} implies equality for~$\AXdb$
under a prefix. In the proof the notion of a concrete process plays a
central role. On the one hand, the lemma aims to reduce the question
of equality for general processes to the question of equality for
concrete processes by appointing a concrete and branching bisimilar
counterpart for each non-deterministic process. On the other hand, the
lemma guarantees equality in the context of a prefix for branching
bisimilar processes.

\begin{lemma}
  \label{lemma-branching-bisim-vs-equal-under-prefix}
  \mbox{}
  \begin{itemize}
  \item [(a)] For all processes $E \in \calE$, a
    concrete process $\Ebar \in \calEconc$ exists such that $E \brbisim
    \Ebar$ and $\AXdb \vdash { \alpha \pref E = \alpha \pref \Ebar }$
    for all $\alpha \in \calA$.
  \item [(b)] For all processes $F,G \in \calE$, if $F \brbisim
    G$ then $\AXdb \vdash { \alpha \pref F = \alpha \pref G }$ for all
    $\alpha \in \calA$.%
  \end{itemize}
\end{lemma}

\begin{proof}
  We prove statements (a) and~(b) by simultaneous induction on 
  $c(E)$ and $\max \lbrace c(F), c(G) \rbrace$, respectively. The
  complexity measure~$c$ is given right after
  Definition~\ref{df-non-det-processes}. 

  Basis, $c(E) = 0$: We have that $E = \bfzero + \cdots +
  \bfzero$. Hence, we can take $\Ebar = \bfzero$. Then clearly, part~(a)
  of the lemma holds as $\bfzero$ is concrete, $E \brbisim \bfzero$,
  and $\AXdb \vdash \alpha \pref E = \alpha \pref \bfzero$ for all
  $\alpha \in \calA$.

  Induction step, $c(E) > 0$: The process~$E$ can be written
  as $\sumiinI \: \alpha_i \pref E_i$ for some finite index set~$I$
  and suitable actions~$\alpha_i \in \calA$ and
  subprocesses~$E_i \in \calE$. We distinguish two cases.

  \vspace{1pt}
  
  \noindent
  \emph{Case~I}. We first assume that for some $i_{\myoh} \in I$ we
  have $\alpha_{i_{\myoh}} = \tau$ and $E_{i_{\myoh}} \brbisim E$.
  Then it holds that $\AXd \vdash E = \tau.E_{i_{\myoh}} + H$, where
  \plat{$H = \sum_{i {\in} I \setminus \{ i_{\myoh} \}}\: \alpha_i
    \pref E_i$}. \vspace{1pt} By the induction hypothesis for part~(a),
  there is a concrete process $\Ebar_{i_{\myoh}} \in \calEconc$ such
  that $E_{i_{\myoh}} \brbisim \Ebar_{i_{\myoh}}$. \vspace{1pt} We
  claim that $\Ebar_{i_{\myoh}} \brbisim E_{i_{\myoh}} + H$.

  For suppose $\Ebar_{i_{\myoh}} \alphaarrow F$.  \vspace*{1pt} Then,
  for suitable processes $E'_{i_{\myoh}}$ and~$G$, we have
  $E_{i_{\myoh}} \Arrow{\:} E'_{i_{\myoh}} \alphahidearrow G$,
  $E'_{i_{\myoh}} \brbisim \Ebar_{i_{\myoh}}$, and~$G \brbisim
  F$. If~$E_{i_{\myoh}} = E'_{i_{\myoh}}$ it follows that
  $E_{i_{\myoh}} \alphahidearrow G$.  Since $\Ebar_{i_o}$ is concrete,
  either $\alpha \neq \tau$ or $F \nbrbisim \Ebar_{i_{\myoh}}$.
  Hence, $\alpha \neq \tau$ or $G \nbrbisim E_{i_{\myoh}}$ and
  $G \neq E_{i_{\myoh}}$. So $E_{i_{\myoh}} \alphaarrow G$.
  Consequently, $E_{i_{\myoh}}+H \alphaarrow G$.
  If~$E_{i_{\myoh}} \neq E'_{i_{\myoh}}$ we have
  $E_{i_{\myoh}}+H \Arrow{\:} E'_{i_{\myoh}} \alphahidearrow G$.

  Now suppose $E_{i_{\myoh}} + H \alphaarrow F$.  Then either
  $E_{i_{\myoh}} \alphaarrow F$ or $H \alphaarrow F$. \vspace{1pt} In
  the first case we have
  $\Ebar_{i_{\myoh}} \Arrow{\:} \Ebar'_{i_{\myoh}} \alphahidearrow G$
  for suitable $\Ebar'_{i_{\myoh}}$ and~$G$ with
  $E_{i_{\myoh}}\brbisim \Ebar'_{i_{\myoh}}$ and $F \brbisim G$,
  \vspace{1pt} while in the latter case we have~$E \alphaarrow F$, and
  since $E \brbisim E_{i_{\myoh}} \brbisim \Ebar_{i_{\myoh}}$ we have
  $\Ebar_{i_{\myoh}} \Arrow{\:} \Ebar'_{i_{\myoh}} \alphahidearrow G$
  for $\Ebar'_{i_{\myoh}}$ and~$G$ with $E\brbisim \Ebar'_{i_{\myoh}}$
  and $F \brbisim G$. Because $\Ebar_{i_{\myoh}}$ is concrete,
  $\Ebar'_{i_o} = \Ebar_{i_{\myoh}}$.  Thus
  $\Ebar_{i_{\myoh}} \alphahidearrow G$ with $F \brbisim G$, which was
  to be shown.

  We conclude that
  $E_{i_{\myoh}} \brbisim \Ebar_{i_{\myoh}} \brbisim E_{i_{\myoh}}+H$
  as claimed. By definition of~$c$ it holds that
  $c(E_{i_{\myoh}}), c(E_{i_{\myoh}}+H) < c(\tau \pref
  E_{i_{\myoh}}+H) = c(E)$.  Therefore, by the induction hypothesis for
  part~(b) for the two processes $E_{i_{\myoh}}$
  and~$E_{i_{\myoh}}+H$, it holds that
  $\AXdb \vdash {\tau \pref E_{i_{\myoh}}} = {\tau \pref (
    E_{i_{\myoh}} + H )}$, which we will use below.  By the induction
  hypothesis for part~(a), there is a process~$\Ebar \in \calEconc$
  such that $\Ebar \brbisim E_{i_{\myoh}} + H$ \vspace{1pt} and
  $\AXdb \vdash {\alpha \pref \Ebar} = {\alpha \pref ( E_{i_{\myoh}} + H )}$ for
  every~$\alpha \in \calA$. Observe that
  $E \brbisim E_{i_{\myoh}} \brbisim E_{i_{\myoh}} + H \brbisim
  \Ebar$. Therefore, we can calculate
  \begin{align*}
    \AXdb \vdash {\alpha \pref E }
    & = \alpha \pref ( \tau \pref E_{i_{\myoh}} + H ) 
    && \\
    & = \alpha \pref ( \tau \pref ( E_{i_{\myoh}} + H ) + H ) 
    && & \text{Since $\AXdb\vdash\tau \pref E_{i_{\myoh}} = \tau \pref
      (E_{i_{\myoh}} + H)$} \phantom{\mkern1mu .} \\ 
    & = \alpha \pref ( E_{i_{\myoh}} + H ) 
    && & \text{Axiom \hyperlink{B}{B}} \phantom{\mkern1mu .} \\
    & = \alpha \pref \Ebar
    && & \text{Choice of $\Ebar$} \mkern1mu .
  \end{align*}
  Hence, for this case we have shown the existence of a
  process~$\Ebar$ with the required properties.

  \vspace{2pt}

  \noindent
  \emph{Case~II}. Now suppose, for all $i \in I$ we have
  $\alpha_i \neq \tau$ or $E_i\nbrbisim E$. Clearly $c(E_i) < c(E)$
  for all $i \in I$.  By the induction hypothesis for part~(a) we can
  find, for all $i \in I$, a concrete process~$\Ebar_i$ such that
  $\Ebar_i \brbisim E_i$ and
  $\AXdb \vdash \alpha \pref \Ebar_i = \alpha \pref E_i$ for every
  action~$\alpha \in \calA$. Define
  $\Ebar = { \sumiinI \: \alpha_i \pref \Ebar_i }$. Then
  $\Ebar \brbisim E$ and $\Ebar$~is concrete too, since
  $\Ebar_i \brbisim E_i \nbrbisim E \brbisim \Ebar$ for~$i \in I$ in
  case $\alpha_i = \tau$.  Moreover, $\AXdb \vdash {E = \Ebar}$, since
  $E = \sumiinI \: \alpha_i \pref E_i = \sumiinI \: \alpha_i \pref
  \Ebar_i = \Ebar$. Hence, for all actions~$\alpha \in \calA$, it holds
  that $\AXdb \vdash \alpha \pref E = \alpha \pref \Ebar$.

  Basis, $\max \lbrace c(F), c(G) \rbrace = 0$: As for the base case
  for~$E$, we have $F,G \brbisim \bfzero$ and $\AXdb \vdash {\alpha
    \pref F} = {\alpha \pref \bfzero} = {\alpha \pref G}$.
  
  Induction step, $\max \lbrace c(F), c(G) \rbrace > 0$: Suppose
  $F \brbisim G$. Choose, by the induction hypothesis for part~(a),
  $\Fbar, \Gbar \in \calEconc$ such that $F \brbisim \Fbar$ and
  $\AXdb \vdash {\alpha \pref F} = {\alpha \pref \Fbar}$ for
  all~$\alpha \in \calA$, and similarly for $G$ and~$\Gbar$. Then we
  have $\Fbar \brbisim \Gbar$. Since $\Fbar$ and~$\Gbar$ are concrete
  it follows that $\AXd \vdash {\Fbar = \Gbar}$, see
  Lemma~\ref{lemma-branching-implies-equal-for-concrete-processes}. Now
  pick any $\alpha \in \calA$. Then we have
  $\AXdb \vdash {\alpha \pref F} = {\alpha \pref \Fbar} = {\alpha
    \pref \Gbar} = {\alpha \pref G}$.
\end{proof}

\noindent
With Lemma~\ref{lemma-branching-bisim-vs-equal-under-prefix} in place,
we have by now gathered sufficient building blocks to provide a
completeness proof for~$\AXdb$ based on concrete processes.

\begin{theorem}[$\AXdb$ complete for $\rootedbrbisim$]
  \label{theorem-completeness}
  For all processes $E, F \in \calE$ it holds that $E \rootedbrbisim
  F$ iff $\AXdb \vdash E = F$.
\pagebreak[3]
\end{theorem}

\begin{proof}
  Suppose $E, F \in \calE$ and $E \rootedbrbisim F$. Let
  $E = \sumiinI \: \alpha_i \pref E_i$ and
  $F = \sumjinJ \: \beta_j \pref F_j$ for suitable index sets $I$
  and~$J$, $\alpha_i, \beta_j \in \calA$, $E_i, F_j \in \calE$ (tacitly
  using theory~$\AXd$). Since $E \rootedbrbisim F$ we have that for all
  $i \in I$ there is a $j \in J$ such that $\alpha_i = \beta_j $
  and~$E_i \brbisim F_j$, and, symmetrically, that for all $j \in J$
  there is an $i \in I$ such that $\alpha_i = \beta_j$ and
  $E_i \brbisim F_j$.  Put
  $K = \lc (i,j) \in I \times J \mid ( \alpha_i = \beta_j ) \land ( E_i
  \brbisim F_j ) \rc$. Define the processes $G, H \in \calE$ by
  \begin{displaymath}
    G = \sumkinK \: \gamma_k \pref G_k
    \quad \text{and} \quad
    H = \sumkinK \: \zeta_{\mkern1mu k} \pref H_k
  \end{displaymath}
  where, for $i \in I$, $\gamma_k = \alpha_i$ and $G_k = E_i$ if
  $k = (i,j)$ for some $j \in J$, and, similarly for $j \in J$,
  $\zeta_{\mkern1mu k} = \beta_j$ and $H_k = F_j$ if $k = (i,j)$ for some $i
  \in I$. Then $G$ and~$H$ are well-defined. Moreover, $\AXd \vdash E
  = G$ and $\AXd \vdash F = H$.

  For $k \in K$, say $k = (i,j)$, it holds that $\gamma_k = \alpha_i =
  \beta_j = \zeta_{\mkern1mu k}$ and $G_k = E_i \brbisim F_j = H_k$, by
  definition of~$K$. By
  Lemma~\ref{lemma-branching-bisim-vs-equal-under-prefix}b we obtain,
  for all $k \in K$, $\AXdb \vdash { \gamma_k \pref G_k = \zeta_{\mkern1mu k}
    \pref H_k }$. From this we get
  \begin{math}
    \AXdb \vdash E 
    = \sumiinI \: \alpha_i \pref E_i
    = \sumkinK \: \gamma_k \pref G_k
    = \sumkinK \: \zeta_{\mkern1mu k} \pref H_k
    = \sumjinJ \: \beta_j \pref F_j
    = F
  \end{math}
  which concludes the proof of the theorem.
\end{proof}

\noindent
In summary, in this section we describe a proof method for
completeness, in particular, of an axiomatisation of branching
{\bisimilarity} in the setting of a simple non-deterministic process
language. The key building blocks of the method are the completeness
of strong {\bisimilarity} and the lemma that every process has a
branching bisimilar concrete counterpart. In the sequel of the paper
we apply the proof method to obtain a complete axiomatisation of
branching probabilistic bisimulation for a process language featuring
both non-deterministic and probabilistic choice.



\section{Probabilistic processes} 
\label{sec-probabilistic}

\noindent
In this section we define a syntax and a transition system semantics
for probabilistic processes.
Following~\cite{BS01:icalp}, we start with adapting the syntax of the
processes of Section~\ref{sec-nondet}, by mixing non-deterministic and
probabilistic choice. Dependent on the top operator, a process is
either a non-deterministic process~$E \in \calE$, for
constant~$\bfzero$, prefix operators~$\alpha \pref {}$, and
non-deterministic choice~$+$, or a probabilistic
process~$P \in \calP$, for the Dirac operator~$\partial$ and
probabilistic choices~$\oplusr$.

\begin{definition}[Syntax]
  \label{def-prob-processes}
  The classes $\calE$ and~$\calP$ of non-deterministic and
  probabilistic processes, respectively, over the set of
  actions~$\calA$, are given by
  \begin{align*}
    E & \bnfeq \bfzero \mid \alpha \pref P \mid E + E \\
    P & \bnfeq \partial(E) \mid P \oplusr P 
  \end{align*}
  with actions~$\alpha$ from~$\calA$ where $r \in ( 0,1)$.
\end{definition}

\noindent
We use $E, F, G, \ldots {}$ to range over~$\calE$ and
$P, Q, R, \ldots {}$ to range over~$\calP$.
The probabilistic process $P_1 \mkern1mu \oplusr P_2$ executes the
behaviour of~$P_1$ with probability~$r$ and the behaviour~$P_2$ with
probability~$1-r$. By convention, $P \probc{1} Q$ denotes~$P$ and
$P \probc{\mkern1mu 0} Q$ denotes~$Q$.

In the sequel we often write $\bigoplus_{i{=}1}^n \: p_i \stardot E_i$ for
$n \geqslant 1$, $p_1, \ldots, p_n \geqslant 0$ such that
$p_1 + \cdots + p_n = 1$ and $E_1, \ldots, E_n \in \calE$ for the
probabilistic process given by
$\textstyle{\bigoplus_{i{=}1}^n} \: p_i \stardot E_i = \partial(E_1)$ if
$p_1=1$ and
$\textstyle{\bigoplus_{i{=}1}^n} \: p_i \stardot E_i = \partial(E_1)
\probc{p_1} \bigl( \textstyle{\bigoplus_{i{=}2}^n} \:
\frac{p_i}{1-p_1} \stardot E_i \bigr)$
otherwise. For example, for non-deterministic processes $E$, $F$
and~$G$, we have that
$\frac12 \stardot E \oplus \frac13 \stardot F \oplus \frac16 \stardot G$ is the probabilistic
process
$\partial(E) \probc{\frac12} ( \partial(F) \probc{\frac23}
\partial(G))$.
Similarly we use the notation $\bigoplus_{i{=}1}^n \: p_i \stardot P_i$ where
$n \geqslant 1$, $p_1, \ldots, p_n > 0$, $p_1 + \cdots + p_n = 1$ and
$P_1, \ldots, P_n \in \calP$, and also the notation
$\bigoplusiinI \: p_i \stardot P_i$ for an ordered finite index set~$I$.
As a consequence, each non-deterministic process $E \in \calE$ and
each probabilistic process $P \in \calP$ can be written as
\begin{displaymath}
  {E = \sumiinI \: \alpha_i \pref P_i} 
  \quad \text{and} \quad
  {P = \bigoplusjinJ \: p_j \stardot E_j}
\end{displaymath}
for suitable index sets $I$ and~$J$, $\alpha_i \in \calA$, $P_i \in \calP$
for~$i \in I$, and $p_j > 0$, $E_j \in \calE$ for~$j \in J$ such that
$\sum_{j\in J} \: p_j = 1$. The proof is straightforward by induction
on the structure of terms.

\begin{figure}
\IfFileExists{SkipFigure}{\vspace{6in}}{
  \centering
  \resizebox{\linewidth}{!}{
  \begin{tabular}{|c|c|c|}
    \hline



\begin{tikzpicture}[> = stealth', semithick, inner sep = 1pt, scale=0.45]
  \tikzstyle{state} = [circle, draw=Black, fill=LightYellow, minimum size=0.0mm]
  \tikzstyle{split} = [circle, draw=black, fill=black]
  \node (top) at (0,9.75) {} ;
  \node (name) at (0,9) {$\mathbf{H_1}$} ;
  \node [state] (t1) at (0,7.75) {}; 
  \node [split] (y1) at (0,6) {}; 
  \coordinate (t3) at (-3.25,4.25); 
  \coordinate (t34) at (-0.75,4.25); 
  \coordinate (t4) at (2,4.25); 
  \node at (-0.5,1.20) (process) { $a \pref \bigl( P \probc{\frac14} (
    P \probc{\frac13} Q ) \bigr)$} ;  
  \path [>=] (t1) edge node [midway, left=0.5pt] {\scriptsize $a$} (y1);
  \path [->, shorten >=0.25mm] (y1) edge node [pos=0.70, above left]
  {\scriptsize $\frac14$} (t3); 
  \path [->, shorten >=0.25mm] (y1) edge node [pos=0.90, above left]
  {\scriptsize $\frac{1}{4}$} (t34);
  \path [->, shorten >=0.25mm] (y1) edge node [pos=0.70, above right]
  {\scriptsize $\frac{1}{2}$} (t4);
  \draw pic[draw, angle eccentricity=1.2, angle radius=4pt] {angle=t3--y1--t4};
  \fill [fill=Pink, draw=Black, >=] (t3) -- ++(-1.0,-2.0) -- ++(2.0,0.0) -- (t3);  
  \node at ([shift={(0,-1.4)}] t3) {P};
  \fill [fill=Pink, draw=Black, >=] (t34) -- ++(-1.0,-2.0) -- ++(2.0,0.0) -- (t34);  
  \node at ([shift={(0,-1.4)}] t34) {P};
  \fill [fill=LightBlue, draw=Black, >=] (t4) -- ++(+1.0,-2.0) -- ++(-2.0,0.0) -- (t4); 
  \node at ([shift={(0,-1.4)}] t4) {Q};
\end{tikzpicture}
    & 



\begin{tikzpicture}[> = stealth', semithick, inner sep = 1pt,
  scale=0.45, , angle eccentricity=1.2, angle radius=4pt]
  \tikzstyle{state} = [circle, draw=Black, fill=LightYellow, minimum size=0.0mm]
  \tikzstyle{split} = [circle, draw=black, fill=black]
  \node (top) at (0,9.50) {} ;
  \node (name) at (0,8.75) {$\mathbf{G_1}$} ;
  \node [state] (t1) at (0,7.75) {}; 
  \node [split] (y1) at (0,6) {}; 
  \coordinate (t3) at (-2,4.25); 
  \coordinate (t4) at (+2,4.25);
  \node (process) at (0,1.20) {$a \pref ( P \probc{\frac12} Q )$} ;
  \path [>=] (t1) edge node [midway, left=0.5pt] {\scriptsize $a$} (y1);
  \path [->, shorten >=0.25mm] (y1) edge node [pos=0.50, above left]
  {\scriptsize $\frac12$} (t3); 
  \path [->, shorten >=0.25mm] (y1) edge node [pos=0.50, above right]
  {\scriptsize $\frac12$} (t4);
  \draw pic[draw]
    {angle=t3--y1--t4};
  \fill [fill=Pink, draw=Black, >=] (t3) -- ++(-1.0,-2.0) --
  ++(2.0,0.0) -- (t3); \node at ([shift={(0,-1.4)}] t3) {P};
  \fill [fill=LightBlue, draw=Black, >=] (t4) -- ++(+1.0,-2.0) --
  ++(-2.0,0.0) -- (t4); \node at ([shift={(0,-1.4)}] t4) {Q}; 
\end{tikzpicture}
    & 



\begin{tikzpicture}[> = stealth', semithick, inner sep = 1pt,
  scale=0.45, angle eccentricity=1.2, angle radius=5pt]
  \tikzstyle{state} = [circle, draw=Black, fill=LightYellow, minimum size=0.0mm]
  \tikzstyle{split} = [circle, draw=black, fill=black]
  \tikzstyle{dtrans} = [>=]
  \tikzstyle{ptrans} = [->, shorten >=0.25mm]
  \node (top) at (2,10.75) {} ;
  \node (name) at (1,9.75) {$\mathbf{I_1}$} ;
  \node [state] (I0) at ( 2,9.75) {}; 
  \node [split] (I1) at ( 2,7.75) {}; 
  \node [state] (I2) at ( 2,5.75) {};
  \coordinate (I3) at ( 0,4.25) {};
  \coordinate (I4) at ( 4,4.25) {};
  \node (process) at (2,1.2) { $a \pref \partial( b \pref P + \tau \pref Q )$} ;
  \path [dtrans] (I0) edge node [midway, left=0.5pt] {\scriptsize
    $a$} (I1); 
  \path [ptrans] (I1) edge node [pos=0.35, left=1pt] {\scriptsize
    $1$} (I2); 
  \path [dtrans] (I2) edge node [pos=0.50, above left] {\scriptsize
    $b$} (I3) ; 
  \path [dtrans] (I2) edge node [pos=0.50, above right] {\scriptsize
    $\tau$} (I4) ; 
  \fill [fill=Pink, draw=Black, >=] (I3) -- ++(-1.0,-2.0) --
  ++(2.0,0.0) -- (I3); \node at ([shift={(0,-1.4)}] I3) {P};
  \fill [fill=LightBlue, draw=Black, >=] (I4) -- ++(+1,-2.0) --
  ++(-2.0,0.0) -- (I4); \node at ([shift={(-0,-1.4)}] I4) {Q};
\end{tikzpicture} \\
    \hline



\begin{tikzpicture}[> = stealth', semithick, inner sep = 1pt, scale=0.45]
  \tikzstyle{state} = [circle, draw=Black, fill=LightYellow, minimum size=0.0mm]
  \tikzstyle{split} = [circle, draw=black, fill=black]
  \node (name) at (0,9) {$\mathbf{H_2}$} ;
  \node [state] (t1) at (0,7.75) {}; 
  \node [split] (y1) at (0,6) {}; 
  \coordinate (t3) at (-2,4.25); 
  \coordinate (t34) at (+0.75,4.25); 
  \coordinate (t4) at (+3.25,4.25); 
  \node at (+0.5,1) (process) {$a \pref \bigl( P
    \probc{\frac12} ( Q \probc{\frac12} Q ) \bigr)$} ; 
  \path [>=] (t1) edge node [midway, left=0.5pt] {\scriptsize $a$} (y1);
  \path [->, shorten >=0.25mm] (y1) edge node [pos=0.70, above left]
  {\scriptsize $\frac12$} (t3); 
  \path [->, shorten >=0.25mm] (y1) edge node [pos=0.90, above right]
  {\scriptsize $\frac{1}{4}$} (t34);
  \path [->, shorten >=0.25mm] (y1) edge node [pos=0.70, above right]
  {\scriptsize $\frac{1}{4}$} (t4);
  \draw pic[draw, angle eccentricity=1.2, angle radius=4pt] {angle=t3--y1--t4};
  \fill [fill=Pink, draw=Black, >=] (t3) -- ++(-1.0,-2.0) -- ++(2.0,0.0) -- (t3);  
  \node at ([shift={(0,-1.4)}] t3) {P};
  \fill [fill=LightBlue, draw=Black, >=] (t34) -- ++(-1.0,-2.0) -- ++(2.0,0.0) -- (t34);  
  \node at ([shift={(0,-1.4)}] t34) {Q};
  \fill [fill=LightBlue, draw=Black, >=] (t4) -- ++(+1.0,-2.0) -- ++(-2.0,0.0) -
- (t4); 
  \node at ([shift={(0,-1.4)}] t4) {Q};
\end{tikzpicture}
    & 



\begin{tikzpicture}[> = stealth', semithick, inner sep = 1pt,
  scale=0.45, angle eccentricity=1.2, angle radius=4pt]
  \tikzstyle{state} = [circle, draw=Black, fill=LightYellow, minimum size=0.0mm]
  \tikzstyle{split} = [circle, draw=black, fill=black]
  \node (top) at (-1,6.50) {} ;
  \node (name) at (-1,5.75) {$\mathbf{G_2}$} ;
  \node [state] (u0) at (0,5.75) {}; 
  \node [split] (u1) at (0,4) {};
  \coordinate   (u3) at (+0.5,2.25) {};
  \coordinate   (u4) at (+4.0,2.25) {};
  \node [state] (u5) at (-2,2.25) {};
  \node [split] (u6) at (-2,0.25) {};
  \coordinate   (u7) at (-4,-1.50); 
  \coordinate   (u8) at (+0,-1.50); 
  \node (process) at (0,-4.55) { $a \pref \bigl( \partial \bigl( \tau \pref ( P
    \probc{\frac12} Q ) \bigl) \probc{\frac13} ( P \probc{\frac12} Q )
    \bigl)$} ;
  \path [>=] (u0) edge node [midway, left=0.5pt] {\scriptsize $a$} (u1);
  \path [->, shorten >=0.25mm] (u1) edge node [pos=0.70, above left] {\scriptsize $\frac13$} (u5); 
  \path [->, shorten >=0.25mm] (u1) edge node [pos=0.50, right=+2pt] {\scriptsize $\frac13$ } (u3);
  \path [->, shorten >=0.25mm] (u1) edge node [pos=0.70, above right] {\scriptsize $\frac13$ } (u4);
  \path [>=] (u5) edge node [midway, left=0.5pt] {\scriptsize $\tau$} (u6);
  \path [->, shorten >=0.25mm] (u6) edge node [pos=0.70, above left] {\scriptsize $\frac12$} (u7); 
  \path [->, shorten >=0.25mm] (u6) edge node [pos=0.70, above right] {\scriptsize $\frac12$} (u8);
  \draw pic[draw] {angle=u5--u1--u4};
  \draw pic[draw] {angle=u7--u6--u8};
  \fill [fill=Pink, draw=Black, >=] (u3) -- ++(-1.0,-2.0) --
  ++(2.0,0.0) -- (u3); \node at ([shift={(0,-1.4)}] u3) {P};
  \fill [fill=LightBlue, draw=Black, >=] (u4) -- ++(+1.0,-2.0) --
  ++(-2.0,0.0) -- (u4); \node at ([shift={(0,-1.4)}] u4) {Q}; 
  \fill [fill=Pink, draw=Black, >=] (u7) -- ++(-1.0,-2.0) --
  ++(2.0,0.0) -- (u7); \node at ([shift={(0,-1.4)}] u7) {P};
  \fill [fill=LightBlue, draw=Black, >=] (u8) -- ++(+1.0,-2.0) --
  ++(-2.0,0.0) -- (u8); \node at ([shift={(0,-1.4)}] u8) {Q}; 

\end{tikzpicture}
    & 



\begin{tikzpicture}[> = stealth', semithick, inner sep = 1pt,
  scale=0.45, angle eccentricity=1.2, angle radius=5pt]
  \tikzstyle{state} = [circle, draw=Black, fill=LightYellow, minimum size=0.0mm]
  \tikzstyle{split} = [circle, draw=black, fill=black]
  \tikzstyle{dtrans} = [>=]
  \tikzstyle{ptrans} = [->, shorten >=0.25mm]
  \node (top) at (3,13.75) {} ;
  \node (name) at (3,13.00) {$\mathbf{I_2}$} ;
  \node [state] (P0) at ( 4,13.00) {};
  \node [split] (P1) at ( 4,11.25) {};
  \node [state] (I0) at ( 4,9.25) {};
  \node [split] (I1) at ( 2,7.75) {}; 
  \node [state] (I2) at ( 2,5.75) {};
  \coordinate (I3) at ( 0,4.25) {};
  \coordinate (I4) at ( 4,4.25) {};
  \coordinate (I5) at ( 5,7.75) {};
  \coordinate (I6) at ( 7.5,7.75) {};
  \node (process) at (3.75,1.20) { $a \pref \partial(
    \tau \pref \partial( b \pref P + \tau \pref Q ) + b \pref P + \tau \pref Q)$} ;
  \path [dtrans] (P0) edge node [pos=0.50, left] {\scriptsize
    $a$} (P1) ; 
  \path [ptrans] (P1) edge node [pos=0.350, left=1pt] {\scriptsize
    $1$} (I0) ; 
  \path [dtrans] (I0) edge node [pos=0.50, above left] {\scriptsize
    $\tau$} (I1) ; 
  \path [ptrans] (I1) edge node [pos=0.35, left=1pt] {\scriptsize
    $1$} (I2); 
  \path [dtrans] (I2) edge node [pos=0.50, above left] {\scriptsize
    $b$} (I3) ; 
  \path [dtrans] (I2) edge node [pos=0.50, above right] {\scriptsize
    $\tau$} (I4) ; 
  \path [dtrans] (I0) edge node [pos=0.80, above right] {\scriptsize
    $b$} (I5) ; 
  \path [dtrans] (I0) edge node [pos=0.50, above right] {\scriptsize
    $\tau$} (I6) ; 
  \fill [fill=Pink, draw=Black, >=] (I3) -- ++(-1.0,-2.0) --
  ++(2.0,0.0) -- (I3); \node at ([shift={(0,-1.4)}] I3) {P};
  \fill [fill=LightBlue, draw=Black, >=] (I4) -- ++(+1,-2.0) --
  ++(-2.0,0.0) -- (I4); \node at ([shift={(-0,-1.4)}] I4) {Q};
  \fill [fill=Pink, draw=Black, >=] (I5) -- ++(-1.0,-2.0) --
  ++(2.0,0.0) -- (I5); \node at ([shift={(0,-1.4)}] I5) {P};
  \fill [fill=LightBlue, draw=Black, >=] (I6) -- ++(-1.0,-2.0) --
  ++(2.0,0.0) -- (I6); \node at ([shift={(0,-1.4)}] I6) {Q};
\end{tikzpicture} \\
    \hline



\begin{tikzpicture}[> = stealth', semithick, inner sep = 1pt,
  scale=0.45, angle eccentricity=1.2, angle radius=5pt]
  \tikzstyle{state} = [circle, draw=Black, fill=LightYellow, minimum size=0.0mm]
  \tikzstyle{split} = [circle, draw=black, fill=black]
  \tikzstyle{dtrans} = [>=]
  \tikzstyle{ptrans} = [->, shorten >=0.25mm]
  \node (name) at (0,9) {$\mathbf{B_1}$} ;
  \node [state] (B1) at ( 0,7.75) {}; 
  \node [split] (B2) at ( 0,5.75) {};
  \coordinate (B3) at ( 0,3.75) {};
  \node (process) at (0,.7) { $a \pref \partial(E + F)$} ;
  \path [dtrans] (B1) edge node [midway, left=0.5pt] {\scriptsize $a$} (B2);
  \path [ptrans] (B2) edge node [pos=0.350, left=1pt] {\scriptsize
    $1$} (B3) ; 
  \fill [fill=LightBlue, draw=Black, >=] (B3) -- ++(+0.25,-2.0) --
  ++(2.0,0.0) -- (B3); \node at ([shift={(0.75,-1.4)}] B3) {F};
  \fill [fill=Pink, draw=Black, >=] (B3) -- ++(-0.25,-2.0) --
  ++(-2.0,0.0) -- (B3); \node at ([shift={(-0.75,-1.4)}] B3) {E};
\end{tikzpicture}
    & 



\begin{tikzpicture}[> = stealth', semithick, inner sep = 1pt,
  scale=0.45, angle eccentricity=1.2, angle radius=5pt]
  \tikzstyle{state} = [circle, draw=Black, fill=LightYellow, minimum size=0.0mm]
  \tikzstyle{split} = [circle, draw=black, fill=black]
  \tikzstyle{dtrans} = [>=]
  \tikzstyle{ptrans} = [->, shorten >=0.25mm]
  \node (top) at (0,13.75) {} ;
  \node (name) at (-1,13) {$\mathbf{T_1}$} ;
  \node [state] (w0) at (0,13) {}; 
  \node [split] (w1) at (0,11.25) {}; 
  \node [state] (w2) at (-2.75,9.50) {};
  \node [split] (w3) at (-2.75,7.50) {}; 
  \coordinate   (w4) at (-4.00,5.75) {};
  \coordinate   (w5) at (-1.50,5.75) {};
  \node [state] (w6) at (+2.75,9.50) {};
  \node [split] (w7) at (+2.75,7.50) {}; 
  \coordinate   (w8) at (+1.50,5.75) {};
  \coordinate   (w9) at (+4.00,5.75) {};
  \node (process) at (0,2.7) { $a \pref {\color{blue}\big(} \mkern1mu
    \partial( b \pref ( P \probc{\frac13} Q ) ) \probc{\frac12}
    \partial( b \pref ( P \probc{\frac23} Q ) ){\color{blue}\big)}$} ; 
  \path [dtrans] (w0) edge node [left=1pt] {\scriptsize $a$} (w1); 
  \path [ptrans] (w1) edge node [above left=0.5pt] {\scriptsize $\frac12$} (w2); 
  \path [ptrans] (w1) edge node [above right=0.5pt] {\scriptsize $\frac12$} (w6);
  \path [dtrans] (w2) edge node [left] {\scriptsize $b$} (w3); 
  \path [ptrans] (w3) edge node [pos=0.70, above left] {\scriptsize $\frac13$} (w4);
  \path [ptrans] (w3) edge node [pos=0.70, above right] {\scriptsize $\frac23$} (w5);
  \path [dtrans] (w6) edge node [right=0.5pt] {\scriptsize $b$} (w7); 
  \path [ptrans] (w7) edge node [pos=0.70, above left] {\scriptsize $\frac23$} (w8);
  \path [ptrans] (w7) edge node [pos=0.70, above right] {\scriptsize $\frac13$} (w9);
  \fill [fill=Pink, draw=Black, >=] (w4) -- ++(-1.0,-2.0) --
  ++(2.0,0.0) -- (w4); \node at ([shift={(0,-1.4)}] w4) {P};
  \fill [fill=LightBlue, draw=Black, >=] (w5) -- ++(-1.0,-2.0) --
  ++(2.0,0.0) -- (w5); \node at ([shift={(0,-1.4)}] w5) {Q};
  \fill [fill=Pink, draw=Black, >=] (w8) -- ++(-1.0,-2.0) --
  ++(2.0,0.0) -- (w8); \node at ([shift={(0,-1.4)}] w8) {P};
  \fill [fill=LightBlue, draw=Black, >=] (w9) -- ++(-1.0,-2.0) --
  ++(2.0,0.0) -- (w9); \node at ([shift={(0,-1.4)}] w9) {Q};
  \draw pic [draw] {angle=w2--w1--w6};
  \draw pic [draw] {angle=w4--w3--w5};
  \draw pic [draw] {angle=w8--w7--w9};
\end{tikzpicture}
    & 



\begin{tikzpicture}[> = stealth', semithick, inner sep = 1pt, scale=0.45]
  \tikzstyle{state} = [circle, draw=Black, fill=LightYellow, minimum size=0.0mm]
  \tikzstyle{split} = [circle, draw=black, fill=black]
  \tikzstyle{dtrans} = [>=]
  \tikzstyle{ptrans} = [->, shorten >=0.25mm]
  \node (name) at (3,5.75) {$\mathbf{E_1}$} ;
  \node [state] (u1) at (4,5.75) {}; 
  \node [split] (z1) at (4,4) {};
  \node [state] (u3) at (+2,2.50) {};
  \node [state] (z6) at (+6,2.50) {}; 
  \node (process) at (2.5,-2.55) { $\tau \pref \mkern1mu (
    \partial( \tau \pref P + a \pref Q + \tau \pref R )
    \probc{\frac34} \partial(\bfzero) )$} ;
  \coordinate (z3) at (-0.25,0.75) {};
  \coordinate (z4) at (+2,0.75) {};
  \coordinate (z5) at (+4.25,0.75) {};
  \path [dtrans] (u1) edge node [midway, left=0.5pt] {\scriptsize $\tau$} (z1);
  \path [ptrans] (z1) edge node [pos=0.50, above right] {\scriptsize
    $\frac14$} (z6);  
  \path [ptrans] (z1) edge node [pos=0.50, above left] {\scriptsize
    $\frac34$} (u3); 
  \path [dtrans] (u3) edge node [pos=0.70, above left] {\scriptsize $\tau$} (z3);
  \path [dtrans] (u3) edge node [pos=0.70, left=2pt] {\scriptsize $a$} (z4);
  \path [dtrans] (u3) edge node [pos=0.70, above right] {\scriptsize $\tau$} (z5);
  \fill [fill=Pink, draw=Black, >=] (z3) -- ++(-1.0,-2.0) --
  ++(2.0,0.0) -- (z3); \node at ([shift={(0,-1.4)}] z3) {P};
  \fill [fill=LightBlue, draw=Black, >=] (z4) -- ++(-1.0,-2.0) --
  ++(2.0,0.0) -- (z4); \node at ([shift={(0,-1.4)}] z4) {Q};
  \fill [fill=LightYellow, draw=Black, >=] (z5) -- ++(+1.0,-2.0) --
  ++(-2.0,0.0) -- (z5); \node at ([shift={(0,-1.4)}] z5) {R}; 
  \draw pic[draw, angle eccentricity=1.2, angle radius=4pt] {angle=u3--z1--z6};
\end{tikzpicture}\\
    \hline



\begin{tikzpicture}[> = stealth', semithick, inner sep = 1pt,
  scale=0.45, angle eccentricity=1.2, angle radius=5pt]
  \tikzstyle{state} = [circle, draw=Black, fill=LightYellow, minimum size=0.0mm]
  \tikzstyle{split} = [circle, draw=black, fill=black]
  \tikzstyle{dtrans} = [>=]
  \tikzstyle{ptrans} = [->, shorten >=0.25mm]
  \node (name) at (0,9) {$\mathbf{B_2}$} ;
  \node [state] (B1) at ( 0,7.75) {}; 
  \node [split] (B2) at ( 0,5.75) {};
  \node [state] (B3) at ( 0,3.75) {};
  \node [split] (B4) at (-2,2.00) {};
  \coordinate (B5) at (-2,0.00) {};
  \node (process) at (-2.125,-3.05) { $a \pref \partial \bigl( \tau \pref \partial(E + F) + F \bigr)$} ;
  \path [dtrans] (B1) edge node [midway, left=0.5pt] {\scriptsize $a$} (B2);
  \path [ptrans] (B2) edge node [pos=0.350, left=1pt] {\scriptsize
    $1$} (B3) ; 
  \path [dtrans] (B3) edge node [pos=0.70, above left] {\scriptsize
    $\tau$} (B4); 
  \path [ptrans] (B4) edge node [pos=0.350, left=1pt] {\scriptsize $1$} (B5); 
  \fill [fill=LightBlue, draw=Black, >=] (B3) -- ++(+0.25,-2.0) --
  ++(2.0,0.0) -- (B3); \node at ([shift={(0.75,-1.4)}] B3) {F};
  \fill [fill=Pink, draw=Black, >=] (B5) -- ++(-0.25,-2.0) --
  ++(-2.0,0.0) -- (B5); \node at ([shift={(-0.75,-1.4)}] B5) {E};
  \fill [fill=LightBlue, draw=Black, >=] (B5) -- ++(+0.25,-2.0) --
  ++(2.0,0.0) -- (B5); \node at ([shift={(0.75,-1.4)}] B5) {F};
\end{tikzpicture}
    & 



\begin{tikzpicture}[> = stealth', semithick, inner sep = 1pt,
  scale=0.45, angle eccentricity=1.2, angle radius=5pt]
  \tikzstyle{state} = [circle, draw=Black, fill=LightYellow, minimum size=0.0mm]
  \tikzstyle{split} = [circle, draw=black, fill=black]
  \tikzstyle{dtrans} = [>=]
  \tikzstyle{ptrans} = [->, shorten >=0.25mm]
  \node (top) at (5.25,14.00) {} ;
  \node (name) at (4.25,13.25) {$\mathbf{T_2}$} ;
  \node [split] (w1) at (0,7.50) {}; 
  \node [state] (w2) at (-2.75,5.75) {};
  \node [split] (w3) at (-2.75,3.75) {}; 
  \coordinate   (w4) at (-4.00,2.00) {};
  \coordinate   (w5) at (-1.50,2.00) {};
  \node [state] (w6) at (+2.75,5.75) {};
  \node [split] (w7) at (+2.75,3.75) {}; 
  \coordinate   (w8) at (+1.50,2.00) {};
  \coordinate   (w9) at (+4.00,2.00) {};
  \node [state] (v0) at (5.25,13.00) {}; 
  \node [split] (v1) at (5.25,11.25) {}; 
  \node [state] (v2) at (5.25,9.25) {};
  \node [split] (v3) at (8.25,7.50) {}; 
  \coordinate   (v4) at (6.50,5.75) {}; 
  \coordinate   (v5) at (10.50,5.75) {}; 
  \node (process) at (3.25,-1.55) { $a \pref \mkern1mu
    \partial \left( \begin{array}{@{}l@{}}\tau \pref \bigl(
    \partial( b \pref ( P \probc{\frac13} 
    Q ) ) \probc{\frac12} \partial( b \pref ( P \probc{\frac23} Q ) ) \bigr) \\
    \mbox{} + b \pref ( P \probc{\frac12} Q ) \end{array}
    \right) $} ;
  \path [ptrans] (w1) edge node [above left=0.5pt] {\scriptsize $\frac12$} (w2); 
  \path [ptrans] (w1) edge node [above right=0.5pt] {\scriptsize $\frac12$} (w6);
  \path [dtrans] (w2) edge node [left] {\scriptsize $b$} (w3); 
  \path [ptrans] (w3) edge node [pos=0.70, above left] {\scriptsize $\frac13$} (w4);
  \path [ptrans] (w3) edge node [pos=0.70, above right] {\scriptsize $\frac23$} (w5);
  \path [dtrans] (w6) edge node [right=0.5pt] {\scriptsize $b$} (w7); 
  \path [ptrans] (w7) edge node [pos=0.70, above left] {\scriptsize $\frac23$} (w8);
  \path [ptrans] (w7) edge node [pos=0.70, above right] {\scriptsize $\frac13$} (w9);
  \path [dtrans] (v0) edge node [left=1pt] {\scriptsize $a$} (v1); 
  \path [ptrans] (v1) edge node [pos=0.35,left=1pt] {\scriptsize $1$} (v2); 
  \path [dtrans] (v2) edge node [above left] {\scriptsize $\tau$} (w1);
  \path [dtrans] (v2) edge node [above right] {\scriptsize $b$} (v3);
  \path [ptrans] (v3) edge node [above left] {\scriptsize $\frac12$} (v4); 
  \path [ptrans] (v3) edge node [above right] {\scriptsize $\frac12$} (v5); 
  \fill [fill=Pink, draw=Black, >=] (w4) -- ++(-1.0,-2.0) --
  ++(2.0,0.0) -- (w4); \node at ([shift={(0,-1.4)}] w4) {P};
  \fill [fill=LightBlue, draw=Black, >=] (w5) -- ++(-1.0,-2.0) --
  ++(2.0,0.0) -- (w5); \node at ([shift={(0,-1.4)}] w5) {Q};
  \fill [fill=Pink, draw=Black, >=] (w8) -- ++(-1.0,-2.0) --
  ++(2.0,0.0) -- (w8); \node at ([shift={(0,-1.4)}] w8) {P};
  \fill [fill=LightBlue, draw=Black, >=] (w9) -- ++(-1.0,-2.0) --
  ++(2.0,0.0) -- (w9); \node at ([shift={(0,-1.4)}] w9) {Q};
  \fill [fill=Pink, draw=Black, >=] (v4) -- ++(-1.0,-2.0) --
  ++(2.0,0.0) -- (v4); \node at ([shift={(0,-1.4)}] v4) {P};
  \fill [fill=LightBlue, draw=Black, >=] (v5) -- ++(-1.0,-2.0) --
  ++(2.0,0.0) -- (v5); \node at ([shift={(0,-1.4)}] v5) {Q};
  \draw pic [draw] {angle=w2--w1--w6};
  \draw pic [draw] {angle=w4--w3--w5};
  \draw pic [draw] {angle=w8--w7--w9};
  \draw pic [draw] {angle=v4--v3--v5};
\end{tikzpicture}
    & 



\begin{tikzpicture}[> = stealth', semithick, inner sep = 1pt, scale=0.45]
  \tikzstyle{state} = [circle, draw=Black, fill=LightYellow, minimum size=0.0mm]
  \tikzstyle{split} = [circle, draw=black, fill=black]
  \tikzstyle{dtrans} = [>=]
  \tikzstyle{ptrans} = [->, shorten >=0.25mm]
  \node (top) at (0,6.50) {} ;
  \node (name) at (-1,5.75) {$\mathbf{E_2}$} ;
  \node [state] (u1) at (0,5.75) {}; 
  \node [split] (z1) at (0,4) {};
  \node [state] (u2) at (-3,2.50) {};
  \coordinate (z3) at (-5.25,0.75) {};
  \coordinate (z4) at (-3,0.75) {};
  \coordinate (z5) at (-0.75,0.75) {};
  \node [state] (u3) at (+3,2.50) {}; 
  \node [split] (z7) at (+3,-0.75) {}; 
  \node [state] (u4) at (+1,-2.75) {}; 
  \coordinate (z10) at (-1.25,-4.5) {}; 
  \coordinate (z11) at (+1.00,-4.5) ; 
  \coordinate (z12) at (+3.25,-4.5) {}; 
  \node [state] (u5) at (+5,-2.75) {};
  \node (process) at (-0.5,-8) { $\tau \pref \left(\begin{array}{@{}l@{}}
    \partial( \tau \pref P + a \pref Q + \tau \pref R )
    \probc{\frac12} {} \\
    \partial( \tau \pref \big( \partial\big( \tau \pref P + a \pref Q + \tau \pref R )
    \probc{\frac12} \partial(\bfzero) \big)\big)\end{array}\right)\!\!\!$} ;
  \path [dtrans] (u1) edge node [midway, left=0.5pt] {\scriptsize $\tau$} (z1);
  \path [ptrans] (z1) edge node [pos=0.50, above left] {\scriptsize $\frac12$} (u2); 
  \path [ptrans] (z1) edge node [pos=0.50, above right] {\scriptsize $\frac12$} (u3);
  \path [dtrans] (u2) edge node [pos=0.70, above left] {\scriptsize $\tau$} (z3);
  \path [dtrans] (u2) edge node [pos=0.70, left=2pt] {\scriptsize $a$} (z4);
  \path [dtrans] (u2) edge node [pos=0.70, above right] {\scriptsize $\tau$} (z5);
  \path [dtrans] (u3) edge node [pos=0.50, right] {\scriptsize $\tau$} (z7);
  \path [ptrans] (z7) edge node [pos=0.50, above left] {\scriptsize $\frac12$} (u4);
  \path [ptrans] (z7) edge node [pos=0.50, above right] {\scriptsize $\frac12$} (u5);
  \path [dtrans] (u4) edge node [pos=0.70, above left] {\scriptsize $\tau$} (z10);
  \path [dtrans] (u4) edge node [pos=0.70, left=2pt] {\scriptsize $a$} (z11);
  \path [dtrans] (u4) edge node [pos=0.70, above right] {\scriptsize $\tau$} (z12);
  \fill [fill=Pink, draw=Black, >=] (z3) -- ++(-1.0,-2.0) --
  ++(2.0,0.0) -- (z3); \node at ([shift={(0,-1.4)}] z3) {P};
  \fill [fill=LightBlue, draw=Black, >=] (z4) -- ++(-1.0,-2.0) --
  ++(2.0,0.0) -- (z4); \node at ([shift={(0,-1.4)}] z4) {Q};
  \fill [fill=LightYellow, draw=Black, >=] (z5) -- ++(+1.0,-2.0) --
  ++(-2.0,0.0) -- (z5); \node at ([shift={(0,-1.4)}] z5) {R}; 
  \fill [fill=Pink, draw=Black, >=] (z10) -- ++(-1.0,-2.0) --
  ++(2.0,0.0) -- (z10); \node at ([shift={(0,-1.4)}] z10) {P};
  \fill [fill=LightBlue, draw=Black, >=] (z11) -- ++(-1.0,-2.0) --
  ++(2.0,0.0) -- (z11); \node at ([shift={(0,-1.4)}] z11) {Q};
  \fill [fill=LightYellow, draw=Black, >=] (z12) -- ++(+1.0,-2.0) --
  ++(-2.0,0.0) -- (z12); \node at ([shift={(0,-1.4)}] z12) {R}; 
  \draw pic[draw, angle eccentricity=1.2, angle radius=4pt] {angle=u2--z1--u3};
  \draw pic[draw, angle eccentricity=1.2, angle radius=4pt] {angle=u4--z7--u5};
\end{tikzpicture} \\
    \hline
  \end{tabular}
  }
  }

  \caption{Pairwise rooted branching probabilistic bisimilar
    non-deterministic processes.}
  \label{fig:examples}
\end{figure}
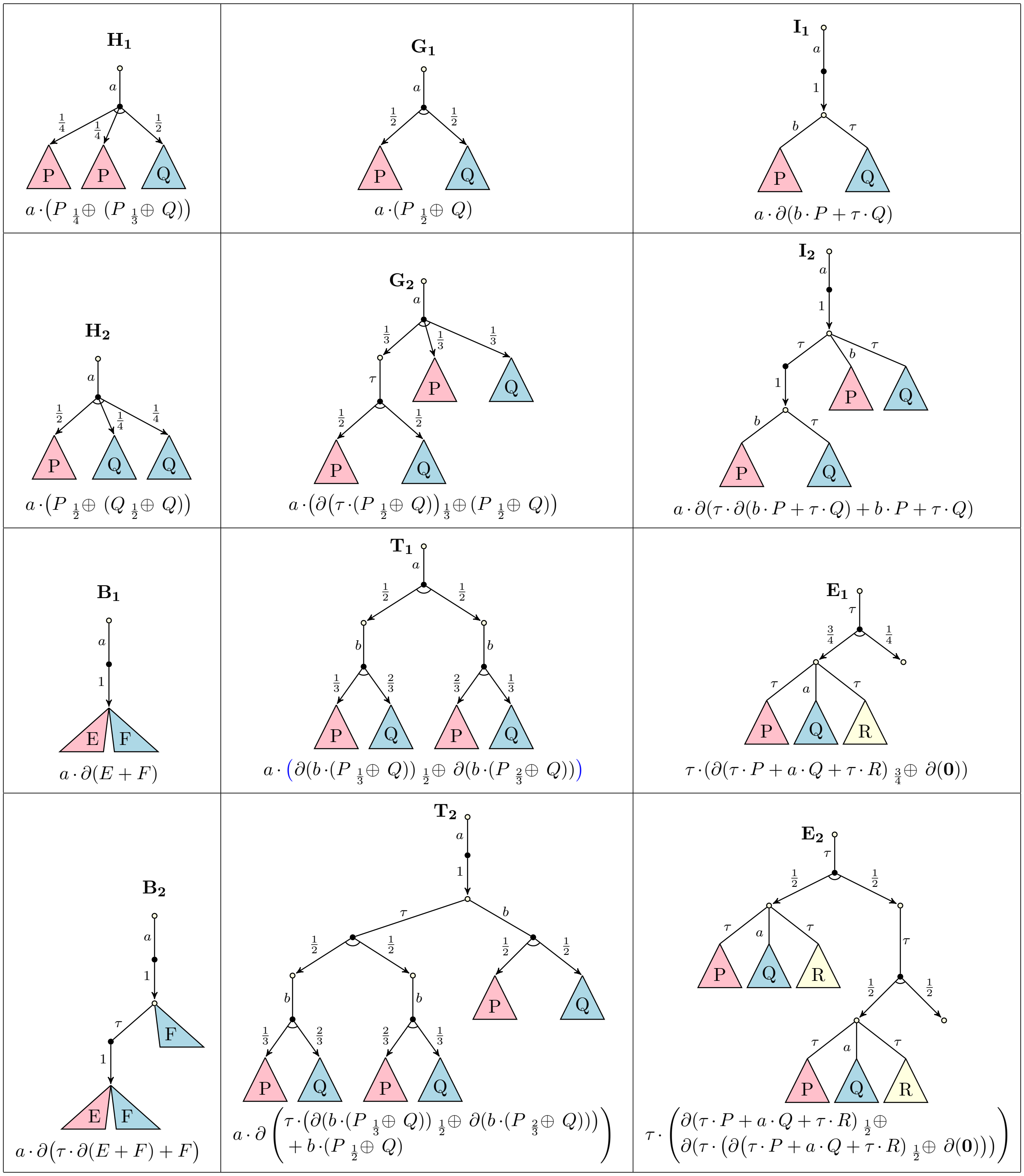

Figure~\ref{fig:examples} presents a number of example pairs of
(rooted branching probabilistic bisimilar) non-deterministic
processes collected from the literature.
The examples $\mathbf{H_1}$ and~$\mathbf{H_2}$ are taken
from~~\cite{Hen12:facj}; $\mathbf{G_1}$ and~$\mathbf{G_2}$ are taken
from~\cite{GGV19:fc}.
Examples $\mathbf{B_1}$ and~$\mathbf{B_2}$ express the axiom called~B
in~\cite{GW96:jacm}, cast to the setting of the present
paper.
The processes $\mathbf{I_1}$ and~$\mathbf{I_2}$ are typical
instances of $\mathbf{B_1}$ and~$\mathbf{B_2}$.
The two examples $\mathbf{E_1}$ and~$\mathbf{E_2}$ are adapted
from~\cite{EHKTZ13:qest}, while the new examples $\mathbf{T_1}$
and~$\mathbf{T_2}$ further illustrate the intricacy of the mingling of
probabilistic choice and action prefix.

\blankline

\noindent
As usual, the SOS semantics for $\calE$ and~$\calP$ makes use of two
types of transition relations~\cite{HJ90:rtss,BS01:icalp}.

\begin{definition}[Operational semantics]
  \label{def-pr-transition-relation} \mbox{}
  \begin{enumerate}[(a)]
    \item \label{transitrel-case_a}
    The transition relations ${\rightarrow} \subseteq
      \calE \times \calA \times \Distr(\calE)$ and ${\mapsto}
      \subseteq \calP \times \Distr(\calE)$ are given by\vspace{-2ex}
      \begin{displaymath}
        \begin{array}{c}
          \sosrule{P \mapsto \mu}{\alpha \pref P \arrow{\alpha} \mu}
          \: \textsc{\small (pref)} \rule{0pt}{22pt} \qquad
          \medskip \\
          \sosrule{E_1 \arrow{\alpha} \mu_1}{E_1 + E_2 \arrow{\alpha}
            \mu_1}
          \: \textsc{\small (nd-choice\,1)}
          \qquad
          \sosrule{E_2 \arrow{\alpha} \mu_2}{E_1 + E_2 \arrow{\alpha}
            \mu_2}
          \: \textsc{\small (nd-choice\,2)}
          \bigskip \\
          \sosrule{}{\partial(E) \mapsto \delta(E)}
          \: \textsc{\small (Dirac)}
          \qquad \quad \ 
          \sosrule{P_1 \mapsto \mu_1 \quad P_2 \mapsto
            \mu_2}{P_1 \oplusr P_2 \mapsto \mu_1 \oplusr \mu_2}
          \: \textsc{\small (p-choice)}
        \end{array}
      \end{displaymath}
    \item \label{transitrel-case_b}
      The transition relation ${\rightarrow}
      \subseteq \DistrE \times \calA \times \DistrE$ is such that $\mu
      \alphaarrow \mu'$ whenever $\mu = \bigoplusiinI \: p_i \stardot E_i$,
      $\mu' = \bigoplusiinI \: p_i \stardot \mu'_i $, and $E_i \alphaarrow
      \mu'_i$ for all $i \in I$.
      \vspace{-.5ex}
  \end{enumerate}
\end{definition}

\medskip

\noindent
With $\den{P}$, for $P\in\calP$, we denote the unique distribution
$\mu$ such that $P\mapsto\mu$. Thus, the probabilistic process
$\bigoplusiinI \: p_i \stardot E_i$ corresponds to the distribution
$\den{ \bigoplusiinI \: p_i \stardot E_i } = \bigoplusiinI \: p_i \cdot
\delta(E_i)$. Similarly, for $\bigoplusiinI \: p_i \stardot P_i \in \calP$ we
have
$\den{\bigoplusiinI \: p_i \stardot P_i} = \bigoplusiinI \: p_i \cdot
\den{P_i}$.

Following~\cite{SL94:concur,Seg95:thesis}, the transition
relation~$\rightarrow$ on distributions as given by
Definition~\ref{def-pr-transition-relation}\ref{transitrel-case_b}
allows for a probabilistic combination of non-deterministic
alternatives resulting in a so-called combined transition. For
example, for the process
$E = a \pref \mkern1mu (P \probc{\mytinyfrac12} Q) + a \pref \mkern1mu
(P \probc{\mytinyfrac13} Q)$ of~\cite{BS01:icalp}, we have that the
Dirac process
$\partial(E) = \partial( a \pref \mkern1mu (P \probc{\mytinyfrac12} Q)
+ a \pref \mkern1mu (P \probc{\mytinyfrac13} Q) )$ provides an
$a$-transition to $\den{P \probc{\mytinyfrac12} Q}$ as well as an
$a$-transition to $\den{P \probc{\mytinyfrac13} Q}$. However, since we
can represent the distribution~$\delta(E)$ by
$\delta(E) = \frac12 \delta(E) \oplus \frac12 \delta(E)$, the
distribution~$\delta(E)$ also has a combined transition
\begin{displaymath}
  \delta(E) 
  = \half \delta(E) \oplus \half \delta(E)
  \arrow{a} \half \den{P \probc{\mytinyfrac12} Q} \oplus \half \den{P
  \probc{\mytinyfrac13} Q} = \den{P \probc{\mytinyfrac{5}{12}} Q} .
\end{displaymath}
As noted in~\cite{Sto02:phd}, the ability to combine transitions is
crucial for obtaining transitivity of probabilistic process
equivalences that take internal actions into account.

\begin{example}
  Referring to some of the processes of Figure~\ref{fig:examples} we
  have, for example,
  \begin{align*}
    \mathbf{H_1} \colon \quad
    & \delta( a \pref \mkern1mu (P \probc{\mytinyfrac14} (P
      \probc{\mytinyfrac13} Q)) ) 
      \arrow{a} \den{P \probc{\mytinyfrac14} (P
      \probc{\mytinyfrac13} Q)} = 
      \den{P \probc{\mytinyfrac12} Q} \\ 
    \mathbf{H_2} \colon \quad
    & \delta( a \pref \mkern1mu (P \probc{\mytinyfrac12} (Q
      \probc{\mytinyfrac12} Q)) )
      \arrow{a} \den{P \probc{\mytinyfrac12} (Q
      \probc{\mytinyfrac12} Q)} = 
      \den{P \probc{\mytinyfrac12} Q} \\
    \mathbf{E_1} \colon \quad
    & \delta( \tau \pref \mkern1mu (
    \partial( \tau \pref P + a \pref Q + \tau \pref R )
      \probc{\frac34} \partial(\bfzero) ) )
      \arrow{\tau}
      \myfrac34 \delta(\tau \pref P + a \pref Q + \tau \pref R) \oplus
      \myfrac14 \delta( \bfzero )
      \mkern1mu .
  \end{align*}
\end{example}

\noindent
In preparation to the definition of the notion of branching
probabilistic {\bisimilarity} in Section~\ref{sec-bpb} we introduce
some notation.

\begin{definition}
  \label{def-weak-arrows}
  For $\mu, \mu' \mathbin\in \DistrE$ and $\alpha \mathbin\in \calA$,
  we write $\mu \alphahidearrow \mu'$ iff (i)~$\mu \alphaarrow \mu'$,
  or (ii)~$\alpha = \tau$ and $\mu'=\mu$, or (iii)~$\alpha = \tau$ and
  there exist $\mu_1, \mu_2, \mu'_1, \mu'_2 \in \DistrE$ such that
  $\mu = \mu_1 \oplusr \mu_2$, $\mu' = \mu'_1 \oplusr \mu'_2$,
  \vspace{1pt} $\mu_1 = \mu_1'$, and $\mu_2  \arrow{\tau} \mu'_2$ for
  some $r \in (0,1)$.
\end{definition}

\noindent
Cases (i) and (ii) in the definition above correspond with the limits
$r = 0$ and $r = 1$ of case (iii).
Following~\cite{DGHM09}, $\mu_1$ and~$\mu_2$ are the fragments of the
distribution~$\mu$ that stay in place, respectively, go forward via
a~$\tau$.
We use $\Arrow{\,}$ to denote the
reflexive transitive closure of $\arrow{(\tau)}$.
A transition $\mu \tauhidearrow \mu'$ is called a partial
transition, and a transition $\mu \Arrow{\;} \mu'$ is called a weak
transition.

\begin{example}
  \begin{itemize}
  \item [(a)] According to Definition~\ref{def-weak-arrows} we have, for
    example,\vspace{-1ex}
    \begin{align*}
      &
      \myfrac13 \delta( \tau \pref \mkern1mu (P \probc{\mytinyfrac12} Q)) \oplus
      \myfrac23 \den{ P \probc{\mytinyfrac12} Q }
      \  \arrow{(\tau)} \
      \myfrac13 \den{P \probc{\mytinyfrac12} Q}
      \oplus
      \myfrac23 \den{P \probc{\mytinyfrac12} Q}
      =
      \den{P \probc{\mytinyfrac12} Q}
      \mkern1mu .
    \end{align*}
  \item [(b)] There are typically multiple ways to construct a weak
    transition $\Arrow{\;}$. Consider the weak transition
    $\myfrac12 \delta( \tau \pref \partial ( \tau \pref P )) \oplus
    \myfrac13 \delta( \tau \pref P ) \oplus \myfrac16 \den{P}
    \Arrow{\:} \den{P}$ which can be obtained, among uncountably many
    other possibilities, via
    \begin{align*}
      & \myfrac12 \delta( \tau \pref \partial ( \tau \pref P ))
        \oplus
        \myfrac13 \delta( \tau \pref P )
        \oplus
        \myfrac16 \den{P}
        \tauhidearrow
        {} \smallskip \\
      & \qquad \qquad
        \myfrac12 \delta ( \tau \pref P ))
        \oplus
        \myfrac13 \delta( \tau \pref P )
        \oplus
        \myfrac16 \den{P}
        =
        \myfrac56 \delta( \tau \pref P )
        \oplus
        \myfrac16 \den{P}
        \tauhidearrow
        \den{P}
      \intertext{or via}
      & \myfrac12 \delta( \tau \pref \partial ( \tau \pref P ))
        \oplus
        \myfrac13 \delta( \tau \pref P )
        \oplus
        \myfrac16 \den{P}
        \tauhidearrow
        \myfrac12 \delta( \tau \pref \partial ( \tau \pref P ))
        \oplus
        \myfrac13 \delta( P )
        \oplus
        \myfrac16 \den{P}
        = {} \smallskip \\
      & \qquad \qquad
        \myfrac12 \delta( \tau \pref \partial ( \tau \pref P ))
        \oplus
        \myfrac12 \delta( P )
        \tauhidearrow
        \myfrac12 \delta ( \tau \pref P )
        \oplus
        \myfrac12 \den{P}
        \tauhidearrow
        \myfrac12 \den{P}
        \oplus
        \myfrac12 \den{P}
        =
        \den{P}
    \end{align*}
  \item [(c)] The distribution
    $\myfrac12 \delta( \tau \pref \partial(\astop + \bstop)) \oplus
    \myfrac12 \delta(a \pref \partial( \cstop ))$ doesn't admit a
    $\tau$-transition $\arrow{\tau}$ nor an $a$-transition
    $\arrow{a}$. However, we have
      \begin{align*}
        & \myfrac12 \delta( \tau \pref \partial(\astop + \bstop))
          \oplus
          \myfrac12 \delta(a \pref \partial( \cstop ))
          \tauhidearrow
          {} \smallskip \\
        & \qquad \qquad
          \myfrac12 \partial(\astop + \bstop)
          \oplus
          \myfrac12 \delta(a \pref \partial( \cstop ))
          \arrow{\mkern2mu a}
          \myfrac12 \delta(\bfzero) \oplus \myfrac12 \delta( \cstop )
          \mkern1mu .
      \end{align*}
    \end{itemize}
\end{example}

\noindent
The next lemma will play a crucial r\^ole in our forthcoming
congruence proof. Its proof can be found in the appendix.

\begin{lemma}
  \label{flexibelDelen}
  Let $I$ and~$J$ be finite index sets, $p_i,q_j\in[0,1]$ and
  $\xi, \mu_i,\nu_j \in \DistrE$, for $i \in I$ and $j \in J$, with
  both $\xi = \bigoplusiinI\: p_i\stardot\mu_i$ and
  $\xi =\bigoplusjinJ \: q_j \stardot \nu_j$. Then there are
  $r_{ij} \in [0,1]$ and $\varrho_{ij} \in \DistrE$ such that
  $\sum_{j\in J} \: r_{ij} = p_i$ and
  $p_i \stardot \mu_i = \bigoplusjinJ \: r_{ij} \stardot \varrho_{ij}$ for
  all~$i \in I$, and $\sum_{i\in I} \: r_{ij}=q_j$ and
  $q_j \stardot \nu_j = \bigoplusiinI \: r_{ij} \stardot \varrho_{ij}$ for
  all~$j \in J$.
\end{lemma}

\noindent
The following lemma says that the transitions $\arrow{\alpha}$,
$\arrow{(\alpha)}$ and $\Arrow{\;}$ of
Definitions~\ref{def-pr-transition-relation}\ref{transitrel-case_b}
and~\ref{def-weak-arrows} can be probabilistically composed.

\begin{lemma}
  \label{lem-composition}
  Let $\mu_i,\mu'_i\in \DistrE$ for a finite index set $I$, and let
  $\sum_{i\in I}p_i=1$. 
  \begin{itemize}
  \item [(a)] If $\mu_i \arrow{\alpha} \mu'_i$ for all $i\in I$, then
    $\bigoplusiinI \: p_i \stardot \mu_i \arrow{\alpha} \bigoplusiinI
    \: p_i \stardot \mu'_i$.
  \item [(b)] If $\mu_i \arrow{(\tau)} \mu'_i$ for all $i\in I$, then
    $\bigoplusiinI \: p_i \stardot \mu_i \arrow{(\tau)} \bigoplusiinI
    \: p_i \stardot \mu'_i$.
  \item [(c)] If $\mu_i \Arrow{\;} \mu'_i$ for all $i\in I$, then
    $\bigoplusiinI \: p_i \stardot \mu_i \Arrow{\;} \bigoplusiinI \:
    p_i \stardot \mu'_i$.
  \end{itemize}
\end{lemma}

\begin{proof}
  Let $\mu := \bigoplusiinI \: p_i \stardot \mu_i$ and
  $\mu' := \bigoplusiinI \: p_i \stardot \mu'_i$.  Without loss of
  generality, we may assume that $p_i>0$ for all $i \in I$.

  (a)~Suppose $\mu_i \arrow{\alpha} \mu'_i$ for all $i \in I$.  Then,
  for all $i \in I$, by
  Definition~\ref{def-pr-transition-relation}\ref{transitrel-case_b},
  exist index sets~$J_i$, $p_{ij} \geqslant 0$, and processes
    $E_{ij} \in \calE$, for $j \in J_i$, such that
  $\mu_i = \bigoplusjinJi \: p_{ij} \stardot E_{ij}$,
  $\mu'_i = \bigoplusjinJi \: p_{ij} \stardot \mu'_{ij} $, and
  $E_{ij} \alphaarrow \mu'_{ij}$ for all $j\in J_i$.  Let
  $K := \lc (i,j) \mid i \in I ,\, j \in J_i \rc$ and
  $q_{(i,j)} := p_i \cdot p_{ij}$ for all $(i,j)\in K$, so that
  $\sum_{k {\in} K} \: q_k=1$.  Then
  $\mu = \bigopluskinK \: q_k \stardot E_{ij}$ and
  $\mu' = \bigopluskinK \: q_k \stardot \mu_{ij}$, so
  $\mu \arrow{\alpha} \mu'$ by
  Definition~\ref{def-pr-transition-relation}\ref{transitrel-case_b}.

  (b)~Let $\mu_i \arrow{(\tau)} \mu'_i$ for all $i\in I$.  Then, for
  all $i \in I$, by Definition~\ref{def-weak-arrows}, there exists
  $r_i \in [0,1]$ and
  $\mu_i^{\rm stay}\!, \mu_i^{\rm go}, \mu''_i
  \mathbin\in \DistrE$ such that
  $\mu_i \mathbin= \mu_i^{\rm stay} \! \probc{r_i} \mu_i^{\rm go}$,
  $\mu'_i \mathbin= \mu_i^{\rm stay} \! \probc{r_i} \mu''_i$, and
  either $r_i \mathbin{=} 1$ or $\mu_i^{\rm go} \arrow{\tau} \mu_i''$.
  In case~$r_i = 0$ for all $i \in I$, we have that
  $\mu_i \arrow{\tau} \mu'_i$ for all $i \in I$, and thus
  $\mu \arrow{\tau} \mu'$ by the first claim of the lemma, and
  $\mu \tauhidearrow \mu'$ by Definition~\ref{def-weak-arrows}i.
  In case $r_i = 1$ for all $i \in I$, we have $\mu' = \mu$ and thus
  $\mu \tauhidearrow \mu'$ by Definition~\ref{def-weak-arrows}ii.
  Otherwise, let $I' := \lc i \in I \mid r_i<1 \rc$,
  $r = \sumiinI \: p_i\mycdot r_i$,
  $\mu^{\rm stay} := \bigoplusiinI \: \frac{p_i\mycdot r_i}{r}
  \stardot \mu_i^{\rm stay}\!$,
  $\mu^{\rm go} := \bigoplus_{i\in I'} \: \frac{p_i\mycdot \mkern2mu
    (1-r_i)}{1-r} \stardot \mu_i^{\rm go}$ and
  $\mu'' := \bigoplus_{i\in I'} \: \frac{p_i\mycdot \mkern2mu (1-r_i)}{1-r}
  \stardot \mu''_i$. Then $\mu^{\rm go} \arrow\tau \mu''$ by the first
  claim of the lemma. Moreover, \vspace{2pt}
  $\mu = \mu^{\rm stay} \! \probc{r} \mu^{\rm go}$,
  $\mu' = \mu^{\rm stay} \! \probc{r} \mu''$ and $r \in (0,1)$.  So
  $\mu \tauhidearrow \mu'$ by Definition~\ref{def-weak-arrows}iii.

  (c)~Let $\mu_i \Arrow{\;} \mu'_i$ for all $i\in I$.  As $I$ is
  finite and $\Arrow{\;}$ is reflexive, there exists an $n \in \IN$
  such that
  $\mu_i = \mu_i^{(0)} \arrow{(\tau)} \mu_i^{(1)} \arrow{(\tau)} \dots
  \arrow{(\tau)} \mu_i^{(n)} = \mu'_i$ for all $i\in I$. Now
  $\mu \Arrow{\;} \mu'$ follows by $n$ applications of the second
  statement of the lemma.
\end{proof} 

\noindent
Likewise, the next lemma allows \emph{probabilistic decomposition} of
 transitions $\arrow{\alpha}$, $\arrow{(\alpha)}$ and $\Arrow{\;}$.

 \begin{lemma}
   \label{lem-decomposition}
   Let $\mu,\mu'\in \DistrE$ and
   $\mu = \bigoplusiinI \: p_i \stardot \mu_i$ with $p_i>0$ for all $i \in I$.
   \begin{itemize}
   \item [(a)] If $\mu \arrow{\alpha} \mu'$, then there are $\mu'_i$
     for $i \in I$ such that $\mu_i \arrow{\alpha} \mu'_i$ for all
     $i \in I$ and $\mu' = \bigoplusiinI \: p_i \stardot \mu'_i$.
   \item [(b)] If $\mu \arrow{(\tau)} \mu'$, then there are $\mu'_i$ for
     $i \in I$ such that $\mu_i \arrow{(\tau)} \mu'_i$ for all
     $i \in I$ and $\mu' = \bigoplusiinI \: p_i \stardot \mu'_i$.
   \item [(c)] If $\mu \Arrow{\;} \mu'$, then there are $\mu'_i$ for
     $i \in I$ such that $\mu_i \Arrow{\;} \mu'_i$ for all $i \in I$
     and $\mu' = \bigoplusiinI \: p_i \stardot \mu'_i$.
   \end{itemize}
 \end{lemma}

\begin{proof}
  (a)~Suppose $\mu \arrow{\alpha} \mu'$.  By
  Definition~\ref{def-pr-transition-relation}\ref{transitrel-case_b}
  $\mu = \bigoplusjinJ \: q_j \stardot E_j$,
  $\mu' = \bigoplusjinJ \: q_j \stardot \nu'_j$, and
  $E_j \alphaarrow \nu'_j$ for all $j \in J$, for suitable
    index set~$J$, $q_j > 0$, $E_j \in \calE$,
    and~$\nu'_j \in \DistrE$. By Lemma~\ref{flexibelDelen} there are
  $r_{ij} \in [0,1]$ and $\varrho_{ij} \in \DistrE$ such that
  $\sum_{j\in J} \: r_{ij} = p_i$ and
  $\mu_i = \bigoplusjinJ \: \frac{r_{ij}}{p_i} \stardot \varrho_{ij}$
  for all~$i \in I$, and $\sum_{i\in I} \: r_{ij}=q_j$ and
  $q_j \stardot \delta(E_j) = \bigoplusiinI \: r_{ij} \stardot
  \varrho_{ij}$ for all~$j \in J$. It follows that
  $\varrho_{ij}=\delta(E_j)$ for all $i \in I$ and $j\in J$.
   
  For all $i \in I$, let
  $\mu'_i := \bigoplusjinJ \: \frac{r_{ij}}{p_i} \stardot \nu'_j$.
  Then $\mu_i \arrow\alpha \mu'_i$, for all $i \in I$, by
  Lemma~\ref{lem-composition}a. Moreover,
  \begin{displaymath}
    \bigoplusiinI \: p_i \stardot \mu'_i =
    \bigoplusiinI \: p_i \stardot \bigoplusjinJ \: \frac{r_{ij}}{p_i}
    \stardot \nu'_j = 
    \bigoplusjinJ \: \bigoplusiinI \: r_{ij} \stardot \nu'_j =
    \bigoplusjinJ \: q_{j} \stardot \nu'_j = \mu'
    \; .
  \end{displaymath}
  (b)~Suppose $\mu \arrow{(\tau)} \mu'$. By
  Definition~\ref{def-weak-arrows}, either (i)~$\mu \tauarrow \mu'$,
  or (ii)~$\mu'=\mu$, or (iii)~there exist
  $\nu_1, \nu_2, \nu'_1, \nu'_2 \in \DistrE$ such that
  $\mu = \nu_1 \oplusr \nu_2$, $\mu' = \nu'_1 \oplusr \nu'_2$,
  $\nu_1 = \nu_1'$, and $\nu_2 \arrow{\tau} \nu'_2$ for some
  $r \in (0,1)$.  In case~(i), the required $\mu'_i$ exist by the
  first statement of this lemma.  In case~(ii) one can simply take
  $\mu'_i := \mu_i$ for all $i\in I$.  Hence assume that case~(iii)
  applies.  Let $J := \{1,2\}$, $q_1:=r$ and $q_2:=1-r$.  By
  Lemma~\ref{flexibelDelen} there are $r_{ij} \in [0,1]$ and
  $\varrho_{ij} \in \DistrE$ with $\sum_{j\in J} \: r_{ij} = p_i$ and
  $\mu_i = \bigoplusjinJ \: \frac{r_{ij} \rule[-3pt]{0pt}{6pt}}{p_i}
  \stardot \varrho_{ij}$ for all~$i \in I$, and
  $\sum_{i \in I} \: r_{ij} = q_j$ and
  $\nu_j = \bigoplusiinI \: \frac{r_{ij} \rule[-3pt]{0pt}{6pt}}{q_j}
  \stardot \varrho_{ij}$ for all~$j \in J$. \vspace{2pt}

  Let $I' := \lc i \in I \mid r_{i2}>0 \rc$. \vspace{2pt} Since
  $\nu_2 = \bigoplus_{i \in I'} \: \frac{r_{i2}
    \rule[-3pt]{0pt}{6pt}}{1-r} \stardot \varrho_{i2}\arrow{\tau}
  \nu_2'$, it follows from the first statement of the lemma, for all
  $i \in I'$ there are $\varrho_{i2}'$ such that
  $\varrho_{i2} \arrow{\tau} \varrho_{i2}'$ and
  $\nu'_2 = \bigoplus_{i \in I'} \: \frac{r_{i2}
    \rule[-3pt]{0pt}{6pt}}{1-r} \stardot \varrho_{i2}'$.\vspace{2pt}
  For all $i \in I{\setminus}I'$, pick
  $\varrho'_{i2} \in \DistrE$ arbitrarily.  It follows that
  \(\mu_i = \varrho_{i1} \probc{\frac{r_{i1} \rule[-3pt]{0pt}{6pt}}{p_i}} \varrho_{i2}
  \tauhidearrow \varrho_{i1} \probc{\frac{r_{i1} \rule[-3pt]{0pt}{6pt}}{p_i}}
  \varrho'_{i2} =: \mu'_i\) for all $i \in I$.  Moreover,
  \begin{displaymath}
    \bigoplusiinI \: p_i \stardot \mu'_i =
    \bigoplusiinI \: p_i \stardot (\varrho_{i1}
    \probc{\frac{r_{i1} \rule[-3pt]{0pt}{6pt}}{p_i}} \varrho'_{i2}) = 
    (\bigoplusiinI \: \frac{r_{i1} \rule[-2pt]{0pt}{6pt}}{r} \stardot \varrho_{i1}) \oplusr
    (\bigoplusiinI \: \frac{r_{i2} \rule[-2pt]{0pt}{6pt}}{1-r} \stardot \varrho'_{i2}) =
    \nu_1 \oplusr \nu'_2 = \mu'
    \; .
  \end{displaymath}
  (c)~The last statement follows by transitivity from the second one.
 \end{proof}



\section{Branching {\bisimilarity} for probabilistic processes}
\label{sec-bpb}

\noindent
We next define the notions of strong probabilistic {\bisimilarity} and
branching probabilistic {\bisimilarity}. Note that the notion of
strong probabilistic {\bisimilarity} is the variant with combined
transitions as defined in \cite{SL94:concur,BS01:icalp}.

\begin{definition}[Strong and branching probabilistic {\bisimilarity}]
  \label{def-probabilistic-branching-bisimilar}
  \begin{itemize}
  \item [(a)] A symmetric relation $\calR \subseteq \DistrE \times
    \DistrE$ is called \textit{decomposable} iff for all $\mu, \nu \in
    \DistrE$ such that $\mu \mopcalR \nu$ and $\mu = \bigoplusiinI \:
    p_i \stardot \mu_i$, there are $\nu_i \in \DistrE$, for~$i \in I$,
    such that
    \begin{displaymath}
      \nu  = \bigoplusiinI \: p_i \stardot \nu_i \ 
      \text{and} \ 
      \mu_i \mopcalR \nu_i ,
      \ \text{for all~$i \in I$.}
    \end{displaymath}

  \item [(b)] A decomposable relation $\calR \subseteq \DistrE \times
    \DistrE$ is called a \emph{strong probabilistic bisimulation
    relation} iff for all $\mu, \nu \in \DistrE$ such that $\mu
    \mopcalR \nu$ and $\mu \arrow{\alpha} \mu'$ there is a $\nu'
    \in \DistrE$ such that
    \begin{displaymath}
      \nu \alphaarrow \nu' \ \text{and} \ 
      \mu' \mopcalR \nu'.
    \end{displaymath}

  \item [(c)] A symmetric relation $\calR \subseteq \DistrE \times
    \DistrE$ is called \textit{weakly decomposable} iff for all $\mu,
    \nu \in \DistrE$ such that $\mu \mopcalR \nu$ and $\mu =
    \bigoplusiinI \: p_i \stardot \mu_i$ there are $\nubar, \nu_i \in
    \DistrE$, for~$i \in I$, such that
    \begin{displaymath}
      \nu \Arrow{} \nubar,\ 
      \mu \mopcalR \nubar,\  
      \nubar = \bigoplusiinI \: p_i \stardot \nu_i,\ 
      \text{and} \ 
      \mu_i \mopcalR \nu_i 
      \ \text{for all~$i \in I$.}
    \end{displaymath}

  \item [(d)] A weakly decomposable relation $\calR \subseteq \DistrE
    \times \DistrE$ is called a \emph{branching} probabilistic
    bisimulation relation iff for all $\mu, \nu \in \DistrE$ such
    that $\mu \mopcalR \nu$ and $\mu \arrow{\alpha} \mu'$, there are
    $\nubar, \nu' \in \DistrE$ such that
    \begin{displaymath}
      \nu \Arrow{} \nubar, \ 
      \nubar \alphahidearrow \nu', \ 
      \mu \mopcalR \nubar, \ \text{and} \ 
      \mu' \mopcalR \nu'.
    \end{displaymath}

  \item [(e)] Strong probabilistic {\bisimilarity}, denoted by~${\bisim}
    \subseteq \DistrE \times \DistrE$, and branching probabilistic
    {\bisimilarity}, written as~${\brbisim} \subseteq \DistrE \times
    \DistrE$, are respectively defined as the largest strong
    probabilistic bisimulation relation on~$\DistrE$ and as the
    largest branching probabilistic bisimulation relation
    on~$\DistrE$.
  \end{itemize}
\end{definition}

\noindent
Strong and branching probabilistic {\bisimilarity} are well-defined by
the usual argument that any union of strong or branching probabilistic
bisimulation relations is again a strong or branching probabilistic
bisimulation relation, respectively. In particular, (weak)
decomposability is preserved under arbitrary unions.

Two non-deterministic processes are considered to be strongly or
branching probabilistic bisimilar iff their Dirac distributions are,
i.e., for $E, F \in \calE$ we have $E \bisim F$ iff
$\delta(E) \bisim \delta(F)$ and $E \brbisim F$ iff
$\delta(E) \brbisim \delta(F)$.
Two probabilistic processes are considered to be strongly or branching
probabilistic bisimilar iff their associated distributions
over~$\calE$ are, i.e., for $P,Q \in \calP$ we have $P \bisim Q$ iff
$\den{P} \bisim \den{Q}$ and $P \brbisim Q$ iff
$\den P \brbisim \den Q$.

The notion of decomposability has been adopted from~\cite{Hen12:facj}
and weak decomposability from~\cite{LV16}. The underlying idea stems
from \cite{DGHM09}. These notions provide a convenient dexterity to
deal with the behaviour of sub-distributions, e.g., to distinguish
$\myfrac12 \partial( a \pref \partial(\bfzero)) \oplus \myfrac12
\partial( b \pref \partial(\bfzero))$ from $\partial(\bfzero)$, and to
handle combined behaviour.

By comparison, on finite processes, as used in this paper, the notion
of branching probabilistic bisimilarity of Segala \&
Lynch~\cite{SL94:concur} can be defined in our framework exactly as in
(d) and~(e) above, but taking a decomposable instead of a weakly
decomposable relation.  This yields a strictly finer equivalence,
distinguishing the processes $\mathbf{G}_0$, $\mathbf{G}_1$,
and~$\mathbf{G}_2$ from the introduction.

\begin{example}
  \label{ex-prob-bisim}
  \begin{itemize}
  \item [(a)] Since the distributions
    $\delta(\mathbf{H_1}) = \delta( a \pref \mkern1mu ( P \probc{\frac14} (
    P \probc{\frac13} Q )))$
    and~$\delta(\mathbf{H_2}) = \delta( a \pref \mkern1mu ( P
    \probc{\frac12} ( Q \probc{\frac12} Q ) ))$ only have the
    transitions
    \begin{displaymath}
      \delta( a \pref \mkern1mu ( P \probc{\frac14} ( P \probc{\frac13} Q
      )))
      \arrow{\alpha} \myfrac12 \den{P} \oplus \myfrac12 \den{Q}
      \quad \text{and} \quad
      \delta( a \pref \mkern1mu ( P \probc{\frac12} ( Q \probc{\frac12} Q
      ))) 
      \arrow{\alpha} \myfrac12 \den{P} \oplus \myfrac12 \den{Q}
      \mkern1mu ,
    \end{displaymath}
    it follows that
    $\calR = \singleton{ \pair{\delta(\mathbf{H_1})}{\delta(\mathbf{H_2})}
    }^\dagger \cup \lc \pair{\mu}{\mu} \mid \mu \in \DistrE \rc$ is a strong
    bisimulation relation.
    Here, for a relation $\calS \subseteq \DistrE \times \DistrE$,
    $\calS^\dagger$~denotes its symmetric closure.
    Since $\calR$~is a strong probabilistic bisimulation relation, it
    follows that $\mathbf{H_1}$ and~$\mathbf{H_2}$ are
    strong probabilistic bisimilar, i.e.,
    $\delta( \mathbf{H_1} ) \bisim \delta( \mathbf{H_2} )$.
  \item [(b)] The distributions
    $\delta( \mathbf{G_1} ) = \delta( a \pref \mkern1mu ( P
    \probc{\mytinyfrac12} Q ))$
    and~$\delta( \mathbf{G_2} ) = \delta( a \pref \mkern1mu ( \partial (
    \tau \pref \mkern1mu ( P \probc{\mytinyfrac12} Q ) ) \probc{\mytinyfrac13} (
    P \probc{\mytinyfrac12} Q ) ))$ are not strong probabilistic
    bisimilar. Although the two distributions both admit an
    $a$-transition,
    \begin{align*}
      \delta( a \pref \mkern1mu ( P \probc{\mytinyfrac12} Q ))
      & \arrow{a}
        \myfrac12 \den{P} \oplus \myfrac12 \den{Q} \\
      \delta( a \pref \mkern1mu ( \partial (
      \tau \pref \mkern1mu ( P \probc{\mytinyfrac12} Q ) )
      \probc{\mytinyfrac13} ( P \probc{\mytinyfrac12} Q ) ))
      & \arrow{a}
        \myfrac13 \delta ( \tau \pref \mkern1mu
        ( P \probc{\mytinyfrac12} Q ) )
        \oplus
        \myfrac13 \den{P} \oplus
        \myfrac13 \den{Q} 
    \end{align*}
    the resulting distributions at the right-hand side are not strong
    probabilistic bisimilar.
    However, the symmetric closure $\calR^\dagger$ of the
    relation $\calR$ containing the pairs
    \begin{displaymath}
      \langle
      \delta( \tau \pref \mkern1mu ( P \probc{\mytinyfrac12} Q ) ),
      \myfrac12 \den{P} \oplus \myfrac12 \den{Q}
      \rangle
      \quad \text{and} \quad
      \langle \mu, \mu \rangle
      \text{ for $\mu \in \DistrE$}
    \end{displaymath}
    clearly is a branching probabilistic bisimulation relation. By
    Lemma~\ref{lm-point-decomposability} below, also its convex
    closure $\mathit{cc}(\calR^\dagger)$ is a branching probabilistic
    bisimulation relation.
    Considering that
    $\calR( \delta( \tau \pref \mkern1mu ( P \probc{\mytinyfrac12} Q )
    ), \myfrac12 \den{P} \oplus \myfrac12 \den{Q} )$ and
    $\calR( \myfrac12 \den{P} \oplus \myfrac12 \den{Q}, \myfrac12
    \den{P} \oplus \myfrac12 \den{Q} )$, we have that
  \begin{displaymath}
    \mathit{cc}(\calR^\dagger)(
    \myfrac13 \delta( \tau \pref \mkern1mu ( P \probc{\mytinyfrac12} Q
    ) \oplus
    \myfrac23 ( \myfrac12 \den{P} \oplus \myfrac12 \den{Q} ),
    \myfrac13( \myfrac12 \den{P} \oplus \myfrac12 \den{Q} ) \oplus
    \myfrac23 ( \myfrac12 \den{P} \oplus \myfrac12 \den{Q} ) )    
    ) {\;.}
  \end{displaymath}
  i.e., the distributions
  $\myfrac13 \delta( \tau \pref \mkern1mu ( P \probc{\mytinyfrac12} Q
  ) ) \oplus \myfrac23 ( \myfrac12 \den{P} \oplus \myfrac12 \den{Q} )$
  and $\myfrac12 \den{P} \oplus \myfrac12 \den{Q}$ are related
  by~$\mathit{cc}(\calR^\dagger)$.
  Adding the pair of processes
  $\langle \delta( a \pref \mkern1mu ( P \probc{\mytinyfrac12} Q )),
  \delta( a \pref \mkern1mu ( \partial ( \tau \pref \mkern1mu ( P
  \probc{\mytinyfrac12} Q ) ) \probc{\mytinyfrac13} ( P
  \probc{\mytinyfrac12} Q ) )) \rangle$ (and closing for symmetry)
  then yields a branching probabilistic bisimulation relation relating
  $\delta( \mathbf{G_1} )$ and $\delta( \mathbf{G_2} )$.
    
\item [(c)] The $a$-derivatives of $\mathbf{I_1}$ and~$\mathbf{I_2}$,
  i.e.\ the processes $I'_1 = \partial( b \pref P + \tau \pref Q )$
  and
  $I_2' = \partial( \tau \pref \partial( b \pref P + \tau \pref Q ) +
  b \pref P + \tau \pref Q)$, are branching probabilistic bisimilar. A
  $\tau$-transition of~$I'_2$ partially based on its left branch, can
  be simulated by~$I'_1$ by a partial stand-still:
    \begin{align*}
      I'_2
      & \myeq r {\cdot} \den{ I'_2 } \oplus (1-r) {\cdot} \den{I'_2}
      & \arrow{~\tau~} \quad
      & r {\cdot} \mkern1mu
        \delta( b \pref P + \tau \pref Q ) \oplus (1-r) {\cdot}
        \den{ Q } \\[2pt] 
      I'_1
      & \myeq r {\cdot} \den{ I'_1 } \oplus (1-r) {\cdot} \den{I'_1}
      & \arrow{(\tau)} \quad
      & r {\cdot} \den{ I'_1 } \oplus (1-r) {\cdot} \den{ Q }
        \; \myeq \;
        r {\cdot}
        \delta( b \pref P + \tau \pref Q ) \oplus (1-r) {\cdot}
        \den{ Q }
        \mkern1mu .
    \end{align*}
    A $\tau$-transition of~$I'_1$ can be directly simulated by~$I'_2$
    of course. It follows that the relation
    $\calR = \singleton{ \langle{ \delta( \mathbf{I_1} ), \delta(
        \mathbf{I_2} ) }\rangle, \langle{ I'_1, I'_2 }\rangle
    }^\dagger \cup \lc \langle \mu, \mu \rangle \mid \mu \in \DistrE
    \rc$, the symmetric relation containing the pairs mentioned and
    the diagonal of~$\DistrE$, constitutes a branching probabilistic
    bisimulation relation containing $\mathbf{I_1}$
    and~$\mathbf{I_2}$.
    
  \item [(d)] The $a$-derivatives of $\mathbf{T_1}$
    and~$\mathbf{T_2}$, say
    $T'_1 = \partial( b \pref \mkern1mu ( P \probc{\frac13} Q ) )
    \probc{\frac12} \partial( b \pref \mkern1mu ( P \probc{\frac23} Q
    ) )$ and
    $T'_2 = \partial ( \tau \pref \mkern1mu ( \partial( b \pref
    \mkern1mu ( P \probc{\frac13} Q ) ) \probc{\frac12} \partial( b
    \pref \mkern1mu ( P \probc{\frac23} Q ) ) ) + b \pref \mkern1mu (
    P \probc{\frac12} Q ) )$, are similarly related as in the above
    example. Here though, it is slightly more difficult to see that
    $T'_1$~can simulate the {$b$}-transition of~$T'_2$ into
    $P \probc{\frac12} Q$. We have
    \begin{align*}
      T'_2 
      & \; \arrow{b} \;  \phantom{\myfrac12}
        \den{ P \probc{\frac12} Q \mkern1mu } \\
      T'_1
      & \; \arrow{b} \;
        \myfrac12 \den{ P \probc{\mytinyfrac13} Q \mkern1mu }
        \oplus
        \myfrac12 \den{ P \probc{\mytinyfrac23} Q\mkern1mu }
        =
        \myfrac16 \den{P} \oplus \myfrac13 \den{Q \mkern1mu }
        \oplus
        \myfrac13 \den{P} \oplus \myfrac16 \den{Q \mkern1mu }
        =
        \den{ P \probc{\mytinyfrac12} Q \mkern1mu }
        \mkern1mu .
    \end{align*}
    In the axiomatisation presented in the sequel, axioms are included
    (the axioms P1 to P3)
    to express the above identification of the distributions
    in~$\DistrE$ in terms of probabilistic processes from~$\calP$.
    
  \item [(e)] A similar recombination of distributions as for case~(d)
    takes place for the examples $\mathbf{E_1}$ and~$\mathbf{E_2}$. In
    this case
    \newcommand{\mmbox}[1]{\makebox[2.5em][c]{\ensuremath{#1}}}
    \begin{align*}
      \delta(\mathbf{E_1})
      & \mmbox{\arrow{\tau}}
        \myfrac34 \delta( \tau \pref P + a \pref Q + \tau \pref R )
        \oplus
        \myfrac14 \delta( \bfzero ) \\
      \delta(\mathbf{E_2})
      & \mmbox{\arrow{\tau}}
        \myfrac12 \delta( \tau \pref P + a \pref Q + \tau \pref R )
        \oplus
        \myfrac12 \delta( \tau \pref (
        \partial( P + a \pref Q + \tau \pref R )
        \probc{\frac12} \partial(\bfzero) ) ) \\
      & \mmbox{\arrow{(\tau)}\,}
        \myfrac12 \delta( \tau \pref P + a \pref Q + \tau \pref R )
        \oplus
        \myfrac12 \den{
        \partial( P + a \pref Q + \tau \pref R )
        \probc{\frac12} \partial(\bfzero) } \\
      & \mmbox{\myeq}
        \myfrac12 \delta( \tau \pref P + a \pref Q + \tau \pref R )
        \oplus
        \myfrac14 \delta( P + a \pref Q + \tau \pref R )
        \oplus
        \myfrac14 \delta(\bfzero) \\
      & \mmbox{\myeq}
        \myfrac34 \delta( \tau \pref P + a \pref Q + \tau \pref R )
        \oplus
        \myfrac14 \delta(\bfzero)
        \mkern1mu .
    \end{align*}
    Because for~$\mathbf{E_2}$ the $\tau$-moves can be combined, the
    branching probabilistic bisimulation relation relating
    $\mathbf{E_1}$ and~$\mathbf{E_2}$ is a bit more extensive. Put
    $S = \tau \pref P + a \pref Q + \tau \pref R$ and
    \begin{align*}
      \calR
      & \, \myeq \!
        \lbrace \,
        \langle{ \delta(\mathbf{E_1}), \delta(\mathbf{E_2}) }\rangle
        \, \rbrace^\dagger \cup {}
      \\ 
      & \phantom{\, \myeq \,} \!
        \lc \langle
        r {\cdot} \den{S} \oplus
        \myfrac{1-r}{2} {\cdot} \den{S} \oplus
        \myfrac{1-r}{2} {\cdot} \delta(\bfzero),
        r {\cdot} \den{S} \oplus
        (1{-}r) {\cdot} \delta( \tau \pref (S \probc{\mytinyfrac12}
        \partial(\bfzero) ) )
        \rangle \mid 0 \leqslant r \leqslant 1 \rc^\dagger \cup {}
      \\
      & \phantom{\, \myeq \,} \!
        \lc \langle \mu, \mu \rangle \mid \mu \in \DistrE \rc
    \end{align*}
    where the pairs
    $\langle r {\cdot} \den{S} \oplus \myfrac{(1-r)}{2} {\cdot}
    \den{S} \oplus \myfrac{1-r}{2} {\cdot} \delta(\bfzero), r {\cdot}
    \den{S} \oplus (1{-}r) {\cdot} \delta( \tau \pref (S
    \probc{\mytinyfrac12} \partial(\bfzero) ) ) \rangle$, for
    $r \neq\frac12$, are needed for~$\calR$ to be weakly decomposable.
  \end{itemize}
\end{example}

\noindent
Similar to the non-deterministic case, transitions $E \arrow{\tau}
\nu$ and $\mu \arrow{\tau} \nu$ are called \emph{inert} iff
$\delta(E) \brbisim \nu$ and $\mu \brbisim \nu$. Typical
cases of inert transitions include
\begin{displaymath}
  \begin{array}{c}
    \tau \pref P \  \arrow{\tau} \  \den{P}
    \quad \text{and} \quad
    E + \tau \pref \partial(E) \  \arrow{\tau} \  \delta(E) \;.
  \end{array}
\end{displaymath}
A transition $E \arrow{\tau} \mu_1\oplusr \mu_2$ with
$r \in (0,1]$ and $\delta(E) \brbisim \mu_1$ is called
\hypertarget{partinert}{\emph{partially inert}}.  An illustrative
instance of partial inertness is
\begin{displaymath}
  \begin{array}{c}
    \tau \pref \mkern1mu (\partial(a \pref P + \tau \pref Q) \oplusr Q)
    + a \pref P + \tau \pref Q
    \  \arrow{\tau} \
    \delta(a \pref P + \tau \pref Q)\oplusr \den{Q} \;.
  \end{array}
\end{displaymath}
Here we have that
$\delta ( \tau \pref \mkern1mu (\partial(a \pref P + \tau \pref Q)
\oplusr Q) + a \pref P + \tau \pref Q ) \brbisim \delta(a \pref P +
\tau \pref Q)$, because in particular
$\delta(a \pref P + \tau \pref Q) = {\delta(a \pref P + \tau \pref Q)}
\probc{r} {\delta(a \pref P + \tau \pref Q)} \tauhidearrow \delta(a
\pref P + \tau \pref Q)\oplusr \den{Q}$.
Partially inert transitions play a central role in the remainder.

\blankline

\noindent
The following lemma is handy for constructing branching probabilistic
bisimulation relations; it was used already in
Example~\ref{ex-prob-bisim}b above. The proof of the lemma,
  because of its technical nature, is delegated to the appendix.

\begin{lemma}\label{lm-point-decomposability}
  If $\calR$ is a branching probabilistic bisimulation relation, then
  so is $\ccR$.
\end{lemma}

\noindent
The next lemma provides a witness for strong
probabilistic bisimilarity; it is used in the proof of
Lemma~\ref{lemma-equal-for-concrete-processes}. The proof of the lemma
itself is given in the appendix.

\begin{lemma}
  \label{state-based}
  Let $\calR \subseteq \DistrE \times \DistrE$ be a decomposable
  relation such that
  \begin{equation}
    \label{cong}
    \mu_1 \mopcalR \nu_1 \ \text{and} \ \mu_2 \mopcalR \nu_2
    \quad  \text{implies} \quad
    (\mu_1 \oplusr \mu_2) \mopcalR (\nu_1 \oplusr \nu_2)
  \end{equation}
  and for each pair $E,F \in \calE$ 
  \begin{equation}
    \label{diamond}
    \delta(E) \mopcalR \delta(F)
    \ \text{and} \
    E \arrow{\alpha} \mu'
    \quad \text{implies} \quad
    \delta(F) \alphaarrow \nu'
    \ \text{and} \ 
    \mu' \mopcalR \nu'
  \end{equation}
  for a suitable $\nu' \in \DistrE$.
  Then $\mu \mopcalR \nu$ implies $\mu\bisim \nu$.
\end{lemma}

\noindent
With the help of Lemma~\ref{flexibelDelen} we show---in the
appendix---that the operator $\probc{r}$ respects branching
probabilistic bisimilarity of distributions.

\begin{lemma}
  \label{unrooted-congruence}
  Let $\mu_1,\mu_2,\nu_1,\nu_2\in \DistrE$ and $r\in(0,1)$. If $\mu_1
  \brbisim \nu_1$ and $\mu_2 \brbisim \nu_2$ then $\mu_1 \probc{r}
  \mu_2 \brbisim \nu_1 \probc{r} \nu_2$.
\end{lemma}

\noindent
A direct consequence of the lemma is that, for probabilistic processes
$P_1, P_2, Q_1, Q_2 \in \calP$, we have that $P_1 \brbisim Q_1$ and
$P_2 \brbisim Q_2$ implies $P_1 \oplusr Q_1 \brbisim P_2 \oplusr Q_2$,
i.e.\ $\den{ P_1 \probc{r} P_2} \brbisim \den{Q_1 \probc{r} Q_2}$ when
$\den{P_1} \brbisim \den{Q_1}$ and $\den P_2 \brbisim \den Q_2$.

As we did for the non-deterministic setting, we introduce a notion of
rooted branching probabilistic {\bisimilarity} for non-deterministic
processes.  We do not need such a notion for probabilistic processes.

\begin{definition}
  \label{rooted-bpb}
  Two nondeterministic processes $E,F \in\calE$ are \emph{rooted
    branching probabilistic bisimilar}, notation $E \rootedbrbisim F$,
  if for each transition $\delta(E) \arrow\alpha \mu$ with $\alpha \in \calA$
  there is a transition $\delta(F) \arrow\alpha \nu$ with
    $\mu \brbisim \nu$, and, 
  vice versa, for each transition $\delta(F) \arrow\alpha \nu$ there is a
  $\delta(E) \arrow\alpha \mu$ with $\mu \brbisim \nu$.
\end{definition}

\noindent
To show that $\brbisim$ is an equivalence relation, we need the following
useful property.

\begin{lemma}
  \label{lemma-weak-transfer}
  Let $\mu, \nu \in \DistrE$ such that $\mu \brbisim \nu$ and
  $\mu \Arrow{\:} \mu'$ for some $\mu' \in \DistrE$. Then exists
  $\nu' \in \DistrE$ such that $\nu \Arrow{\:} \nu'$ and
  $\mu' \brbisim \nu'$.
\end{lemma}

\begin{proof}
  We check that a partial transition $\mu \arrow{(\tau)} \mu'$ can be
  matched by~$\nu$ given $\mu \brbisim \nu$. So, suppose
  $\mu = \mu_1 \probc{r} \mu_2$, $\mu_1 \arrow{\tau} \mu'_1$, and
  $\mu' = \mu'_1 \probc{r} \mu_2$. By weak decomposability
  of~$\brbisim$ we can find distributions~$\nubar, \nu_1, \nu_2$ such
  that $\nu \Arrow{\:} \nubar = \nu_1 \probc{r} \nu_2$ and
  $\mu \brbisim \nubar$, $\nu_1 \brbisim
  \mu_1$,~$\nu_2 \brbisim \mu_2$. Choose
  distributions~$\nubar_1, \nubar'_1$ such that
  $\nu_1 \Arrow{\:} \nubar_1 \arrow{(\tau)} \nu'_1$ and
  $\nubar_1 \brbisim \mu_1$, $\nu'_1 \brbisim \mu'_1$. Put
  $\nu' = \nu'_1 \probc{r} \nu_2$. Then $\nu \Arrow{\:} \nu'$,
  using Lemma~\ref{lem-composition}c, and we
  have by Lemma~\ref{unrooted-congruence}:
  $\nu' = \nu'_1 \probc{r} \nu_2 \brbisim \mu'_1 \probc{r} \mu_2 =
  \mu'$ since $\nu'_1 \brbisim \mu'_1$ and $\nu_2 \brbisim \mu_2$.
\end{proof}

\noindent
We are now in a position to show that the various bisimulation
relations introduced in
Definitions~\ref{def-probabilistic-branching-bisimilar}
and~\ref{rooted-bpb} are equivalence relations.

\begin{lemma} [Equivalence]
  \label{lemma-equivalence-prob}
  The relations $\bisim$, $\rootedbrbisim \mkern1mu$, and~$\brbisim$
  on $\calE$, as well as $\bisim$ and~$\brbisim$ on~$\calP$ are
  equivalence relations. 
\end{lemma}

\begin{proof}
  That these relations are reflexive and symmetric follows immediately
  from the definitions; for reflexivity one takes the identity
  relation between distributions as a witness.  We show that
  $\brbisim$ is transitive by showing that the relation~$\calR$ given
  by
  \begin{displaymath}
    \calR = \lc \langle \mu, \varrho \rangle \mid
    \exists \mkern1mu \nu \colon {\mu \brbisim \nu} \land {\nu
      \brbisim \varrho} \rc
    \mkern1mu ,
  \end{displaymath}
  i.e.\ $\calR = {\brbisim} \mathbin{\cdot} {\brbisim}$, is a
  branching probabilistic bisimulation relation. Symmetry of~$\calR$
  follows from symmetry of~$\brbisim$. As to weak decomposability,
  suppose $\calR(\mu,\varrho)$ for distributions~$\mu$ and~$\varrho$
  such that $\mu = \bigoplusiinI \: p_i \mycdot \mu_i$ for an index
  set~$I$, and $\mu_i \in \DistrE$.  Choose $\nu \in \DistrE$ such
  that $\mu \brbisim \nu$ and~$\nu \brbisim \varrho$.  Since
  $\nu \brbisim \mu$ we can find $\nubar, \nu_i \in \DistrE$
  for~$i \in I$ such that
  $\nu \Arrow{\:} \nubar = \bigoplusiinI \: p_i \mycdot \nu_i$,
  $\mu \brbisim \nubar$, and $\mu_i \brbisim \nu_i$ for~$i \in I$.  By
  Lemma~\ref{lemma-weak-transfer} we can find $\varrhobar \in \DistrE$
  such that $\varrho \Arrow{\:} \varrhobar$
  and~$\varrhobar \brbisim \nubar$. Hence, there exist
  distributions~$\varrhohat$ and~$\varrho_i$ for~$i \in I$ such that
  $\varrhobar \Arrow{\:} \varrhohat = \bigoplusiinI \: p_i \mycdot
  \varrho_i$, $\nubar \brbisim \varrhohat$ and
  $\nu_i \brbisim \varrho_i$ for~$i \in I$. Thus,
  $\varrho \Arrow{\:} \varrhohat = \bigoplusiinI \: p_i \mycdot
  \varrho_i$, where $\mu \brbisim \nubar$ and
  $\nubar \brbisim \varrhohat$, hence $\calR(\mu,\varrhohat)$, and
  likewise $\mu_i \brbisim \nu_i$ and $\nu_i \brbisim \varrho_i$,
  hence $\calR(\mu_i,\varrho_i)$ for~$i \in I$.

  Finally, if $\mu \arrow{\alpha} \mu'$ for some~$\mu' \in \DistrE$,
  then distributions~$\nubar, \nu'$ exist such that
  $\nu \Arrow{\:} \nubar \alphahidearrow \nu'$ and
  $\nubar \brbisim \mu$,~$\nu' \brbisim \mu'$. Again with appeal to
  Lemma~\ref{lemma-weak-transfer} there exists
  $\varrhobar \in \DistrE$ such that $\varrho \Arrow{\:} \varrhobar$
  and~$\varrhobar \brbisim \nubar$. Then there also exist
  $\varrhohat, \varrho' \in \DistrE$ such that
  $\varrhobar \Arrow{\:} \varrhohat \alphahidearrow \varrho'$,
  $\varrhohat \brbisim \nubar$ and $\varrho' \brbisim \nu'$.  So, we
  have $\varrho \Arrow{\:} \varrhohat \alphahidearrow \varrho'$ and
  $\calR(\mu , \varrhohat)$, $\calR(\mu', \varrho')$.
  Thus $\calR$~also satisfies the weak transfer condition and,
  consequently, is a branching probabilistic bisimulation relation.

  The proof of $\bisim$ being transitive is obtained by simplification
  of the above argument.
  The fact that $\rootedbrbisim$ is transitive follows immediately
    from the definition.
\end{proof}

\noindent
For more detail on transitivity proofs, in the setting of a slightly
different notion of branching probabilistic bisimulation,
see~\cite{CT2020:ifp}.

As argued in Section~\ref{sec-nondet}, for non-deterministic processes
branching probabilistic bisimilarity is not a congruence with respect
to non-deterministic choice. The argument holds true in the mixed
non-deterministic and probabilistic setting of this section as well.
This is the reason we use rooted branching bisimilarity instead.

Strong probabilistic bisimulation semantics employs strong
probabilistic bisimilarity $\bisim$ on~$\calE$ as \linebreak[3] well
as~$\calP$. Here we propose a notion of \emph{rooted branching
  probabilistic bisimulation semantics} that employs rooted branching
probabilistic bisimilarity~$\rootedbrbisim$ on~$\calE$, and branching
probabilistic bisimilarity~$\brbisim$ on~$\calP$.

\begin{lemma} [Congruence]
  \label{lemma-congruence-prob}
   Both strong probabilistic bisimulation semantics and rooted
   branching probabilistic bisimulation semantics induce a congruence relation.
\end{lemma}

\begin{proof}
  Regarding rooted branching probabilistic bisimulation semantics we
  have to show four congruence properties:
  \begin{enumerate}
  \item if $P_i \brbisim Q_i$ for $i=1,2$ then
    $P_1 \oplusr Q_1 \brbisim P_2 \oplusr Q_2$,
  \item if $P \brbisim Q$ then $\alpha \pref P \rootedbrbisim a \pref Q$,
  \item if $E_i \rootedbrbisim F_i$ for $i=1,2$ then
    $E_1 + F_1 \rootedbrbisim F_2 + F_2$, and
  \item if $E \rootedbrbisim F$ then
    $\partial(E) \brbisim \partial(F)$.
  \end{enumerate}
  Property~1 was the subject of Lemma~\ref{unrooted-congruence},
  whereas Properties 2 and~4 are trivial with
  Definition~\ref{rooted-bpb}. We are left with Property~3.
  Suppose $\delta(E_1) \rootedbrbisim \delta(F_1)$ and
  $\delta(E_2) \rootedbrbisim \delta(F_2)$.
  If $\delta(E_1+E_2) \arrow{\alpha} \mu'$, then either
  $\delta(E_1) \alphaarrow \mu'$, $\delta(E_2) \alphaarrow \mu'$, or
  $\delta(E_1) \alphaarrow \mu'_1$, $\delta(E_2) \alphaarrow \mu'_2$
  and $\mu' = \mu'_1 \probc{r} \mu'_2$ for suitable
  $\mu'_1, \mu'_2 \in \DistrE$ and~$r \in (0,1)$. We only consider the
  last case, as the first two are simpler. In that case, we can find
  $\nu'_1, \nu'_2 \in \DistrE$ such that
  $\delta(F_1) \alphaarrow \nu'_1$, $\delta(F_2) \alphaarrow \nu'_2$,
  $\mu'_1 \brbisim \nu'_1$, and~$\mu'_2 \brbisim \nu'_2$. From this it
  follows that $\delta(F_1+F_2) \alphaarrow \nu'$
  and~$\mu' \brbisim \nu'$ for~$\nu' = \nu'_1 \probc{r} \nu'_2$ using
  Lemma \ref{unrooted-congruence}.

  The congruence properties for strong probabilistic bisimulation
  semantics are well-known, and can be obtained by simplification of
  the arguments above.
\end{proof}

\noindent
Exactly as for the nondeterministic case \cite{GW96:jacm}, one can
show that rooted branching probabilistic bisimulation semantics yields
the coarsest, or largest, congruence included in $\brbisim \mkern1mu$.

\blankline

\noindent
We close this section with two fundamental properties of branching
probabilistic bisimilarity that we need further on. We discuss the
stuttering property, known from~\cite{GW96:jacm} for non-deterministic
processes, and cancellativity of probabilistic choice with respect
to~$\brbisim \mkern1mu$.

\begin{lemma}[Stuttering Property]
  \label{stuttering}
  For all $\mu, \nu, \varrho \in \DistrE$, if $\mu \Arrow{} \nu$, $\nu
  \Arrow{} \varrho$ and $\mu \brbisim \varrho$ then it holds that $\mu
  \brbisim \nu$.
\end{lemma}

\begin{proof}
  We show that the relation
  $\calR = \lbrace (\mu,\nu), (\nu,\mu) \rbrace \cup
  \mathord{\brbisim}$ is a branching probabilistic bisimulation
  relation.

  Suppose $\mu \alphaarrow \mu'$ for some $\alpha \in \calA$, $\mu'
  \in \DistrE$. Since $\mu \brbisim \varrho$ there exist $\varrhohat,
  \varrho' \in \DistrE$ such that $\varrho \Arrow{} \varrhohat
  \alphahidearrow \varrho'$ matches $\mu \alphaarrow \mu'$. In
  particular, $\varrhohat \brbisim \mu$ and $\varrho' \brbisim
  \mu'$. Then, from $\nu \Arrow{} \varrho$ we obtain $\nu \Arrow{}
  \varrho'$. Hence $\nu \Arrow{} \varrhohat \alphahidearrow \varrho'$
  matches $\mu \alphaarrow \mu'$.

  Now suppose $\nu \alphaarrow \nu'$ for some $\alpha \in \calA$,
  $\nu' \in \DistrE$. Since $\mu \Arrow{} \nu$ by assumption, we see
  that $\mu \Arrow{} \nu \alphaarrow \nu'$ matches $\nu \alphaarrow
  \nu'$.

  To verify weak decomposability of the relation~$\calR$, first
  suppose $\mu = \bigoplusiinI \: p_i \stardot \mu_i$ for some index
  set~$I$ and $\mu_i \in \DistrE$ for~$i \in I$. Since
  $\varrho \brbisim \mu$, we obtain from weak decomposability
  of~$\brbisim$ that
  $\varrho \Arrow{} \bigoplusiinI \: p_i \stardot
  \varrho_i$ for $\varrho_i \in \DistrE$ such that
  $\varrho_i \brbisim \mu_i$ for~$i \in I$. From
  $\nu \Arrow{} \varrho$ we then obtain for $\nu_i = \varrho_i$ for
  $i \in I$ that $\nu \Arrow{} \bigoplusiinI \: p_i \stardot \nu_i$
  and $\nu_i \brbisim \mu_i$ for~$i \in I$. Second, suppose
  $\nu = \bigoplusiinI \: p_i \stardot \nu_i$ for some index set~$I$
  and $\nu_i \in \DistrE$ for~$i \in I$. Since $\mu \Arrow{} \nu$ by
  assumption, we have for $\mu_i = \nu_i$ for $i \in I$ that
  $\mu \Arrow{} \bigoplusiinI \: p_i \stardot \mu_i$ and
  $\mu_i \brbisim \nu_i$ for~$i \in I$.
\end{proof}

\noindent
The so-called \emph{cancellativity law}, proven
in~\cite{GGV23:sos-express}, provides a partial reversal of
Lemma~\ref{unrooted-congruence}: Suppose a convex combination of two
distributions is branching probabilistic bisimilar to a similar convex
combination of two other distributions. If one pair of corresponding
components is branching probabilistic bisimilar, then so is the other.

\begin{lemma}[Cancellativity]
  \label{cancellative}
  Let $\mu, \mu', \nu, \nu' \in \DistrE$. If
  $\mu \oplusr \nu \brbisim \mu'\oplusr \nu'$ with $r \in (0,1]$ and
  $\nu\brbisim\nu'$, then $\mu \brbisim \mu'$.
\end{lemma}

\noindent
  In this section we have introduced notions of strong,
  branching, and rooted branching probabilistic bisimilarity, and we
  derived various properties. In the next section we address how these
  notions of bisimilarity can be captured equationally.


\section{Complete axiomatisation of probabilistic bisimulation}

\label{sec-completeness}

\noindent
In this section we provide a sound and complete equational
characterisation of strong probabilistic bisimulation semantics and
rooted branching probabilistic bisimulation semantics. The
completeness result for the latter is obtained along the same lines as
the corresponding result for branching {\bisimilarity} for
non-deterministic processes in Section~\ref{sec-nondet}.

First, we extend and adapt the non-deterministic theory $\AXd$
of Section~\ref{sec-nondet} to the theory~$\AXp$ to cater for
probabilistic choice and combined transitions.

\begin{definition}[Axiomatization of $\bisim$]
  The equational theory~$\AXp$ is given by the axioms A1
  to~A4, the axioms P1 to~P3, and the axiom~C listed in
  Table~\ref{table-axiomatization-of-probabilistic-branching-bisimulation}.
\end{definition}

\noindent
The axioms A1--A4 for non-deterministic processes are as
before. Regarding probabilistic processes, for the axioms P1 and~P2
dealing with commutativity and associativity, we need to take care of
the probabilities involved. For~P2, it follows from the given
restrictions that it also holds that
$(1{-}r)s = (1{-}\bar{r})\bar{s}$. In particular, the probability for
subprocess~$Q$ to execute is equal for the left-hand and
right-hand side of the equation. Axiom~P3 expresses that a
probabilistic choice between equal processes can be eliminated.
Axiom~C captures that any two non-deterministic
transitions with the same label can be executed in a
combined fashion: assigning probability~$r$
to the one transition and the complementary
probability~$1{-}r$ to the other.
For the mixed setting of non-deterministic and probabilistic choice,
the axioms~A1--A4 and P1--P3, together with axiom~C have been
presented as an equational theory for
probabilistic strong bisimulation elsewhere, e.g.,
in~\cite{BS01:icalp,Hen12:facj,FG19:jlamp}.

The axioms P1 and P2 allow us to write each probabilistic
process~$P$ as 
\begin{displaymath}
  \partial(E_1)\probc{r_1}(\partial(E_2) \probc{r_2} (\partial(E_3)
  \probc{r_3}\cdots))    
\end{displaymath}
for non-deterministic processes $E_i$. As explained in
Section~\ref{sec-bpb}, we often denote such a process \vspace{2pt} by
$\bigoplusiinI \: p_i \stardot E_i$ with
$p_i = r_i\prod_{j{=}1}^{i{-}1}(1{-}r_j)$.\vspace{2pt}
More specifically, if a probabilistic process~$P$ corresponds to a
distribution $\bigoplusiinI \: p_i \stardot E_i$, i.e.\
$\den{P} = \bigoplusiinI \: p_i \stardot E_i$, then we have that
$\AXp \vdash P = \bigoplusiinI \: p_i \stardot E_i$, as can be shown
by induction on the structure of~$P$.

As noted, the equational characterisation of strong probabilistic
{\bisimilarity} has been addressed by various authors. The
theory~$\AXp$ provides a sound and complete equational axiomatisation
of strong probabilistic {\bisimilarity}. For a proof,
see~\cite{Hen12:facj}, for example.

\begin{theorem}
  \label{theorem-completeness-for-strong-probabilistic-bisimulation}
  The theory $\AXp$ is sound and complete for strong
  probabilistic {\bisimilarity}.
  \qed
\end{theorem}

\noindent
Continuing with the axiomatisation of rooted branching probabilistic
bisimulation semantics, we make use of an auxiliary
relation~${\sqsubseteq}$ on~$\calE \times \calP$, referred to as
\emph{inclusion} of a non-deterministic process in a probabilistic
process---see
Table~\ref{table-axiomatization-of-probabilistic-branching-bisimulation}.

\begin{definition}[Probabilistic inclusion]
  \label{def-probabilistic-inclusion}
  The relation ${\sqsubseteq} \subseteq {\calE \times \calP}$ is such that
  \begin{displaymath}
    E \sqsubseteq P
    \ \  \text{iff} \ 
    \begin{array}[t]{l}
      \text{$E \arrow{\alpha} \mu$
      for $\alpha \in \calA$, $\mu \in \DistrE$} 
      \  \text{implies} \  \smallskip \\ \qquad
      \text{$\den{P} \arrow{(\alpha)} \nu $
      for some $\nu \in \Distr$
      such that $\nu \bisim \mu$}
      \, .
    \end{array}
  \end{displaymath}
\end{definition}

\noindent
Thus, for $E \sqsubseteq P$ to hold, we require that every transition
of the non-deterministic process~$E$ can be directly matched by the
probabilistic process~$P$, possibly as a partial transition.
The next lemma confirms that this matching of transitions extends to
the Dirac process of~$E$.

\begin{lemma}
  \label{lem-inclusion}
  If $E \sqsubseteq P$ and $\delta(E) \arrow{\alpha} \mu$, then
  $\den{P} \alphahidearrow \nu$ for some $\nu \in \DistrE$ such that
  $\mu \bisim \nu \mkern1mu$.
\end{lemma}

\begin{proof}
  If $\delta(E) \arrow\alpha \mu$, then
  $\mu = \bigoplusiinI \: p_i \mycdot \mu_i$ and
  $E \arrow\alpha \mu_i$ for suitable index set~$I$,
  $p_i \geqslant 0$, and $\mu_i \in \DistrE$. \vspace{2pt} Since
  $E \sqsubseteq P$, we have for each~$i \in I$ that
  $\den{P} \alphahidearrow \nu_i$ for some $\nu_i \in \DistrE$
  satisfying $\mu_i \bisim \nu_i$. Hence $\den{P} \alphahidearrow \nu$
  for $\nu = \bigoplusiinI \: p_i \mycdot \nu_i$ by
  Lemma~\ref{lem-composition}, and $\mu \bisim \nu$ by
  Lemma~\ref{unrooted-congruence}.
\end{proof}

\noindent
The following lemma tells that the inclusion
  relation~$\sqsubseteq$ is compatible with strong
probabilistic bisimulation semantics.

\begin{lemma}
  \label{lem-inclusion-congruence}
  If $E \sqsubseteq P$, $E \bisim F$, and $P \bisim Q$, then $F
  \sqsubseteq Q$. 
\end{lemma}

\begin{proof}
  Suppose $E \sqsubseteq P$, $E \bisim F$, and $P \bisim Q$.  Let
  $F \arrow{\alpha} \mu$. Then $\delta(F) \arrow{\alpha} \mu$, and by
  Definition~\ref{def-probabilistic-branching-bisimilar}b
  $\delta(E) \arrow{\alpha} \mu'$ for some $\mu'\in\DistrE$ such that
  $\mu \bisim \mu'$. So, by Lemma~\ref{lem-inclusion} exists
  $\nu' \in \DistrE$ such that $\den{P} \alphahidearrow \nu'$ and
  $\mu' \bisim \nu' \mkern1mu$. Again by application of
  Definition~\ref{def-probabilistic-branching-bisimilar}b, exists
  $\nu \in \DistrE$ such that $\den{Q} \alphahidearrow \nu$ and
  $\nu' \bisim \nu$. By transitivity $\mu \bisim \nu$.
\end{proof}

\noindent
However, $\sqsubseteq$ is not compatible with (rooted) branching
probabilistic bisimulation semantics: we have
$a.P \sqsubseteq \partial(a.P) \brbisim \partial(\tau.\partial(a.P))$,
yet $a.P \not\sqsubseteq \partial(\tau.\partial(a.P))$.
Thus, if we have $\AXp \vdash P = Q$ and $E \sqsubseteq P$ we
can conclude that $E \sqsubseteq Q$. However,
with $\AXpb$ the equational theory capturing branching
  probabilistic bisimulation semantics discussed below, it is not
allowed to infer $E \sqsubseteq Q$ from $E \sqsubseteq P$ and
$\AXpb \vdash P = Q$, i.e., after proving
$P = Q$ by means of axioms that are merely sound for
branching probabilistic \bisimilarity.

We must be able to capture inclusion equationally.  The axioms I1--I6,
presented in
Table~\ref{table-axiomatization-of-probabilistic-branching-bisimulation},
together with~$\AXp$, constitute a complete axiomatisation
of~$\sqsubseteq$. We refer to the theory $\AXp,\mbox{I1--I6}$
as~$\AXpI$.

Axioms I1 and~I2 for inclusion take care of matching by a
standard transition and of matching by no transition, respectively.
Matching of a partial transition is captured by axiom~I3.
Note that the variants of axiom~I3 with $r=0$ and $r=1$ follow from I1
and~I2, respectively.
Probabilistic and non-deterministic composition are covered by axioms
I4 and~I5, respectively, deadlock by axiom~I6.

\begin{table}
  \centering
  \def\arraystretch{1.1}
  \begin{tabular}{|@{\;\;}l@{\ \;}l|}
    \hline
    A1 & $E + F = F + E$ \rule{0pt}{12pt} \\
    A2 & $(E + F) + G = E +(F + G)$ \\
    A3 & $E + E = E$ \\
    A4 & $E + \bfzero = E$ 
    \rule[-5pt]{0pt}{12pt}
    \\ \hline
    P1 & $P \oplusr Q = Q \probc{1{-}r} P$ \rule{0pt}{12pt} \\
    P2 & $P \oplusr ( Q \probc{s} R ) = (P \probc{\bar{r}} Q)
    \probc{\bar{s}} R$ \\
    &
    \qquad
    where $r = \bar{r} \mkern1mu \bar{s}$ and $(1{-}r)(1{-}s) =
    1{-}\bar{s}$ \quad \\ 
    P3  & $P \oplusr P = P$
    \rule[-5pt]{0pt}{12pt}
    \\ \hline
    \rule{0pt}{12pt}%
    \,C  & $\alpha \pref P + \alpha \pref Q = \alpha \pref P + \alpha
    \pref ( P \oplusr Q ) + \alpha \pref Q$   
    \rule[-5pt]{0pt}{12pt}
    \\ \hline
    \rule{0pt}{12pt}%
    \hypertarget{BP}{BP} & if ~ $E \sqsubseteq P$ \qquad then ~
    $\partial(E + \tau \pref P) = P$ \\
    \hypertarget{G}{G}  & if ~ $E \sqsubseteq \partial(F)$ ~ then ~
    $\partial(E + F) = \partial(F)$
    \\ \hline
    \rule{0pt}{12pt}%
    I1  & $\alpha \pref P \sqsubseteq \partial(E + \alpha\pref P)$ \\
    I2  & $\tau \pref \partial(E) \sqsubseteq \partial(E)$ \\
    I3  & $\tau \pref ( \partial(E + \tau \pref P) \oplusr P )
          \sqsubseteq \partial(E + \tau\pref P)$ \\ 
    I4  & if $\ \alpha\pref P_1 \sqsubseteq Q_1\ $ and
          $\ \alpha\pref P_2 \sqsubseteq Q_2\ $ then
          $\ \alpha\pref (P_1\oplusr P_2) \sqsubseteq Q_1 \oplusr Q_2\
          $ \\
    I5  & if $\ E_1 \sqsubseteq P\ $ and $\ E_2 \sqsubseteq P\ $ then
          $\ E_1 + E_2 \sqsubseteq P$ \\ 
    I6  & $\bfzero \sqsubseteq P$
    \\ \hline
  \end{tabular}
  \def\arraystretch{1.0}

  \medskip

  \caption{Axioms for strong and rooted branching probabilistic {\bisimilarity}}
  \label{table-axiomatization-of-probabilistic-branching-bisimulation}
\end{table}

\begin{lemma}
  \label{inclusion-completeness}
  Let $E\in\calE$ and $P\in\calP$. Then $E \sqsubseteq P$ iff
  $\AXpI\vdash E \sqsubseteq P$.
\end{lemma}

\begin{proof}
  ``If'', soundness. In view of
  Theorem~\ref{theorem-completeness-for-strong-probabilistic-bisimulation}
  and Lemma~\ref{lem-inclusion-congruence}, it suffices to check
  the soundness of each of the axioms I1--I6.  The soundness of the
  axioms I1, I2, and I6 follows immediately from the definition of
  $\sqsubseteq$.

  For the soundness of I3, the transition
  $\tau\pref (\partial(E + \tau\pref P) \oplusr P) \arrow\tau \delta(E
  + \tau\pref P) \oplusr \den P$ can be mimicked as
  $\den{\partial(E + \tau\pref P)} \arrow{(\tau)} \delta(E + \tau\pref
  P) \oplusr \den P$.  Namely,
  $\den{\partial(E + \tau\pref P)} = \delta(E + \tau\pref P) \oplusr
  \delta(E + \tau\pref P)$ and
  $\delta(E + \tau\pref P) \arrow\tau \den P$.

  For axiom~I4, \vspace{2pt} suppose
  $\alpha\pref P_i \sqsubseteq Q_i$ for $i = 1, 2$.  Since
  $\alpha\pref P_i \arrow\alpha \den{P_i}$, one has
  $\den{Q_i} \arrow{(\alpha)} \den{P_i}$ for
  $i = 1, 2$.  So
  $\den{Q_1 \oplusr Q_2} = \den{Q_1} \oplusr
  \den{Q_2} \arrow{(\alpha)} \den{P_1} \oplusr
  \den{P_2} = \den{P_1 \oplusr P_2}$ by
  Lemma~\ref{lem-composition}.  Hence, the transition
  $\alpha \pref (P_1\oplusr P_2)\arrow\alpha \den{P_1\oplusr
      P_2}$ can be matched by $\den{Q_1 \oplusr Q_2}$.

  For I5, suppose $E_i \sqsubseteq P$ for $i=1,2$, and
  $E_1+E_2\arrow\alpha \mu$.  Then either $E_1 \arrow\alpha \mu$ or
  $E_2 \arrow\alpha \mu$, so $\den{P} \alphahidearrow \nu$ for some
  $\nu \in \DistrE$ such that $\mu \bisim \nu \mkern1mu$.

  ``Only if'', completeness. Suppose $E \sqsubseteq Q$.  Using~$\AXp$,
  process $E$ can be brought in the form
  $\sumiinI \: \alpha_i\pref P_i$, and $Q$ in the form
  $\bigoplusjinJ \: q_j \mycdot F_j$ with $q_j>0$ for all~$j \in J$.

  First consider the situation where $|I| = |J| = 1$, so
  $E = \alpha \pref P$ and~$Q = \partial(F)$ for suitable $\alpha$,
  $P$, and~$F$. Since $E \arrow \alpha \den P$ and
  $E \sqsubseteq \partial(F)$, one has
  $\delta(F) \arrow{(\alpha)} \nu$ for some $\nu \in \DistrE$ such
  that $\nu \bisim \den{P}$. By Definition~\ref{def-weak-arrows}
  either (i)~$\delta(F) \arrow{\alpha} \nu$, or (ii)~$\alpha=\tau$ and
  $\nu = \delta(F)$, or (iii)~$\alpha = \tau$ and
  $\nu = \delta(F) \oplusr \nu_2$ for some $r \in (0,1)$ and
  $\nu_2 \in \DistrE$ with $\delta(F) \arrow{\alpha} \nu_2$.
  (i)~In the first case we have
  $F \bisim F + \alpha \pref P$, so
  $\AXp \vdash F = F + \alpha \pref P$ by
  Theorem~\ref{theorem-completeness-for-strong-probabilistic-bisimulation}
  and $\AXp \vdash \partial(F) = \partial(F + \alpha \pref P)$.
  Application of axiom~I1 yields
  $\AXpI \vdash E = \alpha \pref P \sqsubseteq
  \partial(F + \alpha \pref P ) = \partial(F) = Q$.
  (ii)~In the second case, $\AXp \vdash P = \partial(F)$ by
  Theorem~\ref{theorem-completeness-for-strong-probabilistic-bisimulation}.
  Axiom~I2 yields
  $\AXpI \vdash E = \tau \pref P = \tau \pref \partial(F)
  \sqsubseteq \partial(F) = Q$.
  (iii)~In the third case, decomposability required by
  Definition~\ref{def-probabilistic-branching-bisimilar}a yields
  $\den{P} = \mu_1 \oplusr \mu_2$ with $\mu_1 \bisim \delta(F)$ and
  $\mu_2\bisim \nu_2$. Thus, by
  Theorem~\ref{theorem-completeness-for-strong-probabilistic-bisimulation}
  we have $\AXp \vdash P = \partial(F) \oplusr P_2$ with $P_2$ such
  that $\den{P_2} \bisim \nu_2$. Now $F \bisim F + \tau \pref P_2$,
  since
  $\delta(F) \arrow\tau \nu_2
  \bisim \den{P_2}$. So $\AXp \vdash F = F + \tau \pref P_2$ by
  Theorem~\ref{theorem-completeness-for-strong-probabilistic-bisimulation}.
  Therefore it holds that
  \begin{displaymath}
    \AXpI  \vdash E =
    \tau \pref P =
    \tau \pref \bigl( \partial(F) \oplusr P_2 \bigr) =
    \tau \pref \bigl( \partial(F + \tau \pref P_1) \oplusr P_2 \bigr)
    \sqsubseteq
    \partial(F + \tau \pref P_2) =
    \partial(F) =
    Q
  \end{displaymath}
  by help of axiom~I3. 
  
  Next consider the situation where $|I| = 1$ and $|J| >1
  $. \vspace{1pt} So, $E = \alpha \pref P$ for some $\alpha$
    and~$P$. Since $E \arrow\alpha \den P$ and
  $E \sqsubseteq \bigoplusjinJ \: q_j\mycdot F_j$, one has
  $\bigoplusjinJ \: q_j\mycdot \delta(F_j) \arrow{(\alpha)} \nu$ for
  \vspace{1pt} some $\nu \in \DistrE$ such that
  $\den{P} \bisim \nu \mkern1mu$.  By Lemma~\ref{lem-decomposition},
  $\nu = \bigoplusjinJ \: q_j\mycdot \nu_j$ for $\nu_j \in \DistrE$
  with $\delta(F_j)\arrow{(\alpha)} \nu_{j}$ for all
  $j \in J$. By
  Definition~\ref{def-probabilistic-branching-bisimilar}a,
  $\den{P} = \bigoplusjinJ \: q_j\mycdot \mu_j$ for
  $\mu_j \in \DistrE$ with $\mu_j \bisim \nu_{j}$ for all
  $j \in J$.  We have
    $\AXp \vdash P = \bigoplusjinJ \: q_j\mycdot P_j$ where
    $P_j \in \calP$ is such that $\den{P_j} = \mu_j$ for all~$j \in
    J$. In addition, $\alpha \pref P_j \sqsubseteq \partial(F_j)$,
    hence
    $\AXpI \vdash \alpha \pref P_j \sqsubseteq
    \partial(F_j)$ by the previous case of $|I|=1$ and~$|J|=1$.
  Therefore it follows that
  \begin{displaymath}
      \AXpI \vdash E = \alpha \pref P =
    { \alpha \pref \bigl( \bigoplusjinJ \: q_j \pref P_j \bigr) } \sqsubseteq
    { \bigoplusjinJ \: q_j \pref F_j } = Q
  \end{displaymath}
  by application of axiom~I4 specifically.

  Finally, consider $E = \sumiinI \: \alpha_i\pref P_i$
  with $|I| > 1$. \vspace{1pt} If
  $\alpha_i \pref P_i \arrow{\alpha_i} \mu$ for some $i \in I$ then
  $E \arrow{\alpha_i} \mu$. Since $E \sqsubseteq Q$, this implies
  $Q \arrow{(\alpha_i)} \nu$ for some~$\nu \bisim \mu$. Hence
  $\alpha_i\pref P_i \sqsubseteq Q$.
  Thus, by the above, we obtain $\AXpI \vdash {\alpha_i\pref P_i
    \sqsubseteq Q}$ for all~$i \in I$.
  Axioms I5 and I6 then yield $\AXpI \vdash E \sqsubseteq Q$.
\end{proof}

\noindent
Having introduced the notion of inclusion and its axiomatisation, we
now introduce an axiomatisation for rooted branching probabilistic
bisimulation semantics that we will show to be sound and complete. The
theory involves the inclusion relation. At the end of this section we
convert the conditional axioms \hyperlink{BP}{BP} and \hyperlink{G}{G}
to a purely equational form.

\begin{definition}[Axiomatization of~$\rootedbrbisim$]
  The theory~$\AXpb$ contains $\AXp$ together with the axioms
  \hyperlink{BP}{BP} and \hyperlink{G}{G}. The preconditions
  $E \sqsubseteq P$ of \hyperlink{BP}{BP} and \hyperlink{G}{G} are to
  be obtained from $\AXpI$.
\end{definition}

\noindent
Axiom~\hyperlink{BP}{BP} is an adaptation of axiom~\hyperlink{B}{B} of
the theory~$\AXdb$ presented in Section~\ref{sec-nondet} to
the probabilistic setting of~$\AXpb$.  In the setting of
non-deterministic processes the implication $F \arrow{\alpha} F'$
$\Longrightarrow$ $E \arrow{\alpha} E'$ $\land$ $F' \brbisim E'$ for
some $E'$ is captured by $E + F \mathrel{\rootedbrbisim} E$. If we
reformulate axiom~\hyperlink{B}{B} as $E + F = E$ $\Longrightarrow$
$\alpha \pref ( F + \tau \pref E ) = \alpha \pref E$, then it becomes
more similar to axiom~\hyperlink{BP}{BP} in
Table~\ref{table-axiomatization-of-probabilistic-branching-bisimulation}.
In comparison, the action prefix~$\alpha$ occurring
in~\hyperlink{B}{B}, related to the root condition for rooted
branching bisimulation, has been replaced in~\hyperlink{BP}{BP} by the
Dirac construction $\partial( \cdot )$.

As to~\hyperlink{BP}{BP}, a non-deterministic alternative~$E$ that is
also offered by a probabilistic process~$P$ after a $\tau$-prefix can
be dispensed with, together with the prefix~$\tau$. In a formulation
retaining the $\tau$-prefix on the right-hand side, the
axiom~\hyperlink{BP}{BP} shows similarity with axioms~T2 and~T3
in~\cite{FG19:jlamp} which, in turn, are reminiscent of axioms T1
and~T2 of~\cite{DP07:tcs}; these axioms stem from Milner's second
$\tau$-law~\cite{Mil89:phi}.

\blankline

\noindent
Let us illustrate the working of axiom~\hyperlink{BP}{BP}. Consider the
non-deterministic process $E = \bstop$ and the probabilistic process
$P = \partial( \astop + \bstop ) \probc{1/2} \partial( \bstop)$. Then we
have $E \sqsubseteq P$, i.e.\ 
\begin{displaymath}
  \bstop \sqsubseteq 
  \partial( \astop + \bstop ) \probc{1/2} \partial( \bstop ).
\end{displaymath}
Therefore, we have by application of axiom~\hyperlink{BP}{BP} the
provable equality
\begin{displaymath}
  \begin{array}{l}
    \AXpb \vdash
    \partial \dg( {\bstop} + \tau \pref \mg( 
    \partial\dor( \astop + \bstop \dor) \probc{1/2}
    \rule{0pt}{9.5pt}\partial\dor( 
    \bstop \dor) \mg) 
     \dg)
    = {} \smallskip \\ \hspace{94pt}
     \dg(\rule{0pt}{9.5pt}
    \partial\dor( \astop + \bstop \dor) \probc{1/2} \partial\dor( \bstop \dor)
    \dg) \; .
  \end{array}
\end{displaymath}
Another example is
$a \pref (P_1\oplusr P_2) \sqsubseteq \partial\dg(b\pref R + a\pref
P_1\dg)\oplusr\partial\dg(c\pref S + a\pref P_2\dg)$, 
so
\begin{displaymath}
  \begin{array}{l}
    \AXpb \vdash
   \partial\mg(
    a\pref(P_1\oplusr P_2) + \tau \pref \dor( \partial\dg( b \pref R
    + 
    a \pref P_1\dg)\oplusr\partial\dg(c\pref S + a\pref P_2\dg) \dor) 
     \mg)
    = {} \smallskip \\ \hspace{121pt}
     \dor( \partial\dg(b\pref R + a\pref P_1 \dg) \oplusr
    \partial\dg(c\pref S + a\pref P_2\dg) \dor)
      \;.
  \end{array}
\end{displaymath}
An example illustrating partiality, i.e., the use
of~$(\tau)$, rather than~$\tau$ as label of the matching
transition of~$\den{P}$ in the definition of~$\sqsubseteq$ is
\begin{displaymath}
  \tau \pref \mg(\partial(b \pref P + \tau \pref Q)\oplusr Q\mg)
  \sqsubseteq  \partial(b \pref P + \tau \pref Q)
\end{displaymath}
from which we obtain
\begin{displaymath}
  \begin{array}{l}
    \AXpb \vdash
    \partial\rule{0pt}{9.5pt}\big(
    \tau \pref \mg(\partial(b \pref P + \tau \pref Q)\oplusr Q\mg)
    + \tau \pref \mg( \partial(b \pref P + \tau \pref Q) \mg) 
    \big)
    = 
     \mg( \partial(b \pref P + \tau \pref Q) \mg)\;.
  \end{array}
\end{displaymath}
Roughly, the axiom \hyperlink{G}{G} is a variant of \hyperlink{BP}{BP}
without the $\tau$ prefixing the process $P$. A typical example,
matching the one above, is
\begin{displaymath}
  \begin{array}{l}
    \AXpb \vdash
    \partial\rule{0pt}{9.5pt}\big(
    \tau \pref \mg(\partial(b \pref P + \tau \pref Q)\oplusr Q \mg)
    + \mg( b \pref P + \tau \pref Q \mg) 
    \big)
    =
     \mg( \partial(b \pref P + \tau \pref Q)
    \mg) \; . 
  \end{array}
\end{displaymath}

\blankline

\noindent
Before presenting concrete applications of~$\AXpb$ we first discuss
some derived laws, also as an illustration.
The second equation is the B-axiom of~\cite{GW96:jacm} in the setting
of~$\AXpb$.  The axioms in a non-probabilistic setting are preceded by
an action to deal with the rootedness issue. In the probabilistic
setting it is enough to use only the Dirac operator. See the first
equation. Clearly the second equation follows directly from the first.
The last one is the pendant of Milner's first $\tau$-law.

\begin{lemma} The following equations are derivable from $\AXpb$.
  \label{lemma-derived-laws}
  \begin{enumerate}[(a)]
  \item \label{caseb} 
    $\AXpb \vdash {\partial( E + F )} = {\partial( \tau \pref
      \partial(E+F) + F )}$.
  \item \label{casebx}
    $\AXpb \vdash {\alpha \pref \partial( E + F )} = {\alpha
      \pref \partial( \tau \pref \partial(E+F) + F )}$.    
  \item \label{caseaxx}
    $\AXpb \vdash {P = \partial( \tau \pref P )}$.
  \end{enumerate}
\end{lemma}

\begin{proof}
  \begin{enumerate}[(a)]
  \item The equation follows from axioms \hyperlink{BP}{BP} and~A1 by
    taking $P:=\partial(E+F)$, using that
    $F \sqsubseteq \partial(E+F)$.
  \item The equation follows directly from the previous item using
    congruence.
  \item The equation is yielded by the axioms \hyperlink{BP}{BP}, I6,
    A1, and~A4, taking $E:=\bfzero$.  \qedhere
\end{enumerate}
\end{proof}

\noindent
We further illustrate the working of theory~$\AXpb$ by showing the
equivalence of the remaining example pairs of
Figure~\ref{fig:examples}.
  
\begin{example}
  \begin{enumerate} [(a)]
  \item 



For the two examples taken from~\cite{Hen12:facj} we exploit
associativity and splitting as provided by the axioms P2 and P3.
\begin{flalign*}
  && \AXpb &\vdash \mathbf{H_1} \\
  \qquad
  && & \myeq
    {a \pref \bigl( P \probc{\frac14} ( P \probc{\frac13} Q )
    \bigr)}\\ 
  && & \myeq
    a \pref \bigl( ( P \probc{\frac{1}{2}} P ) \probc{\frac12} Q
    \bigr)
  & && \text{Axiom P2} \\
  & &&\myeq
    a \pref \bigl( P \probc{\frac12} Q \bigr)
  & &&\text{Axiom P3} \\
  && & \myeq
    a \pref \bigl( P \probc{\frac12} ( Q \probc{\frac12} Q )
    \bigr)
  & &&\text{Axiom P3} \\
  && & \myeq
    \mathbf{H_2} \mkern1mu .
  \end{flalign*}    
  \item 



The partial inert expansion of the examples $\mathbf{G_1}$
and~$\mathbf{G_2}$ is covered by a combination of axiom~P3, allowing
to split a distribution, and the axiom~BP to introduce an inert
transition in the proper context.
\begin{flalign*}
  && \AXpb &\vdash\, \mathbf{G_1} \\
  && & \myeq
    a \pref ( P \probc{\frac12} Q ) \\
  && & \myeq
    a \pref \bigl(
    ( P \probc{\frac12} Q )
    \probc{\frac13}
    ( P \probc{\frac12} Q )
    \bigr)
  & &&\text{Axiom P3} \\
  && & \myeq
    a \pref \bigl(
    \partial(
    \tau \pref ( P \probc{\frac12} Q ) )
    \probc{\frac13}
    ( P \probc{\frac12} Q )
    \bigr)
  & &&\text{Lemma~\ref{lemma-derived-laws}\ref{caseaxx}} \\
  && & \myeq
    \mathbf{G_2} \mkern1mu .
  \end{flalign*}
    
  \item 



To show equality of the processes $\mathbf{I_1}$ and $\mathbf{I_2}$ we
make use of Lemma~\ref{lemma-derived-laws}.
We have
\begin{flalign*}
  && \AXpb &\vdash\, \mathbf{I_1} \\
  && & \myeq
    a \pref \partial( b \pref P + \tau \pref Q ) \\
  && & \myeq
    a \pref \partial(
      b \pref P + \tau \pref Q 
    + b \pref P + \tau \pref Q )
  & &&\text{Axiom A3} \\
  && & \myeq
    a \pref \partial( \tau \pref
    \partial( b \pref P + \tau \pref Q + b \pref P + \tau \pref Q )
    + b \pref P + \tau \pref Q )
  & &&\text{Lemma~\ref{lemma-derived-laws}\ref{caseb}} \\
  && & \myeq
    a \pref \partial( \tau \pref
    \partial( b \pref P + \tau \pref Q )
    + b \pref P + \tau \pref Q )
  & &&\text{Axiom A3} \\
  && & \myeq \mathbf{I_2} \mkern1mu .
\end{flalign*}
   
  \item 
    The equality of the process $\mathbf{T_1}
    = a \pref \mkern1mu \partial( b \pref ( P
      \probc{\frac13} Q ) ) \probc{\frac12} \partial( b \pref ( P
      \probc{\frac23} Q ) )$ and the process~$\mathbf{T_2}
    = a \pref \mkern1mu \partial ( \tau \pref \bigl(
      \partial( b \pref ( P \probc{\frac13} Q ) ) \probc{\frac12}
      \partial( b \pref ( P \probc{\frac23} Q ) ) \bigr) + b \pref ( P
      \probc{\frac12} Q )) $ modulo~$\AXpb$ follows immediately from
    axiom~\hyperlink{BP}{BP} once we have
    $b \pref ( P \probc{\frac12} Q ) \sqsubseteq \partial ( b \pref (
    P \probc{\frac13} Q ) ) \probc{\frac12} \partial( b \pref ( P
    \probc{\frac23} Q ) ) $.  This follows from axioms I1 and~I4, by
    writing $b \pref ( P \probc{\frac12} Q )$ as
    $b \pref ( (P \probc{\frac13} Q) \probc{\frac12} (P
    \probc{\frac23} Q) )$, using axioms P1 to~P3 as well.
  \item 

To formally prove that the processes $\mathbf{E_1}$ and~$\mathbf{E_2}$
of~\cite{EHKTZ13:qest} are equal with respect to~$\AXpb$ we observe
\begin{flalign*}
&&  \AXpb &\vdash\, \mathbf{E_1}&\\
&&  & \myeq
    \tau \pref \mkern1mu (
    \partial( \tau \pref P + a \pref Q + \tau \pref R )
    \probc{\frac34} \partial(\bfzero) ) \\
&& & \myeq
    \tau \pref \mkern1mu (
    \partial( \tau \pref P + a \pref Q + \tau \pref R )
    \probc{\frac12 \mkern-1mu}
    (
    \partial( \tau \pref P + a \pref Q + \tau \pref R )
    \probc{\frac12 \mkern-1mu} \partial(\bfzero)
    ) )& &&\text{Axioms P3, P2}\\
&&  & \myeq
    \tau \pref \mkern1mu (
    \partial( \tau \pref P + a \pref Q + \tau \pref R )
    \probc{\frac12 \mkern-1mu}\\
&& &\qquad\qquad    (
    \partial( \tau \pref
    \partial( \tau \pref P + a \pref Q + \tau \pref R ) 
    \probc{\frac12 \mkern-1mu} \partial(\bfzero) ) ) )
  & &&\text{Lemma~\ref{lemma-derived-laws}\ref{caseaxx}}\\
&&  & \myeq
    \mathbf{E_2} \mkern1mu .
  \end{flalign*}

  \end{enumerate}
\end{example}

\noindent
Soundness of
the theory~$\AXpb$ for rooted branching probabilistic {\bisimilarity}
is straightforward.

\begin{lemma} [Soundness]
  \label{lemma-soundness}
  For all $P,Q \in \calP$, if $\AXpb \vdash P = Q$ then
  $P \brbisim Q$.  Likewise, for all $E,F \in\calE$, if
  $\AXpb \vdash E = F$ then $E \rootedbrbisim F$.
\end{lemma}

\begin{proof}
  In view of
  Theorem~\ref{theorem-completeness-for-strong-probabilistic-bisimulation}
  and Lemma~\ref{lemma-congruence-prob} we only need to prove the
  soundness of the axioms \hyperlink{BP}{BP} and \hyperlink{G}{G}.
  For \hyperlink{BP}{BP} we need to show that
  $\delta(E + \tau.P) \brbisim \den P$ when $E \sqsubseteq P$, and for
  \hyperlink{G}{G} that $\delta(E + F) \brbisim \delta(F)$ when
  $E \sqsubseteq \partial(F)$.

  Both statements follow from the observation that the relations
    \begin{displaymath}
      \lc \pair{\delta(E + \tau.P)}{\den P}
      \mid E \sqsubseteq P \rc^\dagger
      \cup \mathord{\brbisim}
    \end{displaymath}
    and 
    \begin{displaymath}
      \lc \pair{\delta(E + F)}{\delta(F)}
      \mid E \sqsubseteq \partial(F) \rc^\dagger
      \cup \mathord{\brbisim}
    \end{displaymath}
    are branching probabilistic bisimulation
    relations, as can be straightforwardly checked.
\end{proof}

\noindent
As for the process language with non-deterministic processes only, we
aim at a completeness proof that is built on the
completeness of strong {\bisimilarity} and the notion of a concrete
process.
In Section~\ref{sec-nondet}, a process is called concrete if it does
not exhibit an inert transition. In the setting with probabilistic
choice we need to be more careful. Not only do we need to consider
\hyperlink{partinert}{partially inert transitions}, but we also want to
exclude, for example, processes of the form
\begin{displaymath}
  \partial( \tau \pref P ) \probc{1/2} \partial( a \pref Q ) )
  \quad \text{and} \quad
  \partial( a \pref P ) \probc{1/2} \partial( b \pref \bigl(
  \partial( \tau \pref Q ) \probc{1/3} Q
  \bigr))
\end{displaymath}
from being concrete, although they cannot perform a transition by
themselves at all. To capture inertness of subprocesses we employ the
following definition.

\begin{definition}
  \label{def-derivatives}
  \begin{itemize}
  \item [(a)] The \emph{derivatives}
    $\der(E), \der(P) \subseteq \calE$ of non-deterministic and
    probabilistic processes $E \in \calE$ and $P \in \calP$ are given
    by
    \begin{displaymath}
      \begin{array}{r@{~=~}l}
        \der( \mkern1mu \sumiinI \: \alpha_i \pref P_i \mkern1mu )
        & \{ \, \sumiinI \: \alpha_i \pref P_i \, \} \cup
          \bigcup_{i \in I} \: \der(P_i)
          \smallskip \\
        \der( \mkern1mu \bigoplusiinI \: p_i \stardot E_i \mkern1mu )
        & \phantom{\{ \,}
          \bigcup_{i \in I} \: \der(E_i) \mkern1mu . \\
      \end{array}
    \end{displaymath}
  \item [(b)] A process $\Ebar \in \calE$ or $\Pbar \in \calP$ is said
    to be \emph{concrete} if none of its derivatives can perform a
    \hyperlink{partinert}{partially inert transition}, i.e., there is
    no $E \in \der(\Ebar)$ or $E \in \der(\Pbar)$, respectively, that
    has a transition of the form $E \arrow{\tau} \mu_1\oplusr \mu_2$
    with $r > 0$ and $\delta(E) \brbisim \mu_1$.
  \item [(c)] A process $E \in \calE$ is called
      \hypertarget{rigid}{\emph{rigid}} if it has no
      \hyperlink{partinert}{inert transition}, i.e., a transition $E
      \arrow\tau \mu$ such that $\delta(E) \brbisim \mu$.
  \item [(d)] The sets $\calEconc$ and $\calPconc$ of concrete
    non-deterministic processes and concrete probabilistic processes
    are given by
    \begin{displaymath}
      \calEconc = \lc \Ebar \in \calE \mid \text{$\Ebar$ is concrete} \rc
      \quad \text{and} \quad \calPconc = \lc \Pbar \in \calP \mid
      \text{$\Pbar$ is concrete} \rc \mkern1mu .
    \end{displaymath}
    The set $\calErigid$ of non-deterministic processes that
      are rigid is given by
      $\calErigid = \lc E \in \calE \mid \text{$E$ is rigid} \rc$.
  \end{itemize}
\end{definition}

\noindent
A concrete process has no derivative that can perform a
\hyperlink{partinert}{partially inert transition}. 
  Thus, a non-deterministic process is concrete if it has
  no partially inert transition and all its derivatives are are
  concrete. A probabilistic process is concrete if it is a
  probabilistic choice of concrete non-deterministic processes.
Clearly, $\calEconc \subseteq \calErigid$.
We use concrete and rigid processes to build the
completeness proof for
rooted branching probabilistic {\bisimilarity} on top of the
completeness proof of strong probabilistic {\bisimilarity}. The
following lemma lists all the properties of concrete and rigid
processes that we need for this.

\begin{lemma}
  \label{cc}
  \begin{enumerate}[(a)]
  \item Let $E = \sumiinI \: \alpha_i \cdot P_i$ with
    $P_i \in \calPconc$ for~$i \in I$. If, for all $i \in I$,
    (i)~$\alpha_i \neq \tau$, or (ii)~$\den{P_i \mkern1mu}$ cannot be
    written as $\mu_1 \oplusr \mu_2$ where~$r > 0$ and
    $\delta(E) \brbisim \mu_1$, then $E \in \calEconc$.

  \item If $P_1,P_2\in\calPconc$, then $P_1 \oplusr P_2\in\calPconc$.

  \item If $\mu = \bigoplusiinI \: p_i \stardot \mu_i \in \Distr(\calEconc)$
    where $p_i > 0$ for all~$i \in I$, then $\mu_i \in \DistrEconc$
    for all~$i \in I$.

  \item If $\mu\in\Distr(\calEconc)$ and $\mu \alphahidearrow \mu'$,
    then $\mu'\in\DistrEconc$. 

  \item If $\mu \in \Distr(\calErigid)$ and $\mu \, \Arrow{\:} \, \mu'$
    with $\mu \brbisim \mu'$, then $\mu = \mu'$.

  \item If $\mu \brbisim \bigoplusiinI \: p_i \stardot \nu_i$ for
    $\mu \in \Distr(\calE_r)$, then
    $\mu = \bigoplusiinI \: p_i\stardot\mu_i$ for certain
    $\mu_i \brbisim \nu_i$ with~$i \in I$.

  \item If $E \in \calErigid$ and $\delta(E) \brbisim \nu$, then
    $E \brbisim F$ for all $F \in \spt(\nu)$.

  \item If $E \in \calEconc$ and
    $\delta(E) \arrow{\tau} \mu_1 \oplusr \mu_2$ with
    $\delta(E) \brbisim \mu_1$, then~$r=0$.

  \item Let $E \mathbin\in \calEconc$ and $F \mathbin\in \calE$ such
    that $\delta(E) \brbisim \delta(F)$. If
    $\delta(E) \arrow{\alpha} \mu$, $\delta(F) \arrow{(\alpha)} \nu$,
    and $\mu \brbisim \nu$ for $\mu, \nu \mathbin\in \DistrE$, then
    $\delta(F) \arrow{\alpha} \nu$ and there is no $G\in\spt(\nu)$
    with $E \brbisim G$.
  \end{enumerate}
\end{lemma}

\begin{proof}
  For property~(a) it suffices to show that $E$ cannot perform a
  \hyperlink{partinert}{partially inert transition}. So suppose, to the
  contrary, that $E \arrow{\tau} \mu_1\oplusr \mu_2$ with $r > 0$ and
  $\delta(E) \brbisim \mu_1$. Then there is an index
  $i \in I$ with $\alpha_i = \tau$ and
  $\den{P_i}=\mu_1 \oplusr \mu_2$, contradicting the assumption.

  Property~(b) follows directly from the observation that following
  Definition~\ref{def-derivatives} it holds that
  $\der(P_1 \oplusr P_2) = \der(P_1) \cup \der(P_2)$.

  To see property~(c), we observe that,
  for~$i \in I$, $\spt(\mu_i) \subseteq \spt(\mu)$, and
  therefore $\spt(\mu) \subseteq \calEconc$ implies
  $\spt(\mu_i) \subseteq \calEconc$.

  As to property~(d), we can assume
  $\spt(\mu) = \lc E_i \mid i \in I \rc$ and
  $\mu'= \bigoplusiinI \: p_i \stardot \mu'_i$ where
  $E_i \arrow{\alpha} \mu'_i$ or $\mu'_i = \delta(E_i)$ for~$i \in
  I$. So, we have
  $\spt(\mu'_i) \subseteq \der(E_i) \subseteq \calEconc$ or
  $\spt(\mu'_i) = E_i \in \spt(\mu) \subseteq \calEconc$,
  respectively, for~$i \in I$. Therefore,
  $\spt(\mu'_i) \subseteq \calEconc$ for~$i \in I$ and
  $\spt(\mu') \subseteq \calEconc$.
  
  For (e), let $\mu \mathbin\in \Distr(\calErigid)$ and
  $\mu \Arrow{\:} \mu'$ with $\mu \brbisim \mu'$. Towards a
  contradiction, suppose $\mu \neq \mu'$. Then there must be a
  distribution $\mubar \neq \mu$ such that $\mu \tauhidearrow \mubar$
  and $\mubar \Arrow{\:} \mu'$. We may even choose $\mubar$ such that
  the transition $\mu \tauhidearrow \mubar$ acts on only one element
  of the support of~$\mu$, i.e.\ there are $E \in \calE$,
  {$r \in (0,1]$} and $\muhat, \varrho \in \DistrE$ such that
  $\mu = \delta(E) \oplusr \varrho$, $E \arrow\tau \muhat$ and
  $\mubar = \muhat \oplusr \varrho$. By the stuttering property of
  Lemma~\ref{stuttering}, $\mu \brbisim \mubar$, thus
  $\delta(E) \oplusr \varrho = \mu \brbisim \mubar = \muhat \oplusr
  \varrho$.  Hence, by cancellativity of Lemma~\ref{cancellative}, we
  obtain $\delta(E) \brbisim \muhat$. So, the transition
  $E \arrow\tau \muhat$ is inert, contradicting $E \in \calErigid$.

  Regarding~(f), if $\mu \brbisim \bigoplusiinI \: p_i \stardot \nu_i$
  for $\mu \mathbin\in \Distr(\calE_r)$, then
  $\mu \Arrow{} \bar\mu := \bigoplusiinI \: p_i \stardot \mu_i$ with
  $\mu \brbisim \mubar$ and $\mu_i \brbisim \nu_i$ by weak
  decomposability of~$\brbisim \mkern1mu$.  By case~(e) of the lemma
  we have $\bar\mu = \mu$.

  For (g), suppose $E \in \calErigid$ and $\delta(E) \brbisim \nu$.
  Let $\nu = \bigoplusiinI \: p_i \mycdot F_i$. Then
  $\delta(E) = \bigoplusiinI \: p_i\stardot \mu_i$ for certain
  $\mu_i \brbisim \delta(F_i)$, using case~(f).  Now $\mu_i=\delta(E)$
  for all $i \in I$ and hence $F_i \brbisim E$ for all~$i \in I$.

  For (h), suppose $E \in \calEconc$ and
  $\delta(E) \arrow{\tau} \mu_1\oplusr \mu_2$ with
  $\delta(E) \brbisim \mu_1$ and $r>0$.  Say,
  $\mu_1 \oplusr \mu_2 = \bigoplusiinI \: p_i \stardot \nu_i$ where
  $E \arrow\tau \nu_i$ for~$i \in I$. Since $r > 0$, for
  some~$i \in I$ we have that
  $\spt(\mu_1) \cap \spt(\nu_i) \neq \emptyset$. By case~(g) of the
  lemma it holds that $F \brbisim E$ for~$F \in \spt(\mu_1)$. Thus,
  the transition $E\arrow{\tau}\nu_i$ is
  \hyperlink{partinert}{partially inert}. This contradicts the
  assumption $E \in \calEconc$.

  To establish~(i), suppose $E \in \calEconc$, $F \in \calE$, and
  $\mu, \nu \in \DistrE$ \vspace{2pt} are such that
  $\delta(E) \brbisim \delta(F)$, $\delta(E) \arrow{\alpha} \mu$,
  $\delta(F) \arrow{(\alpha)} \nu$ and $\mu \brbisim \nu$. Assume, for
  proving the first conclusion of case~(i), that $\alpha = \tau$, for
  otherwise the statement is trivial. Then, for some
  $\nu_1, \nu_2 \in \DistrE$ and~$r \in [0,1]$,
  $\nu = \nu_1 \oplusr \nu_2$, $\nu_1 = \delta(F)$ and either $r=1$ or
  $\delta(F) \tauarrow \nu_2$. Since $\mu \brbisim \nu$ and~$\brbisim$
  is weakly decomposable, there are
  $\bar\mu, \mu_1, \mu_2 \in \DistrE$ such that
  $\mu \Arrow{\:} \bar\mu = \mu_1 \oplusr \mu_2$,
  $\mu \brbisim \bar\mu$, $\mu_1 \brbisim \nu_1$ and
  $\mu_2 \brbisim \nu_1$. Because $\mu \in \Distr(\calEconc)$ by
  case~(d) of the lemma, it follows that $\bar\mu = \mu$ by case~(e)
  of the lemma. Thus $\delta(E) \arrow{\tau} \mu_1\oplusr \mu_2$ and
  $\delta(E) \brbisim \delta(F) = \nu_1 \brbisim \mu_1$.
  Since $E$~is concrete and
  $\delta(E) \arrow{\tau} \mu_1\oplusr \mu_2$, case~(h) of the lemma
  yields~$r = 0$. It follows that $\nu = \nu_{2}$ and
  $\delta(F) \arrow{\tau} \nu$.

  For the second conclusion of case~(i), suppose $E \brbisim G$ for
  some $G \in \spt(\nu)$.  Then $\nu = \delta(G) \oplusr \nu'$ for
  some $\nu' \in \DistrE$ and~$r>0$. Exactly as above we find
  $\mu_1,\mu_2 \in \DistrE$ such that $\mu = \mu_1\oplusr \mu_2$,
  $\delta(E) \arrow{\tau} \mu_1\oplusr \mu_2$ and
  $\delta(E) \brbisim \delta(G) \brbisim \mu_{1}$,
  contradicting case~(h) of the lemma.
\end{proof}

\noindent
Because in concrete probabilistic processes partially inert
transitions are absent, branching bisimilarity and strong bisimilarity
coincide for concrete processes. As
mentioned, this will help us to lift the completeness of the
theory~$\AXp$ for strong probabilistic bisimulation to the
completeness of the theory~$\AXpb$ for branching probabilistic
bisimulation.

\begin{lemma}
  \label{lemma-equal-for-concrete-processes}
  For all $\mu,\nu\in \Distr(\calEconc)$, if $\mu\brbisim\nu$ then
  $\mu\bisim\nu$.
\end{lemma}

\begin{proof}
  Let $\mu,\nu\in \Distr(\calEconc)$ be such that $\mu \brbisim
  \nu$. Define the relation~$\calR$ by
  \[\calR = {\brbisim} \, \cap {(\Distr(\calEconc) \times
      \Distr(\calEconc))}\;.\] Then, by Lemma~\ref{cc}cd,
  $\calR$~is a branching probabilistic bisimulation relation relating
  $\mu$ and~$\nu$. We show that $\calR$ moreover satisfies the
  conditions of Lemma~\ref{state-based}.

  That $\calR$~is decomposable follows by Lemma~\ref{cc}e
  since $\calR$ is weakly
  decomposable.
  Condition~(\ref{cong}) of Lemma~\ref{state-based} is a direct
  consequence of the congruence result of
  Lemma~\ref{unrooted-congruence} and the observation that
    $\mu_1 \oplusr \mu_2, \nu_1 \oplusr \nu_2 \in \Distr(\calEconc)$
    if $\mu_1, \mu_2, \nu_1, \nu_2 \in \Distr(\calEconc)$.
  Now, in order to verify condition~(\ref{diamond}) of the
    lemma, \vspace{2pt} suppose
  $\delta(E), \delta(F) \in \Distr(\calEconc)$ are such that
  $\delta(E) \mopcalR \delta(F)$ and $E\alphaarrow \mu'$. Then
  $\delta(F) \Arrow{} \nubar \alphahidearrow \nu'$ for some
  $\nubar,\nu'\in \DistrE$ such that
  $\delta(E) \mopcalR \nubar$ and $\mu' \mopcalR \nu'$. By
  Lemma~\ref{cc}e we have~$\nubar = \delta(F)$. Thus
  $\delta(F) \arrow{(\alpha)} \nu'$. Using
  Lemma~\ref{cc}i it follows that $\delta(F) \arrow{\alpha} \nu'$.
  With $\calR$ being decomposable and satisfying conditions
  (\ref{cong}) and~(\ref{diamond}), Lemma~\ref{state-based} yields
  $\mu \bisim \nu$.
\end{proof}

\noindent
Lemma~\ref{lemma-equal-for-concrete-processes}
has a partial completeness result as an immediate consequence:
branching bisimilarity for concrete probabilistic processes implies
equality modulo~$\AXpb$, in fact modulo~$\AXp$.

\begin{corollary}
 \label{Corollary5.10}
  For all $\Pbar,\Qbar \in \calPconc$, if $\Pbar \brbisim \Qbar$ then
  $\AXp \vdash \Pbar = \Qbar$. 
\end{corollary}

\begin{proof}
  For $\Pbar,\Qbar \in \calPconc$ such that $\Pbar \brbisim \Qbar$,
  Lemma~\ref{lemma-equal-for-concrete-processes} above
  yields~$\Pbar \bisim \Qbar$. By
  Theorem~\ref{theorem-completeness-for-strong-probabilistic-bisimulation}
  it then follows that $\AXp \vdash \Pbar = \Qbar$, hence
  $\AXpb \vdash \Pbar = \Qbar$.
\end{proof}

\noindent 
The following lemma plays a crucial r\^ole in our completeness proof.
It describes exactly the precondition $E \sqsubseteq \partial(F)$
that is needed for application of the
axiom \hyperlink{G}{G} in the proof.

\begin{lemma}
  \label{inclusion-G}
  Let $F, H \in \calEconc$, $P \in \calPconc$, $r \in (0,1)$, and let
  ${ E := {\tau \pref (\partial(F) \oplusr P) + H }} \in \calE_r$
  be such that $E \brbisim F$.
  Suppose~$G \brbisim E$ for no~$G \in \spt(\den{P})$.  Then
  ${\tau \pref (\partial(F) \oplusr P)} \sqsubseteq
  \partial(H)$.
\end{lemma}

\begin{proof}
  It suffices to show that $\delta(H) \tauhidearrow \nu$ for some
  $\nu \bisim \delta(F) \oplusr \den{P}$. To this end we prove
  three claims. \vspace{-1ex}
  \begin{enumerate} [(a)] \itemsep=0pt
  \item $\delta(F)\arrow\tau \mu$ for some~$\mu \in \DistrE$
    with $\mu \brbisim \den{P}$.
  \item If $\delta(F)\arrow\alpha \varrho$ then
    $\delta(H) \alphaarrow \eta$ for some~$\eta \in \DistrE$
    with $\eta \brbisim \varrho$.
  \item $H \brbisim F$. \vspace{-1ex}
  \end{enumerate}
  From claims (a) and~(b) we derive $\delta(H) \tauarrow \eta$ for
  some~$\eta$ with~$\eta \brbisim \den{P}$. \vspace{2pt} Now take
  $\nu := \delta(H)\oplusr\eta$. Then $\delta(H) \tauhidearrow \nu$.
  By claim~(c) and Lemma~\ref{unrooted-congruence} we obtain
  $\nu \brbisim \delta(F) \oplusr \den{P}$.
  By Lemma~\ref{cc}d we have $\eta \in \DistrEconc$, and thus
  $\delta(H) \,\mathord{\oplusr} \eta, \; \delta(F)
  \,\mathord{\oplusr} \den{P} \in\DistrEconc$ by the
    assumptions for $H$, $F$, and~$P$ being concrete.  Hence
  $\nu \bisim \delta(F) \oplusr \den{P}$ by
  Lemma~\ref{lemma-equal-for-concrete-processes}.

  Proof of claim~(a).
  Since $E \arrow\tau \delta(F) \oplusr \den{P}$ and $E \brbisim F$,
  we have $\delta(F) \Arrow{\;} \mubar \tauhidearrow \mu$ for some
  $\mubar \brbisim \delta(F)$ and
  $\mu \brbisim {\delta(F)\oplusr \den P}$.
  By Lemma~\ref{cc}e it follows that $\mubar = \delta(F)$,
  using that $F$~is rigid.
  By Lemma~\ref{cc}d we have $\mu \in \DistrEconc \subseteq
  \Distr(\calErigid)$.
  So by Lemma~\ref{cc}f, $\mu = \mu_1\oplusr \mu_2$ with
  $\mu_1 \brbisim \delta(F)$ and $\mu_2 \brbisim \den{P}$.
  On the other hand, as $\delta(F) \tauhidearrow \mu$, we obtain
  $\mu = \delta(F) \probc{s} \mu'$ for some $s \in [0,1]$
  and $\mu'$ with either $s=1$ or $\delta(F) \arrow\tau \mu'$.
  Thus, we have two divisions of the same distribution~$\mu$ in two
  parts.
  We proceed to show that these divisions are actually the same.

  By Lemma~\ref{cc}g $F \brbisim G'$ for each $G' \in \spt(\mu_1)$.
  Furthermore, we claim that $F \brbisim G'$ for no
  $G' \in \spt(\mu_2)$.
  Namely, if $G' \in \spt(\mu_2)$ with $F \brbisim G'$, then
  $\mu_2 = \delta(G') \probc{p} \mu_2'$ for some~$p > 0$, and by
  Lemma~\ref{cc}f, using that
  $\den{P} \subseteq \DistrEconc \subseteq \Distr(\calErigid)$,
  $\den{P} = \nu_1 \probc{p} \nu_2$ for some
  $\nu_1 \brbisim \delta(G')$ and $\nu_2 \brbisim \mu_2'$.
  Since $\nu_1 \brbisim \delta(G') \brbisim \delta(F)$ and $F$ is
  rigid, by Lemma~\ref{cc}g one obtains
  $G \brbisim F \brbisim E$ for some, in fact all,
  $G \in \spt(\nu_1) \subseteq \spt(\den{P})$, contradicting
  our assumption.

  Since $\spt(\delta(F)) = F$ it follows that $s \leqslant r < 1$.
  Thus $\delta(F) \arrow{\tau} \mu'$. Since $F \in \calEconc$, by
  Lemma~\ref{cc}h we have that no $G' \in \spt(\mu')$
  satisfies~$F \brbisim G'$.  Hence $s = r$ and
  $\mu' = \mu_2 \brbisim \den{P}$.
  This proves that $\delta(F) \arrow\tau \mu'$ for some $\mu'$ with
  $\mu' \brbisim \den{P}$, as claimed.

  Proof of claim~(b).
  Assume $\delta(F) \arrow\alpha \varrho$.
  Since $E\brbisim F$, we have
  $\delta(E) \Arrow{\;} \etabar \alphahidearrow \eta$ for some
  $\etabar \brbisim \delta(E)$ and $\eta \brbisim \varrho$.
  By Lemma~\ref{cc}e $\etabar=\delta(E)$, using that $E$ is rigid.
  Since $F$~is concrete and $\delta(E) \alphahidearrow \eta$, we have
  $\delta(E) \alphaarrow \eta$ by Lemma~\ref{cc}i.
  Lemma~\ref{cc}i also says that $F\notin\sup(\eta)$.
  Hence no fraction of the transition $\delta(E) \alphaarrow \eta$ can
  stem from the transition $E \arrow\tau \delta(F) \oplusr \den{P}$.
  It follows that $\delta(H) \alphaarrow \eta$ with
  $\eta \brbisim \varrho$.

  Proof of claim~(c).
  Let
  $\calR \mathbin= {\{ \pair{\delta(H)}{\delta(F)} \}^\dagger
    \mathrel\cup {\brbisim}} \mkern1mu$.
  It suffices to show that $\calR$ is a branching probabilistic
  bisimulation relation.
  That $\calR$~is weakly decomposable is straightforward.
  So it suffices to check the transfer condition for the pairs
  $\pair{\delta(H)}{\delta(F)}$ and~$\pair{\delta(F)}{\delta(H)}$.
  Assume $\delta(H) \arrow\alpha \varrho$.
  Then also $\delta(E) \arrow\alpha \varrho$, and since $E \brbisim F$
  this move can be matched by~$\delta(F)$.
  The other case follows by~(b) above.
\end{proof}

\noindent
We have now arrived at the main lemma that we need for a proof of
completeness of the axioms of
Table~\ref{table-axiomatization-of-probabilistic-branching-bisimulation}
with respect to rooted probabilistic branching bisimulation semantics.  In
its proof we reason by induction on the complexity of processes as
captured by the complexity measure~$c$ below. Roughly speaking, the
complexity measure counts the number of transitions of a process.

\begin{definition}
  The complexity function $c : \calE \cup \calP \to \IN$ is
  defined as follows:
  \[\begin{array}{l@{\hspace{1cm}}l@{\hspace{1cm}}l}
    c(\bfzero)=0& c(\alpha\cdot P) = 1+c(P)&c(E_1+E_2)=c(E_1)+c(E_2)\\
    c(\partial(E))=c(E)& c(P_1\oplusr P_2)=c(P_1)+c(P_2). 
  \end{array}\]
\end{definition}
Note, the application of axiom A1, A2, A4, P1 and~P2 does not change the
complexity of a term.

\begin{lemma}
  \label{lemma-concrete-equivalent}
  \begin{itemize}
  \item [(a)] For each non-deterministic process $E \in \calE$ there is
    a concrete probabilistic process $\Pbar \in \calPconc$ such
    that $\den{\Pbar} \brbisim \delta(E)$ and
    $\AXpb \vdash \Pbar = \partial(E)$.

  \item [(b)] For each probabilistic process $P \in \calP$ there is a
    concrete probabilistic process $\Pbar \in \calPconc$ such that
    $\den{\Pbar} \brbisim \den{P}$ and $\AXpb \vdash \Pbar
    = P$.

  \item [(c)] For all probabilistic processes $P,Q \in \calP$, if
    $\den{P} \brbisim \den{Q}$ then $\AXpb \vdash { P = Q }$.
  \end{itemize}
\end{lemma}

\begin{proof}
  We prove items (a) to~(c) by simultaneous induction on $c(E)$,
  $c(P)$, and $\max \lbrace c(P), c(Q) \rbrace$.

  \blankline
  
  \emph{Basis} If $c(E) = 0$, then $E$ is provably equal to
  $\bfzero$. Hence, choosing $\Pbar = \partial(\bfzero) \in \calPconc$
  suffices.
  If $c(P) = 0$, then
  $\AXp\vdash P = \bigoplusiinI \: p_i \stardot \partial(\bfzero)$ for
  a suitable index set~$I$ and $p_i > 0$ for $i \in I$. Hence
  $\AXp \vdash { P = \partial(\bfzero) }$ and choosing
  $\Pbar = \partial(\bfzero) \in \calPconc$ suffices here.
  If $\max \lbrace c(P), c(Q) \rbrace = 0$, then
  $\AXp \vdash {P = \partial(\bfzero) ,\, Q = \partial(\bfzero)}$ by
  the above, from which $\AXpb \vdash {P = Q}$ follows directly.

  \blankline

  \emph{Induction step} As to case~(a), suppose $c(E) > 0$. Using
  axiom A1, A2, and A4 we can rewrite~$E$ such that
  $\AXp\vdash E = \sumiinI \: \alpha_i \pref P_i $ for some index
  set~$I$ and suitable~$\alpha_i \in \calA$ and $P_i \in \calP$. Note
  that, for all $i \in I$, $c(P_i) < c(E)$. Choose for each $i \in I$,
  by the induction hypothesis for part~(b), $\Pbar_i \in \calPconc$
  such that $\den{\Pbar_i} \brbisim \den{P_i}$ and
  $\AXpb \vdash {\Pbar_i = P_i}$.  Now $\AXpb \vdash E = \bar{E}$ for
  $\bar{E} := \sumiinI \: \alpha_i \pref \Pbar_i$. We distinguish two
  cases.

  \medskip

  (i)~First suppose that for some $i_0 \in I$ we have
  $\alpha_{i_0} = \tau$ and $\Pbar_{i_0} \brbisim \partial(\bar E)$, so
  that the summand $\alpha_{i_0} \pref \Pbar_{i_0}$ of $\bar E$
  represents an inert $\tau$-step. Then
  $\AXpb \vdash E = H + \tau \pref \Pbar_{i_0}$, where
  \plat{$H := \sum_{i \in I \setminus \{i_0\}} \: \alpha_i \pref
    \Pbar_i$}. It suffices to show that
  \plat{$H \sqsubseteq \Pbar_{i_0}$}, because then
  axiom~\hyperlink{BP}{BP} yields
  $\AXpb \vdash \partial(E) = \partial(H + \tau \pref
  \Pbar_{i_0}) = \Pbar_{i_0}$. 

  So, suppose $H \arrow{\alpha}\mu$.  Then $\mu = \den{\Pbar_i}$ for
  some $i \in I \setminus \{i_0\}$. Since
  $\Ebar \arrow{\alpha} \den{\Pbar_i}$ and
  $\partial(\Ebar) \brbisim \Pbar_{i_0}$,
  we have $\den{\Pbar_{i_0}} \Arrow{} {\nubar} \alphahidearrow \nu$
  where $\delta(\Ebar) \brbisim {\nubar}$ and
  $\den{\Pbar_i} \brbisim \nu$.  Because $\Pbar_{i_0}$ is concrete,
  $\den{\Pbar_{i_0}} = \nubar$ by Lemma~\ref{cc}e.  Thus
  $\den{\Pbar_{i_0}}\alphahidearrow \nu$.  Since
  $\den{\Pbar_i} \in \Distr(\calEconc)$, $\nu \in \Distr(\calEconc)$
  by Lemma~\ref{cc}d, and $\den{\Pbar_{i}} \brbisim \nu$,
  Lemma~\ref{lemma-equal-for-concrete-processes} yields
  $\den{\Pbar_{i}} \bisim \nu$, which was to be shown.

  \medskip

  (ii)~Next suppose that $\alpha_i\neq \tau$ or
  $\Pbar_i \nbrbisim \partial(\Ebar)$ for all
  $i \in I$, i.e., $\Ebar$~is \hyperlink{rigid}{rigid}.
  We will show that there is a concrete process $C \in \calEconc$
  such that $\AXpb \vdash \partial(\Ebar) = \partial(C)$.
  We proceed by induction on the number of indices $k \in I$ such that
  $\alpha_{k}\mathbin= \tau$ and $\den{\Pbar_{k}}$ can be written as
  $\mu_1 \oplusr \mu_2$ for some $r \mathbin\in (0,1]$ and
  $\delta(\Ebar) \brbisim \mu_1$.
  \emph{Basis} If there are no such~$k$, then
  $C := \Ebar \in \calEconc$ by Lemma~\ref{cc}a.

  \emph{Induction step} Let $i_0 \in I$ be an index such that
  $\alpha_{i_0} = \tau$ and $\den{\Pbar_{i_0}}$ can be written as
  $\mu_1 \oplusr \mu_2$ with $r \in (0,1]$ and
  $\delta(\Ebar) \brbisim\mu_1$.
  Without loss of generality we can assume that
  $\den{\Pbar_{i_0}} = \mu_1 \oplusr \mu_2$ with $r \in (0,1]$,
  $\delta(\Ebar) \brbisim\mu_1$, and $G \brbisim \Ebar$ for no
  $G \in \spt(\mu_2)$, for each such $G \in \spt(\mu_2)$ can be
  shifted to $\spt(\mu_1)$.
  As $\Ebar$ is \hyperlink{rigid}{rigid}, it follows that~$r<1$.

  Let again
  \plat{$H := \sum_{i\in I\setminus\{i_0\}} \: \alpha_i \pref
    \Pbar_i$}.
  Then $\AXp \vdash \bar E = \tau \pref \Pbar_{i_0} + H$.
  By induction, exists $C \in \calEconc$ with
  $\AXpb \vdash \partial(H) = \partial(C)$.
  So, $\AXpb\vdash \bar E = \tau \pref \Pbar_{i_0} + C$.
  It suffices to show that
  $\AXpb \vdash \partial(\bar E) = \partial(C)$.
  This will follow from axiom~\hyperlink{G}{G} once we obtain
  $\tau \pref \Pbar_{i_0} \mathbin\sqsubseteq \partial(C)$.

  By axioms P1--3, there are $P,P' \in \calPconc$, such that
  $\AXp\vdash \Pbar_{i_0} = P' \oplusr P$ with $\den{P'}=\mu_1$ and
  $\den{P}=\mu_2$.
  For all $F \in \mathit{spt}(\den{P'})$ we have $F \brbisim \bar E$
  by Lemma~\ref{cc}g.
  So, for each $F \in \spt( \den{P'} )$, we have
  $P' \brbisim \partial(F)$ and
  $\AXp \vdash P' = \partial(F)$ by Corollary~\ref{Corollary5.10}.
  Hence, $\AXp \vdash \Pbar_{i_0} = \partial(F) \oplusr P$.
  Moreover, $\widehat{E} := \tau \pref ( \partial(F) \oplusr P ) + C
  \brbisim \tau \pref \Pbar_{i_0} + C \brbisim \Ebar$.

  We analyse the $\tau$-transitions of~$\widehat E$ in order to show
  that it is rigid.
  Regarding the transition
  $\widehat{E} = \tau \pref (\partial(F) \oplusr P) + C \arrow\tau
  \den{\partial(F) \oplusr P}$ we observe that
  $\partial(F) \oplusr P \brbisim \Pbar_{i_0}$,
  $\Pbar_{i_0} \nbrbisim \partial(\Ebar)$ by assumption, and
  $\Ebar \brbisim$ $\widehat E$.
  So, $\den{\partial(F) \oplusr P} \nbrbisim \widehat{E}$. 
  Now assume, to arrive at a contradiction, that
  $C \arrow\tau \gamma$ and $\gamma \brbisim \delta(\widehat{E})$ for
  some $\gamma \in \DistrE$.
  By Lemma~\ref{cc}d $\gamma \in \Distr(\calEconc)$.
  We claim that, given the assumption, it holds that~$C \brbisim \widehat{E}$:
  Clearly, any transition $\delta(C) \arrow\alpha \varrho$ can be
  matched by
  $\delta(\widehat E) = \delta(\tau\pref (\partial(F) \oplusr P) +
  C)$.  Conversely, if $\delta(\widehat E) \arrow\alpha \varrho$, then
  $\gamma \Arrow{} \bar\nu \arrow{(\alpha)} \nu$ for some
  $\bar\nu,\nu \in \Distr(\calE)$ with
  $\bar\nu \brbisim \delta(\widehat E)$ and $\nu \brbisim\varrho$, and
  therefore $\delta(C) \Arrow{} \bar\nu \arrow{(\alpha)} \nu$.
  So, we have $C \arrow\tau \gamma$ and
  $\gamma \brbisim \delta(\widehat E) \brbisim \delta(C)$, which
  contradicts $C$~being concrete.
  We conclude that $\tau \pref (\partial(F) \oplusr P)$ nor~$C$ yields
  an inert transition and that
  $\widehat{E} = \tau \pref (\partial(F) \oplusr P) + C$ is rigid.

  Since $F$, $P$, and~$C$ are concrete, $r \in (0,1)$,
  $\widehat{E} = \tau \pref (\partial(F) \oplusr P) + C$ is rigid,
  $F \brbisim \Ebar \brbisim \widehat{E}$ and $G \brbisim \widehat{E}$ for
  no~$G \in \spt(\den{P})$, we obtain
  $\tau \pref (\partial(F) \oplusr P) \mathbin\sqsubseteq \partial(C)$
  by Lemma~\ref{inclusion-G}.
  Above we have shown that $\AXp \vdash \Pbar_{i_0} = \partial(F)
  \oplusr P$, hence
  $\partial(F) \oplusr P \bisim \Pbar_{i_0}$. Therefore
  Lemma~\ref{lem-inclusion-congruence} gives
  $\tau \pref \Pbar_{i_0} \mathbin\sqsubseteq \partial(C)$ as desired,
  closing the induction step for part~(a).

  \vspace{2pt}

  As to the induction step for part~(b), let
  $P = \bigoplusiinI \: p_i \stardot E_i$ where $p_i > 0$,
  $E_i \in \calE$ for~$i \in I$. Note, $c(E_i) \leqslant c(P)$, for
  $i \in I$. By part~(a), for all $i \in I$, concrete
  processes~$\Pbar_i \in \calPconc$ exist such that
  $\den{\Pbar_i} \brbisim \delta(E_i)$ and
  $\AXpb \vdash \Pbar_i = \partial(E_i)$. Put
  $\Pbar = \bigoplusiinI \: p_i \stardot \Pbar_i$. Then
  $\Pbar \in \calPconc$ by Lemma~\ref{cc}b. We have that
  $\den{\Pbar} = \den{\mkern1mu \bigoplusiinI \: p_i \stardot \Pbar_i
    \mkern1mu} \brbisim \den{\mkern1mu \bigoplusiinI \: p_i \stardot
    E_i \mkern1mu} = \den{P}$, since
  $\den{\Pbar_i} \brbisim \delta(E_i)$ for~$i \in I$. Moreover, it
  holds that
  $\AXpb \vdash \Pbar = \bigoplusiinI \: p_i \stardot \Pbar_i=
  \bigoplusiinI \: p_i \stardot E_i = P$, because
  $\AXpb \vdash \Pbar_i = \partial(E_i)$ for~$i \in I$. This concludes
  the induction step for part~(b).

  \vspace{2pt}

  For the induction step of part~(c), let $P, Q \in \calP$ be such
  that $\max \lbrace c(P), c(Q) \rbrace > 0$ and assume
  $\den{P} \brbisim \den{Q}$. By part~(b) exist concrete processes
  $\Pbar, \Qbar \in \calPconc$ such that
  $\den{\Pbar} \brbisim \den{P}$ and $\AXpb \vdash \Pbar = P$, on the
  one hand, and $\den{\Qbar} \brbisim \den{Q}$ and
  $\AXpb \vdash \Qbar = Q$, on the other hand. Thus
  $\den{\Pbar} \brbisim \den{P} \brbisim \den{Q} \brbisim
  \den{\Qbar}$. Since $\Pbar, \Qbar \in \calPconc$, it follows that
  $\AXp \vdash \Pbar = \Qbar$ using
  Corollary~\ref{Corollary5.10}. Combination of this gives
  $\AXpb \vdash P = \Pbar = \Qbar = Q$ and settles the induction step
  for part~(c) and the proof of the lemma.
\end{proof}

\noindent
By now we have gathered all ingredients to prove that the
theory~$\AXpb$ is a complete axiomatisation of
rooted probabilistic branching bisimulation semantics.
In the proof of this theorem we exploit axiom~C.

\begin{theorem}[$\!\AXpb$ is sound and complete for $\rootedbrbisim$]\!
  \label{theorem-prob-completeness}
  For all non-deterministic processes $E, F \in \calE$ and all
  probabilistic processes $P,Q \in \calP$ it holds that
  $E \rootedbrbisim F$ iff $\AXpb \vdash {E = F}$ and
  $\den{P} \brbisim \den{Q}$ iff $\AXpb \vdash {P = Q}$.
\end{theorem}

\begin{proof}
  Lemma~\ref{lemma-soundness} settles the soundness
  of~$\AXpb$. Therefore it remains to show that the
  theory~\plat{$\AXpb$} is complete. The completeness result for
  probabilistic processes was already delivered as
  Lemma~\ref{lemma-concrete-equivalent}c.

  So, choose non-deterministic processes $E, F \in \calE$ such that
  $E \rootedbrbisim F$. Suppose
  $E = \sumiinI \: \alpha_i \pref P_i$ and
  $F = \sumjinJ \: \beta_j \pref Q_j$ for suitable index sets~$I, J$,
  \vspace{2pt} actions~$\alpha_i, \beta_j$, and probabilistic
  processes~$P_i, Q_j$, for $i \in I$, $j \in J$.
  For each $j \in J$ we have $F \arrow{\beta_j}
  \den{Q_j}$. \vspace{2pt} Since
  $E \rootedbrbisim F$, we can find, for
  each~$j \in J$, a non-empty subset $I_j \subseteq I$ and
  $p_{ij} \geqslant 0$ such that
  $\alpha_i = \beta_j$ for~$i \in I_j$,
  $\delta(E) \arrow{\beta_j} \bigoplus_{i {\in} I_j} \: p_{ij}
  \mathord{\mkern1mu \stardot \mkern1mu} P_i$ and
  $\den{ \mkern1mu \bigoplus_{i {\in} I_j} \: p_{ij}
    \mathord{\mkern1mu \stardot \mkern1mu} P_i \mkern1mu} \brbisim
  \den{Q_j}$. \vspace{2pt} By Lemma~\ref{lemma-concrete-equivalent}c
  we have
  $\AXpb \vdash \bigl( \bigoplus_{i {\in} I_j} \: p_{ij}
  \mathord{\mkern1mu \stardot \mkern1mu} P_i \bigr) = Q_j$. Thus, for
  each $j \in J$, it holds that
  \begin{flalign*}
  &&    \AXpb & \vdash \, \sumiinI \: \alpha_i \pref P_i\\
  && & = 
      \sumiinI \: \alpha_i \pref P_i \, + \, \textstyle\sum_{i {\in} I_j} \:
      \alpha_i \pref P_i
      & &&\text{Axioms A1--A3} \\
  && & = 
      \sumiinI \: \alpha_i \pref P_i \, + \, \textstyle\sum_{i {\in} I_j} \:
      \beta_j \pref P_i
      & &&\text{$\alpha_i = \beta_j$ for $i \in I_j$}\\
  && & = 
      \sumiinI \: \alpha_i \pref P_i \, + \, \textstyle\sum_{i {\in} I_j} \:
      \beta_j \pref P_i \, + \, \beta_j \pref \bigl(
      \textstyle\bigoplus_{i {\in} I_j} \: p_{ij} \stardot P_i \bigr)
      & &&\text{Axiom~C}\\
  && & =  
      \sumiinI \: \alpha_i \pref P_i \, + \, \textstyle\sum_{i {\in} I_j} \:
      \beta_j \pref P_i \, + \, \beta_j \pref Q_j
      & &&\text{Since $\AXpb \vdash \textstyle\bigoplus_{i {\in} I_j}
      \: p_{ij} \mathord{\mkern1mu \stardot \mkern1mu} P_i  = Q_j$}\\
  && & = 
      \sumiinI \: \alpha_i \pref P_i \, + \, \beta_j \pref Q_j
      \mkern1mu .
      & &&\text{Reversing two steps above}
  \end{flalign*}
  Exploiting this equality for each~$j \in J$, as well as axioms A1
  and~A2, we obtain
  \begin{displaymath}
    E \, = \,
    \sumiinI \: \alpha_i \pref P_i \, = \, \sumiinI \: \alpha_i \pref
    P_i \, + \, \sumjinJ \: \beta_j \pref Q_j \, .
  \end{displaymath}
  Symmetrically, we derive
  \begin{displaymath}
    F \, = \,
    \sumjinJ \: \beta_j \pref Q_j \, = \, \sumjinJ \: \beta_j \pref
    Q_j \, + \, \sumiinI \: \alpha_i \pref P_i \, .
  \end{displaymath}
  Combination of the two equations yields, again with help of axioms A1
  and~A2,
  \begin{displaymath}
    \AXpb \vdash E 
    = \sumiinI \: \alpha_i \pref P_i
    \begin{array}[t]{@{}l}
      {} =
      \sumiinI \: \alpha_i \pref P_i \, + \, \sumjinJ \: \beta_j
      \pref Q_j 
      \smallskip \\ {}
      = \sumjinJ \: \beta_j \pref Q_j \, + \, \sumiinI \: \alpha_i \pref
      P_i
      = \sumjinJ \: \beta_j \pref Q_j
      = F 
    \end{array}
  \end{displaymath}
  which was to be shown.
\end{proof}

\blankline

\noindent
The axiomatisation~$\AXpb$, whose soundness and completeness
has been proven above, involves the auxiliary relation
${\sqsubseteq} \subseteq \calE \times \calP$. This
  relation does not satisfy the law
\begin{equation}
  \label{inclusion-congruence}
  \frac {F=E \qquad E \sqsubseteq P \qquad P=Q}{F \sqsubseteq Q}
\end{equation}
when equality denotes provability with respect to~$\AXpb$. However, it
satisfies~(\ref{inclusion-congruence}) when equality denotes
provability with respect to~$\AXp$. See
Lemma~\ref{lem-inclusion-congruence}. This could be considered better,
as it tells us that we never need to use the axioms
\hyperlink{BP}{BP} and~\hyperlink{G}{G} to prove the inclusions
$E \sqsubseteq P$ that are needed as premises of \hyperlink{BP}{BP}
and~\hyperlink{G}{G}.  However, if seen as a shortcoming of the
relation $\sqsubseteq$, this can be easily repaired. Define the
relation ${\sqsubseteq'} \subseteq \calE \times \calP$ by
\begin{displaymath}
  F \sqsubseteq' Q ~~~\mbox{iff}~~~
  \exists E \in \calE, P \in \calP.~
  F \rootedbrbisim E \land E \sqsubseteq P \land
  P \brbisim Q \;.
\end{displaymath}
As can be directly obtained, this relation
satisfies~(\ref{inclusion-congruence}), i.e.\ $F \rootedbrbisim E$,
$E \sqsubseteq' P$, and~$P \brbisim Q$ implies~$F \sqsubseteq' Q$. It
also satisfies axioms I1--I6, as can be straightforwardly verified.
In addition, the axioms \hyperlink{BP}{BP} and~\hyperlink{G}{G}
trivially remain sound when using $\sqsubseteq'$ in the precondition
instead of $\sqsubseteq$.  Thus, replacing $\sqsubseteq$ by
$\sqsubseteq'$ in
Table~\ref{table-axiomatization-of-probabilistic-branching-bisimulation}
yields the same axiomatisation, which is therefore sound and complete
for rooted branching probabilistic bisimulation semantics, and now
also satisfies~(\ref{inclusion-congruence}).

\blankline

\noindent
We now proceed to convert the axioms \hyperlink{BP}{BP} and
\hyperlink{G}{G} into equations.

\begin{lemma}
  \label{equational inclusion}
  Through application of~$\AXp$, each true statement of the form
  $E \sqsubseteq P$ can be written in the form
  \begin{displaymath}
    \textstyle\sum_{i\in I} \tau \pref \mg{\Bigl(}
    \bigopluskinK \: p_k\mycdot \dor{\bigr(}
    \partial(F_k) \probc{r_{ik}\!} P_{ik} \dor{\bigl)}
    \mg{\Bigl)} 
    \, + \,
    \textstyle\sum_{j \in J} \: a_j \pref
    \mg(\bigopluskinK \: p_k\mycdot P_{jk}\mg)
    \: \sqsubseteq \;
    \bigopluskinK \: p_k\mycdot F_k  
  \end{displaymath}
  with\vspace{-1ex} 
  \begin{displaymath}
    F_k = E_k + \textstyle{\sum_{i \in I_k}} \: \tau \pref P_{ik} \;
    + \textstyle{\sum_{j\in J}} \: a_j \pref P_{jk} \;,
  \end{displaymath}
  where $I,J,K$ are finite index sets, $I \cap J \mathbin= \emptyset$,
  $E_k\in\cal E$ and $p_k \mathbin\in (0,1]$ for all $k \mathbin\in K$
  such that $\sum_{k\in K} \: p_k \mathbin= 1$,
  $P_{ik} \mathbin\in \cal P$ and $r_{ik} \mathbin\in [0,1]$ for
  $i \mathbin\in I$ and $k \mathbin\in K$, $P_{jk} \mathbin\in \cal P$
  for $j \mathbin\in J$ and $k \mathbin\in K$,
  $a_j\in\calA{\setminus}\{\tau\}$ for $j \in J$, and
  $I_k \mathbin= \lc i \mathbin\in I \mid r_{ik}\mathbin<1 \rc$.
\end{lemma}

\begin{proof}
  Suppose that $E$ and~$P$ have the above form. We need to show
  that~$E \sqsubseteq P$.  So, suppose $E \arrow\alpha \mu$ for some
  $\alpha \in \calA$ and~$\mu \in \DistrE$.

  First assume that $\alpha \neq \tau$.  Then
  $\mu = \bigopluskinK \: p_k\mycdot \den{P_{jk}}$ for some
  $j \in J$ with $a_j = \alpha$. For each~$k \in K$ one has
  $F_k \arrow\alpha \den{P_{jk}}$.  Hence
  $\den{P} = \bigopluskinK \: p_k\mycdot \delta(F_k) \arrow\alpha
  \bigopluskinK \: p_k\mycdot\den{P_{jk}} = \mu$ by
  Lemma~\ref{lem-composition}a.
  Next assume that $\alpha = \tau$.  Then
  $\mu = \bigopluskinK \: p_k\mycdot \dor( \delta(F_k)
  \probc{r_{ik}} \den{P_{ik}} \dor)$ for some~$i \in I$.\vspace{1pt}
  For each~$k \in K$ one has
  $\delta(F_k) \tauhidearrow \delta(F_k) \probc{r_{ik}} \den{P_{ik}}$.
  Hence $\den{P} \tauhidearrow \mu$ by Lemma~\ref{lem-composition}b.

  \medskip

  Now suppose $E\sqsubseteq P$. We need to show that using~$\AXp$ the
  expressions $E$ and~$P$ can be written in the
  required form. Let $E$ have the form
  $\sum_{i\in I} \: \tau \pref P_i + \sum_{j\in J} \: a_j \pref P_j$
  for finite index sets~$I,J$ with $I \cap J \mathbin= \emptyset$,
  $a_j \in \calA{\setminus}\{\tau\}$ for~$j \in J$, and
  $P_i \in \calP$ for $i \in I\cup J$. Moreover, let $P$ have the form
  $\bigopluskinK \: p_k\mycdot F_k$ for a finite index set~$K$,
  $F_k \in \calE$ and $p_k \mathbin\in (0,1]$ for
  all~$k \mathbin\in K$, with $\sum_{k\in K} \: p_k \mathbin= 1$. \vspace{2pt}

  For each $j \in J$ one has $E\arrow{a_j} \den{P_{j}}$ \vspace{2pt}
  and thus $P \arrow{a_j} \nu_j$ for some $\nu_j \in \DistrE$ with
  $\nu_j \bisim \den{P_j}$.  Lemma~\ref{lem-decomposition}a implies
  that $\delta(F_k) \arrow{a_j} \nu_{jk}$ for certain
  $\nu_{jk}\in \DistrE$ such that
  $\nu_j = \bigopluskinK \: p_k \mydot \nu_{jk}$.
  Let $P_{jk} \in \calP$ be such that $\den{P_{jk}}=\nu_{jk}$ for
  all~$j \in J$ and $k\in K$.  Then $F_k \bisim F_k + a_j\pref P_{jk}$
  for all~$j\in J$ and~$k \in K$, and thus
  $\AXp \vdash F_k = F_k + a_j\pref P_{jk}$ for all~$j \in J$
  and~$k \in K$, by
  Theorem~\ref{theorem-completeness-for-strong-probabilistic-bisimulation}.
  Moreover,
  $P_j \bisim \bigopluskinK \: p_k \mydot P_{jk}$,
  and thus \vspace{2pt}
  $\AXp \vdash P_j = \bigopluskinK \: p_k \mydot P_{jk}$ by
  Theorem~\ref{theorem-completeness-for-strong-probabilistic-bisimulation}.

  For each~$i \in I$ one has $E \arrow{\tau} \den{P_{i}}$ and thus
  $\den{P} \tauhidearrow \nu_i$ for some~$\nu_i \in \DistrE$ with
  $\nu_i \bisim \den{P_i}$.\vspace{2pt} Lemma~\ref{lem-decomposition}b
  implies that $\delta(F_k) \tauhidearrow \nu_{ik}$ for certain
  $\nu_{ik} \in \DistrE$ such that
  $\nu_i = \bigopluskinK \: p_k \mydot \nu_{ik}$. By
  Definition~\ref{def-weak-arrows}, for each~$i \in I$ and~$k \in K$,
  either (i)~$\delta(F_k) \tauarrow \nu_{ik}$, or
  (ii)~$\nu_{ik}=\delta(F_k)$, or (iii)~there exists a
  $\nu_{ik}' \in \DistrE$ such that
  $\nu_{ik} = \delta(F_k) \probc{r_{ik}\!} \nu_{ik}'$ and
  $\delta(F_k) \arrow{\tau} \nu_{ik}'$ for some $r_{ik} \in (0,1)$.
  Let $P_{ik}\in\calP$ be such that $\den{P_{ik}}=\nu_{ik}$ for all
  $(i,k)\in I \times K$ for which possibility (i) applies, and
  $\den{P_{ik}}=\nu'_{ik}$ for all $(i,k)\in I \times K$ for which
  possibility (iii) applies. Pick $P_{ik}\in\calP$ arbitrary in case
  possibility (ii) applies. Let $r_{ik} := 0$ for all
  $(i,k)\in I \times K$ for which possibility (i) applies, and
  $r_{ik} := 1$ for all $(i,k)\in I \times K$ for which possibility
  (ii) applies. Then
  $P_i \bisim \bigopluskinK \: p_k \mydot \bigl(
  \partial(F_k) \probc{r_{ik}\!}  P_{ik} \bigr)$ for all
  $(i,k) \in I \times K$, and thus
  $\AXp \vdash P_i = \bigopluskinK(\partial(F_k) \probc{r_{ik}\!}
  P_{ik})$ for all $(i,k)\in I \times K$ by
  Theorem~\ref{theorem-completeness-for-strong-probabilistic-bisimulation}.
  Moreover, putting $I_k = \lc i \in I \mid r_{ik} < 1 \rc$,
  $F_k \bisim F_k + \tau \pref P_{ik}$ for all~$k \in K$ and
  $i \in I_k$, and thus $\AXp \vdash F_k = F_k + \tau \pref P_{ik}$
  for all~$k\in K$ and~$i \in I_k$, by
  Theorem~\ref{theorem-completeness-for-strong-probabilistic-bisimulation}.

  Thus, for all~$k \in K$ there are $E_k \in \calE$
  such that
  $\AXp \vdash F_k = E_k + \sum_{i\in I_k} \:
  \tau \pref P_{ik} + \sum_{j\in J}\:  a_j \pref P_{jk}$.  It follows
  that using $\AXp$ the expressions $E$ and $P$ can be written in the
  form given by the lemma.
\end{proof}

\noindent
It follows that an equational form of axiom \hyperlink{BP}{BP} can be
obtained by skipping the precondition $E \sqsubseteq P$, and instead
filling in the general forms of Lemma~\ref{equational inclusion} for
$E$ and $P$.  Thus, \hyperlink{BP}{BP} can be rendered as
\begin{displaymath}
  \def\arraystretch{1.2}
  \begin{array}{l}
    \partial \Bigl( \sum_{i\in I} \: \tau \pref 
    \mg(\bigopluskinK \: p_k\mycdot
    \dor( \partial(F_k) \probc{r_{ik}} P_{ik} \dor) \mg)
    +
    {} \\ \qquad \qquad \sumjinJ \: a_j \pref \mg(
    \bigopluskinK \: p_k\mycdot P_{jk}\mg) +
    \tau \pref \bigopluskinK \: p_k \mycdot F_k
    \Bigr) 
    \; = \;
    \bigopluskinK \: p_k \mycdot F_k
    \end{array}
    \def\arraystretch{1.2}
  \end{displaymath}
  with
  \begin{displaymath}
    F_k = E_k + \textstyle{\sum_{i {\in} I_k} \: \tau \pref P_{ik} \; 
      + \sumjinJ \: a_j \pref P_{jk} } \;,
  \end{displaymath}
  and where $I,J,K$ are finite index sets,
  $I \cap J \mathbin= \emptyset$, $E_k\in\cal E$ and
  $p_k \mathbin\in (0,1)$ for all $k \mathbin\in K$, 
  $\sum_{k\in K} p_k \mathbin= 1$, $P_{ik} \mathbin\in \cal P$ and
  $r_{ik} \mathbin\in [0,1]$ for $i \mathbin\in I$ and
  $k \mathbin\in K$, $P_{jk} \mathbin\in \cal P$ for $j \mathbin\in J$
  and $k \mathbin\in K$, $a_j\in\calA{\setminus}\{\tau\}$ for
  $j \in J$, and
  $I_k \mathbin= \lc i \mathbin\in I \mid r_{ik}\mathbin<1 \rc$.

In the same way, axiom \hyperlink{G}{G} can be
transformed to a purely equational form.
Since its precondition has the form $E \sqsubseteq \partial(F)$, we
may restrict attention to the special case of the form for
Lemma~\ref{equational inclusion} with $|K|=1$. This yields
\begin{displaymath}
  \partial \Bigl( \sumiinI \: \tau \pref \mg( \partial(F)
  \probc{r_{i}\!} P_{i} \mg) + \sumjinJ \: a_j \pref P_{j} + F
  \Bigr) = \partial(F)
\end{displaymath}
with\vspace{-1ex} 
\begin{displaymath}
  F = E + \textstyle\sum_{i {\in} I'} \: \tau \pref P_{i} \;
  + \sumjinJ \: a_j \pref P_{j} \;,
\end{displaymath}
where $I,J$ are finite index sets, $I \cap J \mathbin= \emptyset$,
$E\in\cal E$, $P_{i} \mathbin\in \cal P$ and $r_{i} \mathbin\in [0,1]$
for $i \mathbin\in I$, $P_{j} \mathbin\in \cal P$ and
$a_j\in\calA{\setminus}\{\tau\}$ for $j \in J$, and
$I' \mathbin= \lc i \mathbin\in I \mid r_{i}\mathbin<1 \rc$.  By
application of axiom~A3 the above axiom scheme can be simplified by
restricting to the case $J=\emptyset$. Another application of A3
allows restriction to the cases that $I'=I$ or $|I{\setminus}
I'|=1$. The latter case can already be obtained as an instance of
\hyperlink{BP}{BP}.  Hence \hyperlink{G}{G} can be rendered
equationally as
\begin{displaymath}
  \partial \Bigl( \, \sumiinI \: \tau \pref 
  \mg(P_i \probc{r_i} \partial
  \dor(E + \sumiinI \: \tau \pref P_i\dor)
  \mg)
  + E + \sumiinI \: \tau \pref P_i \Bigr)
  \; = \;
  \partial \dor( E + \sumiinI \: \tau \pref P_i \dor)
\end{displaymath}
with $I$ a finite index set, $E \in \cal E$, $P_i \in \cal P$
and~$r_i \in [0,1]$ for all~$i\in I$. 



\section{Concluding remarks}

\label{sec-concl}
\noindent%
We presented an axiomatisation of rooted branching
probabilistic {\bisimilarity} and proved its soundness and
completeness. In doing so, we aimed to stay close to a straightforward
completeness proof for the axiomatisation of rooted branching
{\bisimilarity} for non-deterministic processes that employed concrete
processes, also presented in this paper. In particular, the route via
concrete processes guided us to find the right formulation of the
axioms \hyperlink{BP}{BP} and \hyperlink{G}{G} for
branching {\bisimilarity} in the probabilistic case.

Future work will include the study of the extension
of the setting of the present paper with a parallel
operator~\cite{DPP05:klop}. In particular a congruence result for the
parallel operator should be obtained, which for the mixed
non-deterministic and probabilistic setting can be challenging.
Also the inclusion of recursion\,\cite{DP07:tcs,FG19:jlamp}
is a clear direction for further research. 

Also, we want to develop a minimisation algorithm for probabilistic
processes modulo probabilistic branching {\bisimilarity}.  Eisentraut
et al.\ propose in~\cite{EHKTZ13:qest} an algorithm for deciding
equivalence with respect to weak distribution {\bisimilarity} relying
on a state-based characterisation, a result presently not available in
our setting.  Other work and proposals for weak {\bisimilarity}
include~\cite{CS02:concur,FHHT16:fac,TH15:ic}, but these do not fit
well with the installed base of our toolset~\cite{Bun19:tacas}. For
the case of strong probabilistic {\bisimilarity} without combined
transitions we recently developed in~\cite{GRV18:algorithms} an
algorithm improving upon the early results of~\cite{BEM00:jcss}. In
\cite{TH15:ic} a polynomial algorithm for Segala's probabilistic
branching {\bisimilarity}, which differs from our notion of
probabilistic branching {\bisimilarity}, is defined. We hope to arrive
at an efficient algorithm by combining ideas
from~\cite{Val10:fi,VF10:tacas,TH15:ic} and
of~\cite{GV90:icalp,GJKW17:tcl}.

\bibliographystyle{plain}
\bibliography{infcomp}
\newpage

\section*{Appendix}


\noindent
\textbf{Lemma \ref{flexibelDelen}} Let $I$ and~$J$ be finite index
sets, $p_i,q_j\in[0,1]$ and $\xi, \mu_i,\nu_j \in \DistrE$, for
$i \in I$ and $j \in J$, with both
$\xi = \bigoplusiinI\: p_i\stardot\mu_i$ and
$\xi =\bigoplusjinJ \: q_j \stardot \nu_j$. Then there are
$r_{ij} \in [0,1]$ and $\varrho_{ij} \in \DistrE$ such that
$\sum_{j\in J} \: r_{ij} = p_i$ and
$p_i \stardot \mu_i = \bigoplusjinJ \: r_{ij} \stardot \varrho_{ij}$
for all~$i \in I$, and $\sum_{i\in I} \: r_{ij}=q_j$ and
$q_j \stardot \nu_j = \bigoplusiinI \: r_{ij} \stardot \varrho_{ij}$
for all~$j \in J$.

\begin{proof}
  Put
  $\textstyle r_{ij} = \sum_{E\in \mathit{spt}(\xi)} \:
  \displaystyle{\frac{\rule[-3pt]{0pt}{10pt} p_i \mkern1mu \mu_i(E) \cdot
      q_j \mkern1mu \nu_j(E)}{\rule{0pt}{7pt} \xi(E)}}$ for all
  $i \in I$ and $j \in J$.  In case $r_{ij} = 0$, choose
  $\varrho_{ij} \in \DistrE$ arbitrarily.  In case $r_{ij} \neq 0$,
  define $\varrho_{ij} \in \DistrE$, for $i \in I$ and $j \in J$, by
  \begin{displaymath}
    \varrho_{ij}(E) =
    \left\lbrace
      \begin{array}{cl}
        \displaystyle\frac
        {\rule[-3pt]{0pt}{10pt} p_i \mkern1mu \mu_i(E) \cdot
        q_j \mkern1mu \nu_j(E)}
        {\rule{0pt}{7pt} r_{ij} \, \xi(E)} \quad &
        \textrm{if } \xi(E)>0, \smallskip \\
        0 & \textrm{otherwise}
      \end{array}
    \right.
  \end{displaymath}
  for all $E \in \cal{E}$. By definition of $r_{ij}$
  and~$\varrho_{ij}$ it holds that $\sum \lc \varrho_{ij}(E) \mid E
  \in \calE \rc = 1$. So, $\varrho_{ij} \in \DistrE$ indeed.

  We verify $\sumjinJ \: r_{ij} = p_i$ and
  $p_i \stardot \mu_i = \bigoplusjinJ \: r_{ij} \stardot \varrho_{ij}$ for~$i \in
  I$.
  \begin{align*}
    \sumjinJ \: r_{ij}
    & = \sumjinJ \: \sum_{E \in \mathit{spt}(\xi)} \: p_i \mkern1mu
      \mu_i(E) \cdot q_j \mkern1mu \nu_j(E) / \xi(E) \\
    & = \textstyle \sum_{E \in \mathit{spt}(\xi)} \: p_i \mkern1mu
      \mu_i(E) \cdot \sumjinJ \: q_j \mkern1mu \nu_j(E) / \xi(E) \\
    & = \textstyle \sum_{E \in \mathit{spt}(\xi)} \: p_i \mkern1mu \mu_i(E)
    & \text{Since $\xi = \bigoplusjinJ \: q_j \stardot \nu_j$.} \\
    & = p_i \textstyle \sum_{E \in \mathit{spt}(\xi)} \: \mu_i(E) \\
    & = p_i \mkern1mu .
  \end{align*}
  Next, pick $F \in \calE$ and~$i \in I$. If $\xi(F) = 0$, then
  $p_i \mkern1mu \mu_i(F) = 0$, since
  $\xi(F) = \sumiinI \: p_i \mkern1mu \mu_i(F)$, and $r_{ij} = 0$ or
  $\varrho_{ij}(F) = 0$ for all~$j \in J$, by the various definitions,
  thus $\sumjinJ \: r_{ij} \mkern1mu \varrho_{ij}(F) = 0$ as well.

  Suppose $\xi(F) > 0$. Put $J_i = \lc j \in J \mid r_{ij} > 0
  \rc$. Note, if $j \in J \backslash J_i$, i.e.\ if $r_{ij} = 0$, then
  $p_i \mkern1mu \mu_i(F) q_j \mkern1mu \nu_j(F) / \xi(F) = 0$ by
  definition of~$r_{ij}$. Therefore we have
  \begin{align*}
    \sumjinJ \: r_{ij} \varrho_{ij}(F) \,
    & = \, \textstyle \sum_{j {\in} J_i} \: r_{ij} \mkern1mu \varrho_{ij}(F) \\
    & = \, \textstyle \sum_{j {\in} J_i} \: r_{ij} \mkern1mu p_i \mkern1mu 
      \mu_i(F) \cdot q_j \mkern1mu \nu_j(F) / (r_{ij} \mkern1mu \xi(F) ) \\
    & = \, \textstyle \sum_{j {\in} J_i} \: p_i \mkern1mu \mu_i(F) \cdot q_j
      \mkern1mu \nu_j(F) / \xi(F) \\
    & = \, \sumjinJ \: p_i \mkern1mu \mu_i(F) \cdot q_j \mkern1mu
      \nu_j(F) / \xi(F)
    & \text{Summand zero for $j \in J \backslash J_i$.} \\
    & = \, p_i \mkern1mu \mu_i(F) / \xi(F) \cdot
      \sumjinJ \: q_j \mkern1mu \nu_j(F)
    & \\
      & = \, p_i \mkern1mu \mu_i(F) \mkern1mu
    & \text{Since $\xi = \bigoplusjinJ \: q_j \stardot \nu_j$}.
  \end{align*}
  The statements $\sum_{i\in I} \: r_{ij}= q_j$ and
  $q_j \stardot \nu_j = \bigoplusiinI \: r_{ij} \stardot \varrho_{ij}$
  for~$j \in J$, follow by symmetry.
\end{proof}


\pagebreak[3]

\noindent
\textbf{Lemma \ref{lm-point-decomposability}}
  If $\calR$ is a branching probabilistic bisimulation relation, then
  so is $\ccR$.

  \begin{proof}
    We first verify that $\ccR$ is weakly decomposable. Suppose
    $\mu \ccR \nu$ for $\mu, \nu \in \DistrE$. Say
    $\mu = \bigoplusiinI \: p_i \cdot \mu_i$ and
    $\nu = \bigoplusiinI \: p_i \cdot \nu_i$ for a suitable index
    set~$I$ and $p_i > 0$, $\mu_i, \nu_i \in \DistrE$ such that
    $\mu_i \mopcalR \nu_i$ for~$i \in I$.
    For~$i \in I$, we have
    $\mu_i = \bigoplus_{E {\in} \spt(\mu)} \: \mu_i(E) \stardot E$.
    Since $\calR$ is 
    weakly decomposable and $\mu_i \mopcalR \nu_i$, there exist
    distributions $\nubar_i$ and~$\nu_{i,E}$ for~$i \in I$,
    $E \in \spt(\mu)$, such that $\nu_i \Arrow{} \nubar_i$,
    $\nubar_i = \bigoplus_{E {\in} \spt(\mu)} \: \mu_i(E) \cdot \nu_{i,E}$, and
    $\delta(E) \mopcalR \nu_{i,E}$. 
    Let $\nubar := \bigoplusiinI \: p_i \cdot \nubar_i$
    and $\nu_E := \bigoplusiinI \: p_i \cdot \nu_{i,E}$.
    Then  $\nu = \bigoplusiinI \: p_i \cdot
    \nu_i \Arrow{} \bigoplusiinI \: p_i \cdot \nubar_i = \nubar$
    because $\nu_i \Arrow{} \nubar_i$ for~$i \in I$.
    Moreover, $\delta(E) \ccR \nu_{E}$, because  $\delta(E) \mopcalR
    \nu_{i,E}$ for~$i \in I$. 

    Let $\bigoplusjinJ \: q_j \cdot \mu'_j$ be an arbitrary decomposition
    of~$\mu$.  It suffices to find distributions~$\nubar'$ and $\nu'_j$
    for~$j \in J$ such that (i)~$\nu \Arrow{} \nubar'$,
    (ii)~$\nubar' = \bigoplusjinJ \: q_j \cdot \nu'_j$, and
    (iii)~$\mu'_j \ccR \nu'_j$ for~$j \in J$.
    In fact, we choose $\nubar' := \nubar$, thus independently of the
    given decomposition of~$\mu$.  Without limitation of
    generality we may assume that $q_i>0$ for all~$j \in J$.
 
    Because, for each~$E\in\spt(\mu)$,
    $\sumiinI \: p_i \cdot \mu_i(E) = \mu(E) = \sumjinJ \: q_j
    \cdot \mu'_j(E)$ there are $\rijE \geqslant 0$ for $i \in I$ and
    $j \in J$, such that $\sumjinJ \: \rijE = p_i \cdot \mu_i(E)$
    and $\sumiinI \: \rijE = q_j \cdot \mu'_j(E)$. Define
    $\nu'_j = \bigoplus_{E {\in} \spt(\mu)} \bigoplusiinI \: \rijE / q_j \cdot
    \nu_{i,E}$ for~$j \in J$. Note that $\nu'_j$ is well-defined since
    \begin{displaymath}
      \textstyle\sum_{E {\in} \spt(\mu)}
      \sumiinI \: \rijE/q_j =
      \textstyle\sum_{E {\in} \spt(\mu)} \: q_j \mu'_j(E) / q_j =
      \textstyle \sum_{E {\in} \spt(\mu)} \: \mu'_j(E) = 1
      \mkern1mu .
    \end{displaymath}
    We verify the remaining requirements (ii) and~(iii) listed above.
    Regarding requirement~(ii) we calculate
    \def\arraystretch{1.2}
    \begin{flalign*}
      \qquad \bigoplusjinJ \: q_j \cdot \nu'_j
      & \myeq
      \bigoplusjinJ \: q_j \cdot \bigl( \mkern1mu
      \bigoplus_{E {\in} \spt(\mu)} \bigoplusiinI \: \rijE / q_j \cdot \nu_{i,E}
      \mkern1mu \bigr)
      & \\
      & \myeq
      \textstyle\bigoplus_{E {\in} \spt(\mu)} \bigoplusiinI \bigl( \mkern1mu
      \sumjinJ \: \rijE \mkern1mu ) \cdot \nu_{i,E}
      \\
      & \myeq
      \textstyle\bigoplus_{E {\in} \spt(\mu)} \bigoplusiinI
      p_i \mycdot \mu_i(E) \cdot \nu_{i,E}
      \\
      & \myeq
      \textstyle\bigoplusiinI \: p_i \cdot \bigl( \mkern1mu
      \bigoplus_{E {\in} \spt(\mu)} \: \mu_i(E) \cdot \nu_{i,E}
      \mkern1mu \bigr)
      \\
      & \myeq
      \bigoplusiinI \: p_i \cdot \nubar_i
      \\
      & \myeq
      \nubar
    \end{flalign*}
    \def\arraystretch{1.0}
    Regarding requirement~(iii) we have
    \def\arraystretch{1.2}
    \begin{flalign*}
        && \mu'_j
        & \makebox[1.2cm][c]{=}
        \textstyle\bigoplus_{E {\in} \spt(\mu)} \: \mu'_j(E) \stardot E \\
        &&
        & \makebox[1.2cm][c]{=}
        \textstyle\bigoplus_{E {\in} \spt(\mu)} \: \bigl( \mkern1mu
          \sumiinI \: \rijE/q_j ) \stardot E
        & && \text{Choice of $\rijE$ for $i \in I$, $j \in J$, $E \in
          \spt(\mu)$}
        \\ &&
        & \makebox[1.2cm][c]{=}
        \textstyle\bigoplus_{E {\in} \spt(\mu)} \bigoplusiinI \:
        \rijE / q_j \cdot \delta(E) 
        \\ &&
        & \makebox[1.2cm][c]{\ccR}
        \textstyle\bigoplus_{E {\in} \spt(\mu)} \bigoplusiinI \:
        \rijE / q_j \cdot \nu_{i,E} 
        & && \text{$\delta(E) \mopcalR \nu_{i,E}$ for $i \in I$}
        \\ &&
        & \makebox[1.2cm][c]{=} \nu'_j
        && & \text{Definition of $\nu'_j$}
    \end{flalign*}
    \def\arraystretch{1.0}
    Thus $\mu = \bigoplusjinJ \: q_j \cdot \mu'_j \ccR \bigoplusjinJ \:
    q_j \cdot \nu'_j = \nubar$ as was to be checked.

    Next we check that $\ccR$ satisfies the transfer condition of
    Definition~\ref{def-probabilistic-branching-bisimilar}d.  Suppose
    $\mu \ccR \nu$ for $\mu, \nu \in \DistrE$. Say
    $\mu = \bigoplusiinI \: p_i \cdot \mu_i$ and
    $\nu = \bigoplusiinI \: p_i \cdot \nu_i$ for a suitable index
    set~$I$ and $p_i > 0$, $\mu_i, \nu_i \in \DistrE$ such that
    $\mu_i \mopcalR \nu_i$ for~$i \in I$.  Furthermore, suppose
    $\mu \arrow{\alpha} \mu'$, that is,
    $\mu = \bigoplusjinJ \: q_j \cdot E_j$,
    $\mu' = \bigoplusjinJ \: q_j \cdot \mu'_i$ and
    $E_j \alphaarrow \mu'_j$ for all $j \mathbin\in I$.  By
    Lemma~\ref{flexibelDelen} there are $r_{ij} \in [0,1]$ and
    $\varrho_{ij} \in \DistrE$ such that
    $\sum_{j\in J} \: r_{ij} = p_i$ and
    $p_i \stardot \mu_i = \bigoplusjinJ \: r_{ij} \stardot
    \varrho_{ij}$ for all~$i \in I$, and $\sum_{i\in I} \: r_{ij}=q_j$
    and
    $q_j \stardot E_j = \bigoplusiinI \: r_{ij} \stardot \varrho_{ij}$
    for all~$j \in J$; naturally $\varrho_{ij} = \delta(E_j)$ for all
    $i\in I$ and $j\in J$.  Write $s_{ij}$ for ${r_{ij}}/{p_i}$, so
    that $\sum_{j\in J} \: s_{ij}=1$ and
    $\mu_i = \bigoplusjinJ \: s_{ij} \stardot E_j$ for all $i\in I$.
    For all $i \in I$ let
    $\hat\mu'_i := \bigoplusjinJ \: s_{ij} \stardot \mu'_j$, so that
    $\bigoplusiinI \: p_i\stardot\hat\mu'_i = \bigoplusjinJ \:
    (\sumiinI \: r_{ij}) \stardot \mu'_j = \bigoplusjinJ \: q_j \cdot
    \mu'_i = \mu'$.

  Since $\calR$ is weakly decomposable, $\mu_i \mopcalR \nu_i$ and
  $\mu_i = \bigoplus_{j\in J} s_{ij} \stardot E_j$, there exist
  distributions $\nubar_i$ and~$\nu_{ij}$ for~$i \in I$, $j \in J$,
  such that $\nu_i \Arrow{} \nubar_i$,
  $\nubar_i = \bigoplus_{j\in J} s_{ij} \cdot \nu_{ij}$, and
  $\delta(E_j) \mopcalR \nu_{ij}$.  Moreover, since
  $E_j \alphaarrow \mu'_j$, for each $i \in J$ and $j \in J$ there are
  $\nubar_{ij}', \nu'_{ij} \in \DistrE$ such that
  \begin{displaymath}
    \nu_{ij} \Arrow{} \nubar'_{ij}, \ 
    \nubar'_{ij} \alphahidearrow \nu'_{ij}, \ 
    \delta(E_j) \mopcalR \nubar'_{ij}, \ \text{and} \ 
    \mu'_j \mopcalR \nu'_{ij} \mkern1mu .
  \end{displaymath}
  Let $\nubar'_i := \bigoplusjinJ\: s_{ij} \cdot \nubar'_{ij}$ and
  $\nu'_i := \bigoplusjinJ\: s_{ij} \cdot \nu'_{ij}$.\vspace{1pt} Then
  $\nubar_{i} \Arrow{} \nubar'_i$ and
  $\nubar'_{i} \alphahidearrow \nu'_i$ because
  $\nu_{ij} \Arrow{} \nubar'_{ij}$ and
  $\nubar'_{ij} \alphahidearrow \nu'_{ij}$ for~$j \in J$.  Moreover,
  $\mu_i \ccR \nubar'_i$ and 
  $\hat\mu'_i \ccR \nu'_i$, because
  $\delta(E_j) \mopcalR \nubar'_{ij}$ and $\mu'_j \mopcalR \nu'_{ij}$
  for~$j \in J$.  Let
  $\nubar' := \bigoplusiinI \: p_i \cdot \nubar'_i$ and
  $\nu' := \bigoplusiinI \: p_i \cdot \nu'_{i}$.  Then
  $\nu = \bigoplusiinI \: p_i \cdot \nu_i \Arrow{} \bigoplusiinI \:
  p_i \cdot \nubar'_i = \nubar'$ because
  $\nu_i \Arrow{} \nubar_i \Arrow{} \nubar'_i$ for~$i \in I$.  Also,
  $\mu \ccR \nubar'$ and $\mu' \ccR \nu'$, as $\mu_i \ccR \nubar'_{i}$
  and $\hat\mu'_i \ccR \nu'_i$
  for~$i \in I$, which finishes the proof.
\end{proof}
  


\noindent
\textbf{Lemma \ref{state-based}} 
  Let $\calR \subseteq \DistrE \times \DistrE$ be a decomposable
  relation such that
  \begin{equation*}
    \tag{\ref{cong}}
    \mu_1 \mopcalR \nu_1 \ \text{and} \ \mu_2 \mopcalR \nu_2
    \quad  \text{implies} \quad
    (\mu_1 \oplusr \mu_2) \mopcalR (\nu_1 \oplusr \nu_2)
  \end{equation*}
  and for each pair $E,F \in \calE$ 
  \begin{equation}
    \tag{\ref{diamond}}
    \delta(E) \mopcalR \delta(F)
    \ \text{and} \
    E \arrow{\alpha} \mu'
    \quad \text{implies} \quad
    \delta(F) \alphaarrow \nu'
    \ \text{and} \ 
    \mu' \mopcalR \nu'
  \end{equation}
  for a suitable $\nu' \in \DistrE$.
  Then $\mu \mopcalR \nu$ implies $\mu\bisim \nu$.

\begin{proof}
  We show that $\calR$ is a strong probabilistic bisimulation
  relation.
  So, let $\mu, \nu \in \DistrE$ be such that $\mu \mopcalR \nu$ and
  $\mu \arrow{\alpha} \mu'$.  By
  Definition~\ref{def-pr-transition-relation}\ref{transitrel-case_b}
  we have $\mu = \bigoplusiinI \: p_i \stardot E_i$,
  $\mu' = \bigoplusiinI \: p_i \stardot \mu'_i $, and
  $E_i \alphaarrow \mu'_i$ for all $i \in I$.  Since $\calR$ is
  decomposable, there are $\nu_i \in \DistrE$, for~$i \in I$, such
  that
    \begin{displaymath}
      \nu  = \bigoplusiinI \: p_i \stardot \nu_i
      \quad \text{and} \quad
      \delta(E_i) \mopcalR \nu_i 
      \ \text{for all~$i \in I$.}
    \end{displaymath}
    Let, for each $i\in I$,
    $\nu_i = \bigoplus_{j\in J_i} \: p_{ij} \stardot F_{ij}$.  Since $\calR$
    is decomposable, there are $\mu_{ij} \in \DistrE$,
    for~$j \in J_i$, such that
    \begin{displaymath}
      \delta(E_i)  = \textstyle\bigoplus_{i \in J_i} \: p_{ij} \stardot \mu_{ij}
      \quad \text{and} \quad 
      \mu_{ij} \mopcalR \delta(F_{ij})
      \ \text{for all~$j \in J_i$.}
    \end{displaymath}
    Here $\mu_{ij} = \delta(E_i)$.  Writing $E_{ij} \, {:=} \, E_i$,
    $q_{ij} \, {:=} \, p_i\cdot p_{ij}$ and
    $K = \lc (i,j) \mid i \in I ,\, j \in J_i \rc$ we
    obtain (with $q_k = q_{ij}$, $E_k = E_{ij}$, and
      $F_k = F_{ij}$ for~$k \in K$ such that $k = (i,j)$) that
    \begin{displaymath}
      \mu = \bigopluskinK \: q_k \stardot E_k \ ,\
      \nu = \bigopluskinK \: q_{k} \stardot F_{k}
      \quad \text{and} \quad
      \delta(E_k) \mopcalR \delta(F_k)
      \ \text{for all~$k \in K$.}
    \end{displaymath}
    Let $\mu'_{ij} {:=} \mu'_i$ for all $i \in I$ and $j \in
    J_i$. Then $\mu' = \bigopluskinK \: q_{k} \stardot \mu'_k$. Using
    that $E_k \arrow{\alpha} \mu'_k$ for all $k \in K$,
    by~(\ref{diamond}) there must be distributions~$\nu'_k$ for
    $k \in K$ such that
    \begin{displaymath}
      \delta(F_k) \alphaarrow \nu'_k
      \quad \text{and} \quad
      \mu'_k \mopcalR \nu'_k.
    \end{displaymath}
    By
    Definition~\ref{def-pr-transition-relation}\ref{transitrel-case_b}
    this implies $\nu \alphaarrow \nu'$, for
    $\nu' \, {:=} \, \bigopluskinK \: q_{k} \stardot \nu'_k$.
    Moreover, property~(\ref{cong})~yields $\mu' \mopcalR \nu'$.
\end{proof}



\noindent
\textbf{Lemma \ref{unrooted-congruence}}
  Let $\mu_1,\mu_2,\nu_1,\nu_2\in \DistrE$ and $r\in(0,1)$. If $\mu_1
  \brbisim \nu_1$ and $\mu_2 \brbisim \nu_2$ then $\mu_1 \probc{r}
  \mu_2 \brbisim \nu_1 \probc{r} \nu_2$.

\begin{proof}
  Suppose $\mu_1 \brbisim \nu_1$ and $\mu_2 \brbisim \nu_2$ through
  branching probabilistic bisimulation relations $\calR_1$
  and~$\calR_2$. We show that the relation $\calR = \lc \langle \xi'
  \probc{s} \xi'' , \eta' \probc{s} \eta'' \rangle \mid \xi'
  \mathop{\calR_1} \eta' ,\, \xi'' \mathop{\calR_2} \eta'' ,\, {s \in
    (0,1)} \rc$ is a branching probabilistic bisimulation relation
  relating $\mu_1 \probc{r} \mu_2$ with $\nu_1 \probc{r} \nu_2$.
  
  Symmetry is straightforward. We first show that $\calR$ is weakly
  decomposable. So, assume $\xi \mopcalR \eta$ and
  $\xi = \bigoplusiinI \: p_i \cdot \xi_i$ with all $p_i>0$. Thus,
  $\xi = \xi' \probc{s} \xi''$ and $\eta = \eta' \probc{s} \eta''$
  with $\xi' \mathop{\calR_1} \eta'$ and
  $\xi'' \mathop{\calR_2} \eta''$ for suitable
  $\xi', \xi'', \eta', \eta'' \in \DistrE$. Since
  $\xi' \probc{s} \xi'' = \bigoplusiinI \: p_i \cdot \xi_i$, we have
  by Lemma~\ref{flexibelDelen} that there must be
  $s'_i, s''_i \geqslant 0$ and $\xi'_i, \xi''_i \in \DistrE$, for
  $i \in I$, such that
  \begin{displaymath}
    \begin{array}{rcl}
      s \cdot \xi' \phantom{{}'} & = & \bigoplusiinI \: s'_i \cdot \xi'_i  
      \smallskip \\ 
      (1 {-} s) \cdot \xi''  & = & \bigoplusiinI \: s''_i \cdot \xi''_i
      \smallskip \\ 
      p_i \cdot \xi_i & = & s'_i \cdot \xi'_i\, \oplus \, s''_i \cdot \xi''_i
      \quad \text{for all $i\in I$}
    \end{array}
  \end{displaymath}
  and
  $\sum_{i\in I}s'_i = s$, $\sum_{i\in I}s''_i = 1{-}s$,
  and $s'_i + s''_i = p_i$ for all $i \in I$.
  Thus $\xi' \phantom{{}'} = \bigoplusiinI \: s'_i/s \cdot \xi'_i$ and
  $\xi'' = \bigoplusiinI \: s''_i/(1 {-} s) \cdot \xi''_i$.
  Since the relations $\calR_1$ and~$\calR_2$ are weakly decomposable,
  there are $\etabar', \etabar'', \eta'_i, \eta''_i \in \DistrE$
  for~$i \in I$ such that
  \begin{displaymath}
    \begin{array}{lclclcl}
      \eta' \Arrow{} \etabar' & \quad &
      \phantom{{}'} \xi' \mathop{\calR_1} \etabar' & \quad &
      \etabar' \, = \bigoplusiinI \:
      {s'_i}/{s} \cdot \eta'_i & \quad &
      \phantom{{}'} \xi'_i \mathop{\calR_1} \eta'_i
      \smallskip \\ 
      \eta'' \Arrow{} \etabar'' &&
      \xi'' \mathop{\calR_2} \etabar'' &&
      \etabar'' = \bigoplusiinI \:
      {s''_i}/{(1{-}s)} \cdot \eta''_i &&
      \xi''_i \mathop{\calR_2} \eta''_i 
  \end{array}
  \end{displaymath}
  for all $i \in I$. Therefore, we can conclude that 
  \begin{displaymath}
    \eta' \probc{s} \eta'' \Arrow{} \etabar' \probc{s} \etabar''
    \quad \text{and} \quad
    ( \xi' \probc{s} \xi'' ) \mopcalR \mkern1mu ( \etabar'
    \probc{s} \etabar'' )
    \mkern1mu .
  \end{displaymath}
  Moreover, 
  \begin{displaymath}
    \def\arraystretch{1.3}
    \begin{array}{rcl}
      \etabar' \probc{s} \etabar''
      & = &
      \bigl( \bigoplusiinI \: {s'_i}/{s} \cdot \eta'_i \bigr)
      \probc{s} 
      \bigl( \bigoplusiinI \: {s''_i}/{(1{-}s)} \cdot \eta''_i \bigr) 
      \\
      & = &
      \bigl( \bigoplusiinI \: s'_i \cdot \eta'_i \bigr)
      \oplus 
      \bigl( \bigoplusiinI \: s''_i \cdot \eta''_i \bigl)
      \\ & = & \phantom{\bigl(}
      \bigoplusiinI \: p_i \cdot ( \eta'_i \probc{s'_i/p_i} \eta''_i )
      \mkern1mu .
    \end{array}
    \def\arraystretch{1.0}
  \end{displaymath}
  Thus, for $\etabar = \etabar' \probc{s} \etabar''$ and
  $\eta_i = \eta'_i \probc{s'_i/p_i} \eta''_i$ for $i \in I$, we have
  that
  $\eta = \eta' \probc{s} \eta'' \Arrow{} \etabar' \probc{s} \etabar''
  = \etabar$,
  $\xi = ( \xi' \probc{s} \xi'' ) \, \calR \,( \etabar' \probc{s}
  \etabar'' ) = \etabar$, $\etabar = \bigoplusiinI \: \eta_i$, and
  $\xi_i = ( \xi'_i \probc{s'_i/p_i} \xi''_i ) \mopcalR \mkern1mu (
  \eta'_i \probc{s'_i/p_i} \eta''_i ) = \eta_i$ for all $i \in I$.
  This finishes the argument that $\calR$ is decomposable.
  
  Next, we show that $\calR$ satisfies the transfer property for
  branching bisimulations. Suppose $\xi \mathop{\calR} \eta$. We have
  $\xi = \xi_1 \probc{r} \xi_2$, $\eta = \eta_1 \probc{r} \eta_2$ for
  distributions $\xi_1, \xi_2, \eta_1, \eta_2$ such that
  $\xi_1 \mathop{\calR}_1 \eta_1$ and $\xi_2 \mathop{\calR}_2
  \eta_2$. If $\xi \alphaarrow \xi'$, then
  $\xi' = \xi'_1 \probc{r} \xi'_2$ for distributions $\xi'_1, \xi'_2$
  with $\xi_1 \alphaarrow \xi'_1$, $\xi_2 \alphaarrow \xi'_2$ and
  $\xi' = \xi'_1 \probc{r} \xi'_2$.
  \vspace{1pt}
  By assumption, $\etabar_1$,
  $\eta'_1$ and~$\etabar_2, \eta'_2$ exist such that
  $\eta_1 \Arrow{} \etabar_1 \alphahidearrow \eta'_1$,
  $\eta_2 \Arrow{} \etabar_2 \alphahidearrow \eta'_2$,
  $\xi_1 \mathop{\calR_1} \etabar_1$,
  $\xi_2 \mathop{\calR_2} \etabar_2$,
  $\xi'_1 \mathop{\calR_1} \eta'_1$,
  and~$\xi'_2 \mathop{\calR_2} \eta'_2$. From this we obtain for
  \vspace{1pt}
  $\etabar = \etabar_1 \probc{r} \etabar_2$,
  $\eta' = \eta'_1 \probc{r} \eta'_2$ that
  $\xi = ( \xi_1 \probc{r} \xi_2 ) \mopcalR ( \etabar_1 \probc{r}
  \etabar_2 ) = \etabar$,
  $\eta = \eta_1 \probc{r} \eta_2 \Arrow{} \etabar_1 \probc{r}
  \etabar_2 = \etabar \alphahidearrow \eta'$,
  and~$\xi' = ( \xi'_1 \probc{r} \xi'_2 ) \mopcalR ( \beta'_1
  \probc{r} \eta'_2 ) = \eta'$, as required for~$\calR$ to satisfy the
  transfer property of
  Definition~\ref{def-probabilistic-branching-bisimilar}d. Since
  $\calR$ clearly relates $\mu_1 \probc{r} \mu_2$ and $\nu_1 \probc{r}
  \nu_2$, this finishes the proof.
\end{proof}


\end{document}